\documentclass[12pt, a4paper]{book}		
\usepackage[utf8x]{inputenc}

\usepackage[T1]{fontenc}

\usepackage[english]{babel}
\usepackage{graphicx}				
\usepackage{epsf}
\usepackage{amsfonts}
\usepackage{slashed}
\usepackage{amssymb}
\usepackage{amsmath}
\usepackage{mathtools}
\usepackage{cancel}
\usepackage{amssymb}
\usepackage{titlesec}
\usepackage{titletoc}
\usepackage[titletoc]{appendix}
 \usepackage{pdfpages}
 
 \usepackage{mciteplus}
 \usepackage{tocloft}
 \usepackage{microtype}
 \usepackage{scrextend}
 
\usepackage[labelsep=period,justification=RaggedRight, tableposition=top,font=small,labelfont=bf]{caption}
\usepackage[caption=false]{subfig}

\usepackage{mciteplus}
\usepackage[comma,square,numbers,compress]{natbib}

\usepackage{chngcntr}
\usepackage{hyperref}
\hypersetup{%
  pdfauthor={Heikki M\"antysaari},%
  pdftitle={Scattering off the Color Glass Condensate},%
  pdfsubject={PhD thesis}
}

\newcommand{\der}{\mathrm{d}}  % Derivaatta-d

\newcommand{\A}{\mathcal{A}}
\newcommand{\N}{\mathcal{N}}

\newcommand{\nc}{N_\mathrm{c}}
\newcommand{\nf}{n_\mathrm{f}}
\newcommand{\cf}{C_\mathrm{f}}
\newcommand{\as}{\alpha_s}

\newcommand{\alphaem}{\alpha_\text{em}}

\newcommand{\lqcd}{\Lambda_\text{QCD}}
\newcommand{\qso}{Q_{s,0}}
\newcommand{\qs}{Q_s}

\newcommand\tr{\operatorname{Tr}}
\newcommand\PE{\operatorname{P} \exp}
\newcommand\PO{\operatorname{P}}  % Path ordering

\newcommand{\xt}{{x_T}}
\newcommand{\yt}{y_T}
\newcommand{\ut}{u_T}
\newcommand{\vt}{v_T}
\newcommand{\zt}{{z_T}}
\newcommand{\rt}{r_T}
\newcommand{\bt}{b_T} % jso vaihdat niin vaihda diffraktionotaatio myös
\newcommand{\kt}{k_T}
\newcommand{\pt}{p_T}
\newcommand{\qt}{q_T}
\newcommand{\st}{s_T}

\newcommand{\fig}{Fig.~}
\newcommand{\figs}{Figs.~}
\newcommand{\eq}{Eq.~}
\newcommand{\se}{Sec.~}

\newcommand{\paper}{paper~}
\newcommand{\papers}{papers~}
\newcommand{\re}{Ref.~}
\newcommand{\res}{Refs.~}
\newcommand{\eqs}{Eqs.~}
\newcommand{\ch}{Chapter~}
\newcommand{\chs}{Chapters~}
\newcommand{\tab}{Table~}
\newcommand{\gev}{\ \textrm{GeV}}
\newcommand{\tev}{\ \textrm{TeV}}
\newcommand{\mev}{\ \textrm{MeV}}
\newcommand{\ptrig}{{p_{T}^\textrm{trig}}}
\newcommand{\pass}{{p_{T}^\textrm{ass}}}
\newcommand{\xpom}{{x_\mathbb{P}}}

\newcommand{\ylatila}{\mbox{}\\ \mbox{}\\\mbox{}\\ \mbox{}\\ \mbox{}\\ \mbox{}\\ \mbox{}\\ \mbox{}\\ }

\begin{document}
\pagestyle{empty}

%\vspace*{-40mm}

\centerline{DEPARTMENT OF PHYSICS}
\centerline{UNIVERSITY OF JYV\"ASKYL\"A}
\centerline{RESEARCH REPORT No. 6/2015}

\vspace{25mm} 

%\centerline{\bf A UNIFIED DESCRIPTION OF SCATTERING PROCESSES }
%\centerline{\bf FROM THE COLOR GLASS CONDENSATE}

%\centerline{\bf APPLICATIONS OF THE COLOR GLASS CONDENSATE}
%\centerline{\bf TO SCATTERING PROCESSES}

%\centerline{\bf PARTICLE PRODUCTION FROM}
%\centerline{\bf THE COLOR GLASS CONDENSATE}
 
%\centerline{\bf COLOR GLASS CONDENSATE IN SCATTERING PROCESSES}

\centerline{\bf SCATTERING OFF THE COLOR GLASS CONDENSATE}

\vspace{13mm}

\centerline{\bf BY}
\centerline{\bf HEIKKI MÄNTYSAARI}

\vspace{13mm}

\centerline{Academic Dissertation}
\centerline{for the Degree of}
\centerline{Doctor of Philosophy}

\vspace{13mm}

\centerline{To be presented, by permission of the}
\centerline{Faculty of Mathematics and Natural Sciences}
\centerline{of the University of Jyv\"askyl\"a,}
\centerline{for public examination in Auditorium FYS 1 of the}
\centerline{University of Jyv\"askyl\"a on June 12th, 2015}
\centerline{at 12 o'clock noon}

\vspace{20mm}

\begin{figure}[!h]
\center
\includegraphics[height=27mm]{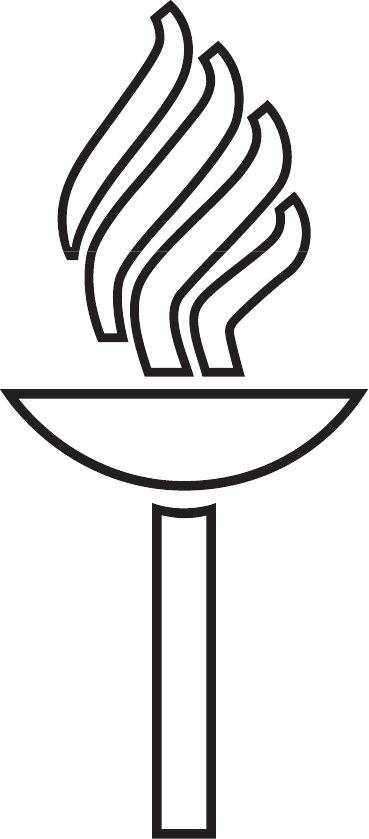}
\end{figure}

%\centerline{\picture{19mm}{50mm}{bwteksti.EPS}}

\centerline{Jyväskylä, Finland}
\centerline{June 2015}

\pagebreak

%DRAFT \today

\frontmatter
% näytä sivunumerointi myös tyhjillä sivuilla
\KOMAoptions{cleardoublepage=plain}

\chapter*{Preface}

The work presented in this thesis has been carried out from 2012 to 2015 at the Department of Physics of the University of Jyväskylä, and is partially based on research done as an MSc student at the same department in 2010--2011.

This work could not be done entirely by myself, and I want to give credit to people who really deserve it. First of all I want to express my gratitude to  Dr. Tuomas Lappi for excellent and friendly supervision and guidance during all these years. Tuomas, it has been a privilege to work with you!

Secondly, I wish to thank Prof. Kari J. Eskola for teaching me the secrets of quantum mechanics and particle physics, and taking me into his excellent research group. Thanks for all the support I have got.

My friends and colleagues  have created a friendly atmosphere in the Physics department and have made these years in Jyväskylä so memorable and enjoyable. There are too many of you who would deserve to be mentioned here. In particular I want to thank my previous and current fellow PhD students, including Dr. Risto Paatelainen, Dr. Ilkka Helenius, Mr. Jarkko Peuron, and Ms. Andrecia Ramnath, and everyone who has shared FL347 with me. The office personnel, especially the former department coordinator Ms. Soili Leskinen, deserve credit for running the bureaucracy very smoothly. I also express my thanks to Prof. Raju Venugopalan and Dr. Bertrand Ducloué for fruitful collaboration.

I thank Prof. Nestor Armesto and Dr. Cyrille Marquet for reviewing the manuscript and providing useful comments and Prof. Jamal Jalilian-Marian for promising to be my opponent. Financial support from the Graduate School of Particle and Nuclear Physics (PANU) and the Academy of Finland are gratefully acknowledged.

Finally and most importantly I want to express my deepest gratitude to my family, and especially to Kaisa, for love and support.

\vfill
\noindent Jyväskylä, June 2015 \\
Heikki Mäntysaari

%\newpage

%\begin{titlepage}
%\cleardoublepage

\chapter*{Abstract}

In this thesis the Color Glass Condensate (CGC) framework, which describes  quantum chromodynamics (QCD) at high energy, is applied  to  various scattering processes. Higher order corrections to the CGC evolution equations, known as the BK and JIMWLK equations, are also considered.

It is shown that the leading order CGC calculations describe the experimental data from electron-proton deep inelastic scattering (DIS), proton-proton and proton-nucleus collisions. The initial condition for the BK evolution equation is obtained by performing a fit to deep inelastic scattering data. The fit result is used as an input to calculations of single particle spectra and nuclear suppression in proton-proton and proton-nucleus collisions, which are shown to be in agreement with  RHIC and LHC measurements. In particular, the importance of a proper description of the nuclear geometry consistently with the DIS data fits is emphasized, as it results in a nuclear suppression factor $R_{pA}$ which is consistent with the available experimental data.

In addition to single particle production, the correlations between two hadrons at forward rapidity are computed. The RHIC measurements are shown to be naturally explainable in the CGC framework, and the previous CGC calculations are improved by including the so called inelastic and double parton scattering contributions. This improvement is shown to be required in order to get results compatible with the experimentally measured correlations.

Exclusive vector meson production, which can be a powerful tool to study the gluonic structure of nuclei at small Bjorken-$x$, is also considered. The cross sections are calculated within the CGC framework in the context of a future electron-ion collider. In particular, the cross section for incoherent diffractive vector meson production is derived and a centrality estimator for this process is proposed. Exclusive processes are also studied in ultraperipheral heavy ion collisions.

\newpage
\thispagestyle{plain}

\ylatila

\begin{tabular}{ll}

{\bf Author} & Heikki Mäntysaari \\ 

 & Department of Physics \\ 

 & University of Jyväskylä \\ 

 & Finland \\
 
 & \\ 

{\bf Supervisor} & Dr. Tuomas Lappi \\ 

 & Department of Physics \\ 
 
 & University of Jyväskylä \\ 
 
 & Finland \\ 
 & \\
 
{\bf Reviewers}  & Prof. Nestor Armesto \\ 
 
 & Departamento de F\'\i sica de Part\'\i culas \\ 
 
 & Universidade de Santiago de Compostela \\ 
 
 & Spain \\ 
 
 &  \\ 

& Dr. Cyrille Marquet \\ 
 
 & Centre de Physique Theorique \\ 
 
 & Ecole Polytechnique \\ 
 
 & France \\ 
 
 &  \\

{\bf Opponent} & Prof. Jamal Jalilian-Marian \\ 
 
 & Department of Natural Sciences \\ 
 
 & Baruch College, City University of New York \\ 
 
 & USA \\ 
 
\end{tabular} 

%\end{titlepage}

%\setcounter{page}{1}
%\pagenumbering{roman}

\chapter*{List of Publications}
\vspace*{-0.15cm}
This thesis consists of an introductory part and the following
publications:

\begin{itemize}
\item[{\bf \cite{Lappi:2010dd}}]	\textbf{T. Lappi and H. Mäntysaari}  \href{http://link.aps.org/doi/10.1103/PhysRevC.83.065202}{\textit{Phys. Rev. C.} \textbf{83} (2011) 065202}, \href{http://arxiv.org/abs/1011.1988}{\texttt{arXiv:1011.1988 [hep-ph]}}: \emph{Incoherent diffractive $J/\Psi$ production in high-energy nuclear deep-inelastic scattering}.

\item[{\bf \cite{Lappi:2012nh}}]	\textbf{T. Lappi and H. Mäntysaari}  \href{http://dx.doi.org/10.1016/j.nuclphysa.2013.03.017}{\textit{Nucl. Phys.} \textbf{A908} (2013) 51-72}, \href{http://arxiv.org/abs/1209.2853}{
	 \texttt{arXiv:1209.2853 [hep-ph]}}: \emph{Forward dihadron correlations in deuteron-gold collisions with a Gaussian approximation of JIMWLK}.  
	 
\item[{\bf \cite{Lappi:2012vw}}]	\textbf{T. Lappi and H. Mäntysaari}
	\href{http://dx.doi.org/10.1140/epjc/s10052-013-2307-z}{\textit{Eur. Phys. J.} \textbf{C73} (2013) 2307}, 
	\href{http://arxiv.org/abs/arXiv:1212.4825}{\texttt{arXiv:1212.4825 [hep-ph]}}: \emph{On the running coupling in the JIMWLK equation}.	
	
\item[{\bf \cite{Lappi:2013am}}] \textbf{T. Lappi and H. Mäntysaari} 
	 \href{http://dx.doi.org/10.1103/PhysRevC.87.032201}{\textit{Phys. Rev. C.} \textbf{87} (2013) 032201}, \href{http://arxiv.org/abs/1301.4095}{\texttt{arXiv:1301.4095 [hep-ph]}}: \emph{$J/\Psi$ production in ultraperipheral Pb+Pb and p+Pb collisions at energies available at the CERN Large Hadron Collider}. 

\item[{\bf \cite{Lappi:2013zma}}] 	\textbf{T. Lappi and H. Mäntysaari} 
	\href{http://dx.doi.org/10.1103/PhysRevD.88.114020}{\textit{Phys. Rev. D.} \textbf{88} (2013) 114020},  \href{http://arxiv.org/abs/arXiv:1309.6963}{\texttt{arXiv:1309.6963 [hep-ph]}}: \emph{Single inclusive particle production at high energy from HERA data to proton-nucleus collisions}.

\item [{\bf \cite{Lappi:2014foa}}] \textbf{T. Lappi, H. Mäntysaari and R. Venugopalan} \href{http://dx.doi.org/10.1103/PhysRevLett.114.082301}{\textit{Phys. Rev. Lett. 114 (2015) 082301}},
	\href{http://arxiv.org/abs/arXiv:1411.0887}{\texttt{arXiv:1411.0887 [hep-ph]}}: \emph{Ballistic protons in Incoherent Exclusive Vector Meson Production as a Measure of Rare Parton Fluctuations at an Electron-Ion Collider}.

\item [{\bf \cite{Lappi:2015fma}}] \textbf{T. Lappi and H. Mäntysaari} 
	\href{http://dx.doi.org/10.1103/PhysRevD.91.074016}{\textit{Phys. Rev. D.} \textbf{91} (2015) 074016},	
	\href{http://arxiv.org/abs/arXiv:1502.02400}{\texttt{arXiv:1502.02400 [hep-ph]}}: \emph{Direct numerical solution of the coordinate space Balitsky-Kovchegov equation at next-to-leading order}.

\end{itemize}

The author has performed all numerical calculations in all publications except in \paper\cite{Lappi:2012vw} where the author did the BK reference calculations.
% \cite{Lappi:2010dd,Lappi:2012nh, Lappi:2013am,Lappi:2013zma, Lappi:2014foa,Lappi:2015fma} using the codes written in scratch and performed the BK reference calculations for paper \cite{Lappi:2012vw}.
The author has written the original drafts of the manuscripts for papers \cite{Lappi:2013am,Lappi:2013zma,Lappi:2015fma}.

\pagebreak

%\newpage
%\phantom{Tyhjaa}
%\newpage

%\pagenumbering{arabic}

\cleardoublepage

% Ei sivunumeroa viimeiselle tyhjälle sivulle sisällysluettelon jälkeen
\KOMAoptions{cleardoublepage=empty}
\setcounter{tocdepth}{1}
\vspace*{-1.9cm}	% Tämä tarvitaan kun paperit on mukana jotta mahtuu 1 sivulle
%\vspace*{-1.1cm}	% tämä verkkoversioon, kun ei ole papereita
\tableofcontents
\pagestyle{plain}

\mainmatter

\pagenumbering{arabic}
\chapter{Introduction}
\label{intro}

Quantum Chromodynamics  (QCD) is a theory describing the strong interactions between the constituents of hadrons, called quarks and gluons. Thanks to thorough experimental tests in a variety of collider experiments, QCD has been established as the right theory to describe the structure of matter. QCD has, however, turned out to be extremely difficult to solve in many situations and many interesting questions related to strong interactions remain unanswered. 

Experimentally the QCD dynamics can be studied in many scattering processes. For example, the partonic structure of a proton has been studied precisely in electron/positron-proton collisions at the DESY-HERA accelerator in Germany. In current particle accelerators, such as Relativistic Heavy Ion Collider (RHIC) at Brookhaven, USA and the Large Hadron Collider (LHC) at CERN in France and Switzerland, QCD plays a dual role: it can be the subject of the research efforts itself or the QCD processes may be a large source of background reactions for other measurements (such as Higgs boson production) that must be subtracted.

When colliding the heavy nuclei with each other QCD predicts~\cite{Collins:1974ky} that a new state of matter called the quark-qluon plasma (QGP) is formed. The goal of the heavy-ion program at RHIC and at the LHC is to study the properties of this QCD matter in order to probe the details of QCD dynamics. These collisions can be described using relativistic hydrodynamics, but the initial condition for these simulations must be obtained using another approach.

Atomic nuclei are more complex objects than a collection of protons and neutrons. For example, the partonic structure of the nucleus 
is not just an incoherent superposition of the bound nucleons. 
This makes the interpretation of the heavy ion collisions very challenging, as one has to take into account both \emph{initial state} (the structure of the high-energy nucleus) and \emph{final state} (the formation of the QCD matter and the interaction of particles with the plasma) effects.

In order to probe separately the initial state effects, the RHIC and the LHC experiments have performed proton-nucleus collisions in which one may not expect QGP formation. Even though these experiments were initially designed as reference measurements, several unexpected phenomena have been observed during the past couple of years. These include signs of collectivity  for example in rapidity and azimuthal angle correlations in two-particle production~\cite{Abelev:2012ola,Aad:2013fja,Chatrchyan:2013nka}.

%In this Thesis an effective theory of QCD, called the Color Glass Condensate (CGC), which is applicable at high energy is discussed. 
In this thesis an effective theory of high-energy QCD, the Color Glass Condensate (CGC), is discussed. 
In \ch \ref{ch:cgc} we summarize how high-energy scattering processes, where strong color fields are relevant, can be described in QCD using the CGC. The framework is then applied in \ch\ref{ch:dis} to describe deep inelastic lepton-hadron scattering (DIS) and non-perturbative input for the phenomenological applications of the framework is obtained. The CGC model results for exclusive vector meson production are presented in \ch\ref{ch:ddis}. Finally, in \chs\ref{ch:sinc} and \ref{ch:dihad} we present calculations for single and double inclusive particle production in proton-proton and proton-nucleus collisions and give an outlook for future directions of the research in \ch\ref{ch:outlook}.

\chapter{The Color Glass Condensate}
\label{ch:cgc}

\section{QCD at high energy}

The partonic structure of the proton can be studied, for example, in lepton-proton scattering experiments like deep inelastic scattering (as discussed in more detail in \ch \ref{ch:dis}). The most precise measurements of the quark and gluon structure of the proton come from the HERA particle accelerator, which collided electrons and positrons with protons in 1992--2007.

What has been seen in these experiments is that the proton looks very different depending on at which scale it is measured. 
When the proton structure is probed with a photon that has a long wavelength compared with the proton size, a charged particle with electric charge $+1e$ is seen. The inner structure of the proton becomes visible when the photon wavelength is decreased to be of the order of the proton radius. First one observes three valence quarks having a fractional electric charge and carrying a fraction $\sim \frac{1}{3}$ of the proton longitudinal momentum, viewed in the frame where the proton energy is very large.

\begin{figure}[tb]
\begin{center}
%\begin{minipage}{0.45\textwidth}
\includegraphics[width=0.7\textwidth]{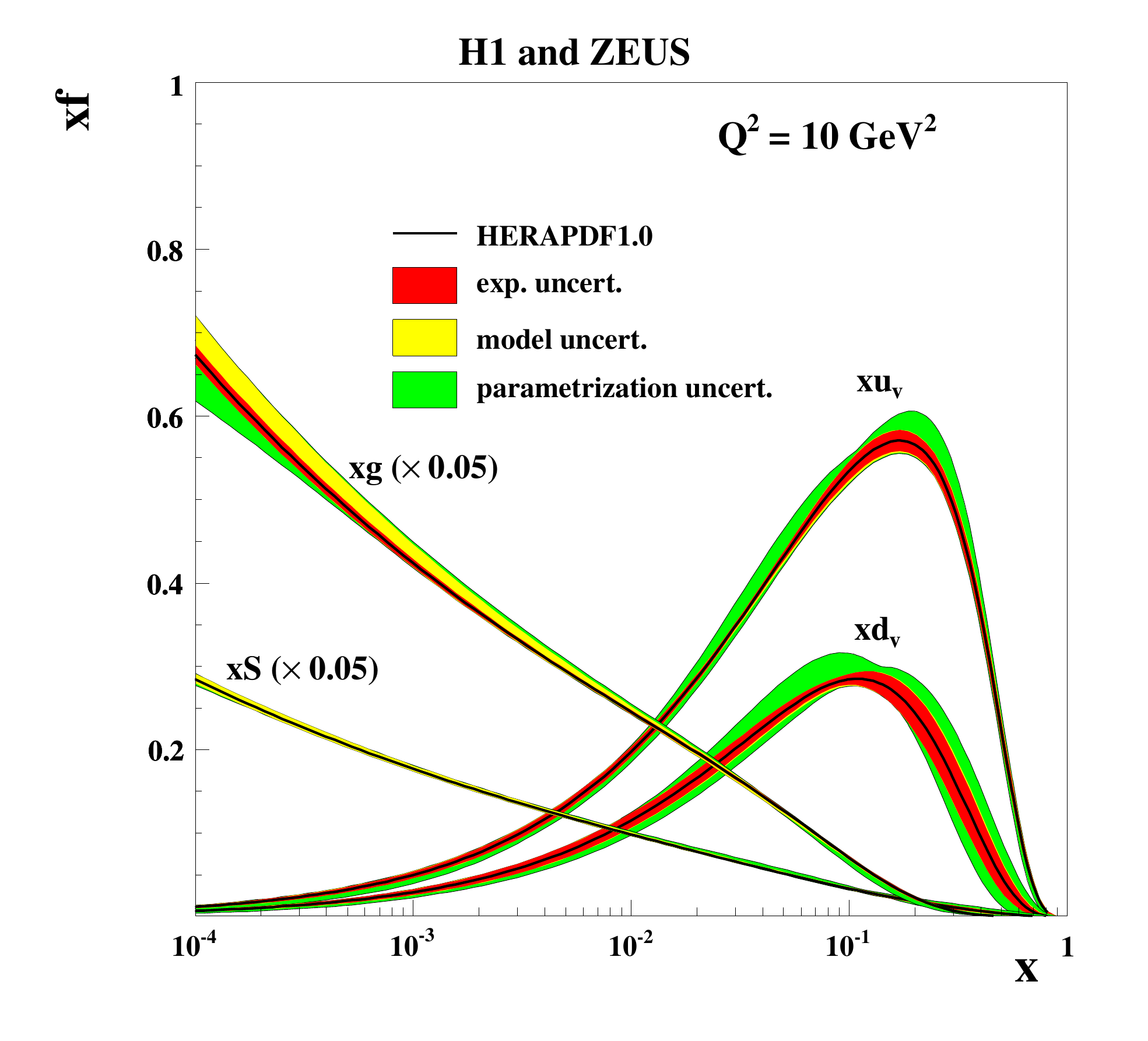}
\caption{The parton density of the proton extracted from the HERA data. $u_v$ and $d_v$ are u and d quark densities, $S$ is the sea quark and $g$ the gluon density. Figure from \re \cite{Aaron:2009aa}.}
\label{fig:herapdf}
\end{center}
\end{figure}

When the wavelength of the photon is decreased more (or the virtuality $Q^2$ is increased), a richer structure becomes visible.
The photon starts to see a large number of sea quarks and antiquarks
%The quark density increases, as a large number of sea quarks becomes visible. 
that carry a small fraction of the proton longitudinal momentum, denoted by Bjorken-$x$. These quarks originate from gluon splittings to quark-antiquark pairs, which makes it possible to also determine the distribution of electrically neutral gluons that can not be directly probed with a photon. When the proton structure is measured at smaller and smaller $x$, more and more sea quarks (and thus, gluons) are seen. The extracted quark and gluon densities from the HERA lepton-proton data~\cite{Aaron:2009aa} are shown in \fig \ref{fig:herapdf}.

%\end{minipage}
%\begin{minipage}{0.45\textwidth}
\begin{figure}[tb]
\begin{center}
\includegraphics[width=0.5\textwidth]{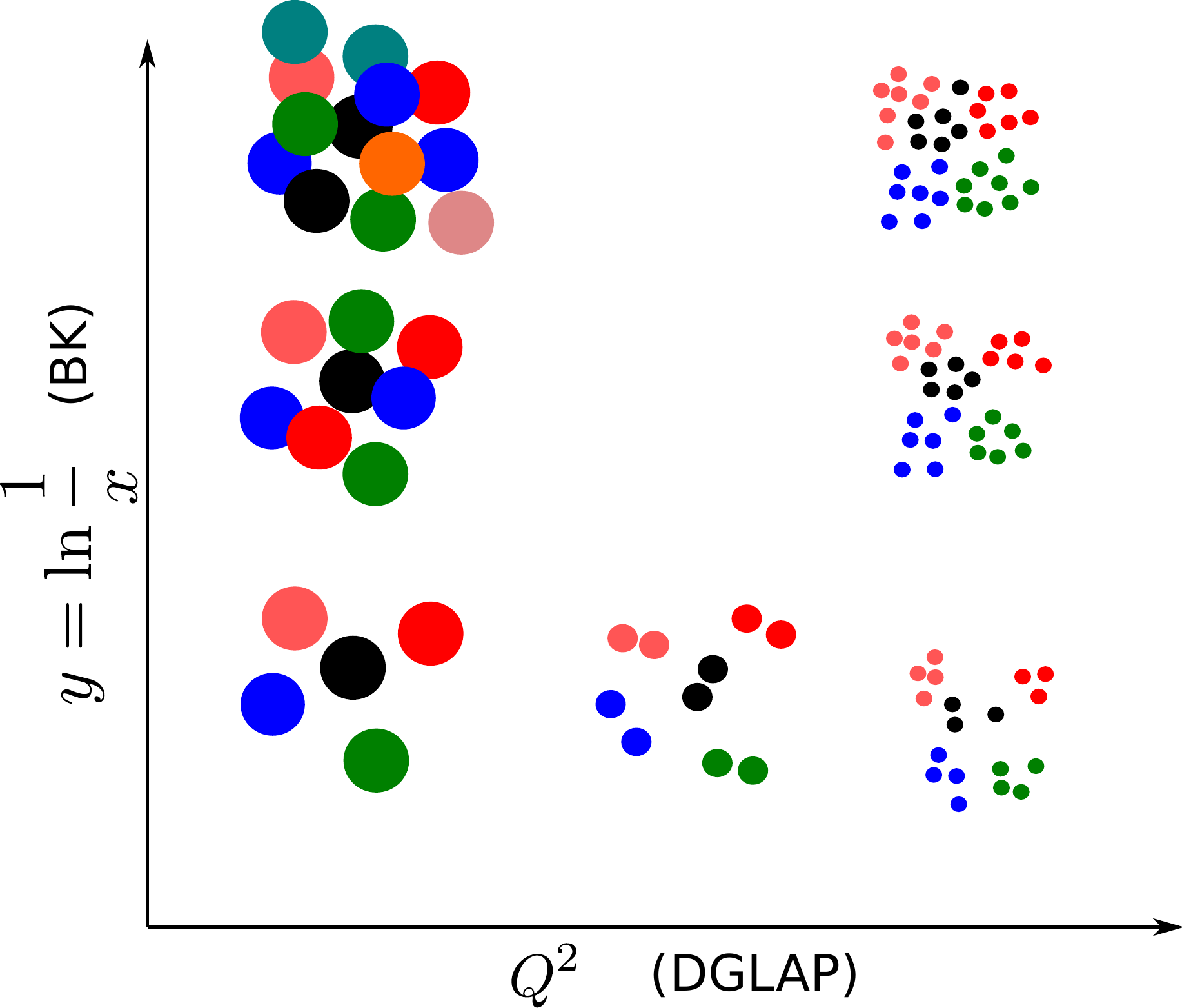}
\caption{Evolution of the gluonic structure of the proton.}
\label{fig:partons}
%\end{minipage}
\end{center}
\end{figure}

The large gluon densities at small $x$ are expected, as the QCD splitting functions for the emission of a soft gluon from a quark or a gluon have a singularity in the limit where the gluon momentum fraction vanishes. 
This evolution, in the linear regime where gluon densities are not very large, is driven by the Balitsky-Fadin-Kuraev-Lipatov (BKFL) equation~\cite{Kuraev:1977fs,Balitsky:1978ic}.
On the other hand, this growth towards small momentum fractions can not continue indefinitely without breaking the unitarity of the theory, because the cross section (or interaction probability) increases when the parton density increases and the total interaction probability is limited by unity. This limit is the so called Froissart bound~\cite{Froissart:1961ux}. The growth of the gluon density can be limited by other partonic processes. Namely, when the gluon densities become sufficiently large (of the order of the inverse strong coupling constant $1/\as$), in an appropriate gauge, the gluon recombination processes $gg \to g$ become important as the probability for the recombination is proportional to $g^2\sim \as$. This phenomenon is called the \emph{saturation} of gluon distribution, which takes place at small $x$. The scale at which these non-linear effects tame the growth of the gluon density is referred to  the \emph{saturation scale} $Q_s(x)$.

As an illustration let us consider the evolution of the gluonic structure of the proton as shown in \fig \ref{fig:partons} where the colorful circles represent gluons. The apparent size of the gluons is set by the scale at which they are probed, given by $1/Q^2$, where $Q^2$ is the virtuality of the probe. The change of the gluon density as a function of the probe virtuality can be computed from QCD using perturbative techniques. When $Q^2$ is increased, more gluons are seen, but as their apparent size is also smaller, the proton remains \emph{dilute}. The evolution equations describing the $Q^2$ evolution are known as the Dokshitzer-Gribov-Lipatov-Altarelli-Parisi (DGLAP) equations~\cite{Gribov:1972ri,Gribov:1972rt,
Altarelli:1977zs,Dokshitzer:1977sg}. 

On the other hand, when $Q^2$ is kept fixed and the longitudinal momentum fraction of the gluon (denoted by $x$) is decreased, more and more soft gluons are seen as discussed earlier. As the apparent size of the gluons remains the same, at some point the gluons start to overlap. Eventually the gluon fusion $gg\to g$ starts to compensate the gluon splitting $g \to gg$, and the saturation regime has been reached. This evolution in $x$  is given by the Balitsky-Kovchegov equation~\cite{Balitsky:1995ub,Kovchegov:1999yj} discussed in \se\ref{sec:bk}. As we will show in \ch \ref{ch:dis}, in electron-proton scattering the Bjorken-$x$ probed goes like $\sim 1/s$, where $s$ is the center-of-mass energy of the process squared, and the saturation effects should manifest themselves in scattering experiments at sufficiently high energy. 
What is argued in this thesis is that the experimental data from many different scattering processes is consistent with the saturation picture, suggesting that the energies available at HERA, RHIC and the LHC are large enough for the gluon saturation effects to be visible.

When the proton is replaced by a heavy nucleus with mass number $A$, and the nucleus is accelerated to high energy, due to the Lorentz contraction there will be $\sim A^{1/3}$ overlapping nucleons. Thus the gluon densities probed at same momentum fraction $x$ are considerably different between the proton and the nucleus, and the nonlinear phenomena should be visible at larger $x$. In other words, the saturation scale of the nucleus is larger than that of the proton at the same $x$. In scattering experiments with nuclei saturation phenomena should then be visible at lower center-of-mass energies compared to collisions between protons, which makes proton-nucleus and electron-nucleus experiments very interesting for saturation physics.

The saturation phenomena are taken into account very naturally within the Color Glass Condensate framework, which is the subject of this thesis and presented in more detail in \se\ref{sec:cgc-intro}. It should however be noted that it is not the only possibility to describe saturation physics. For example, in \re\cite{Paatelainen:2012at} an initial condition for hydrodynamical description of heavy ion collisions is calculated from perturbative QCD using collinear factorization and including saturation effects. Note that the different views of saturation are not automatically excluding each other, as CGC is an approximation of QCD and other QCD based calculations can also encapsulate the same physics.

\section{QCD on the light cone}
\label{sec:lightcone}

In high-energy scattering processes the particles have velocities close to the speed of light and travel along the positive and negative light cone axes defined as
\begin{equation}
 	x^\pm = \frac{1}{\sqrt{2}} \left(x^0 \pm x^3\right).
 \end{equation} 
Describing the high-energy limit of scattering processes in a quantum field theory becomes easier when the field theory is written using the light cone coordinates $(x^+,x^-,\xt)$, where the inner product is $u\cdot v = u^+v^- + u^-v^+ - \ut \cdot \vt$. For a detailed discussion of the quantization of field theories on the light cone we refer the reader to \re\cite{Brodsky:1997de}, and a more pedagogical description can be found e.g. in \re\cite{Kovchegov:2012mbw}.

Let us first consider a single quark propagating along the light cone with momentum $p$, color $i$ and spin $\alpha$. This quark state can be created by operating on the vacuum by a creation operator $b_{i,\alpha}^\dagger(p)$ as
\begin{equation}
\label{eq:incoming}
	|\text{in}\rangle = \N b_{i,\alpha}^\dagger(p)|0\rangle = \N |p, i, \alpha\rangle.
\end{equation}
The normalization factor $\N$ sets the correct normalization for the single particle states. The fermionic operators anticommute as
\begin{equation}
\label{eq:b-anticom}
	\{b_{i,\alpha}(p),b_{j,\beta}^\dagger(k)\} = \delta_{ij} \delta_{\alpha \beta} \delta^{(3)}(p-k),
\end{equation}
and $\delta^{(3)}(p-k) = \delta(p^+-k^+)\delta^{(2)}(\pt-\kt)$. Fourier transforming the creation operator to transverse coordinate space we obtain
\begin{equation}
	b_{i,\alpha}^\dagger(p^+,\xt) = \int \der^2 \bt e^{-i \pt \cdot \xt} b^\dagger_{i,\alpha}(p^+,\pt),
\end{equation}
which, inverted, gives
\begin{equation}
	b_{i,\alpha}^\dagger(p^+,\pt) = \int \frac{\der^2 \xt}{(2\pi)^2} e^{i \pt \cdot \xt} b_{i,\alpha}^\dagger(p^+,\xt).
\end{equation}
Note that in the mixed transverse coordinate-longitudinal momentum space one obtains
\begin{equation}
	\label{eq:fermion_anticommutator_coordinate}
	\{b_{i,\alpha}(p^+,\xt),b_{j,\beta}^\dagger(k^+,\yt)\} = (2\pi)^2 \delta_{ij} \delta_{\alpha \beta} \delta^{}(p^+-k^+)\delta^{(2)}(\xt-\yt).
\end{equation}
To evaluate the normalization constant $\N$ we require that the incoming free particle state \eqref{eq:incoming} is normalized to unity. Using the anticommutator \eqref{eq:b-anticom} we obtain
\begin{equation}
	\langle \text{in} | \text{in} \rangle = \N^2 \langle  k, i, \alpha |  k, i, \alpha \rangle  = \N^2 \delta^{(3)}(0) = \N^2 \int \frac{\der^3 x}{(2\pi)^3} e^{i x \cdot 0} = \N^2 \frac{S_T L^-}{(2\pi)^3},
\end{equation}
where $L^- = \int \der x^-$ is the size of the box in the $x^-$ direction and $S_T = \int \der^2 \xt$ is the size of the transverse space, and we have chosen $\N$ to be real. These, in principle infinite, factors are an artefact from the usage of plane waves instead of finite-size wave packets. As we require that the single quark states are normalized to unity, we get
\begin{equation}
\label{eq:state-norm}
	\N = \sqrt{ \frac{(2\pi)^3}{S_T L^-} }.
\end{equation}
The correct normalization for the single particle states is essential e.g. when calculating the single inclusive cross section in \se\ref{sec:hybrid}.

An important quantity in light cone perturbation theory is the light cone wave function which is used to expand a state as a superposition of Fock states. For example, when considering the photon-hadron interaction in case of deep inelastic scattering (see \ch\ref{ch:dis}) the photon state is written as
\begin{equation}
	|\gamma^*\rangle = |\gamma^*\rangle_0 + \Psi^{\gamma^*\to q\bar q} |q\bar q\rangle_0 + \dots
\end{equation}
Here the subscript $0$ refers to non-interacting theory states and we neglect states like $|q\bar q g\rangle$ which would be higher order in QCD coupling $\as$ (or QED coupling $\alphaem)$. Applying the Feynman rules of the light cone perturbation theory (LCPT) from \re\cite{Brodsky:1997de} 
one can obtain the transversally polarized photon wave function
\begin{multline}
\label{eq:photon-wf}
	\Psi^{\gamma^*\to q\bar q}_{ss'} = 
 	\frac{e_f e}{\sqrt{(2\pi)^3}} \Big[
		i \sqrt{2}  \frac{\varepsilon^{\pm 1} \cdot \rt}{|\rt|} K_1(\epsilon r_T) ( z \delta_{s, \mp 1} - (1-z) \delta_{s, \pm 1})\delta_{s,-s'} \\
		 + m_f K_0(\epsilon r_T) \delta_{s, \pm 1}\delta_{s', \pm 1} \Big].
\end{multline}
Similarly for the longitudinal photon the wave function is
\begin{equation}
	\label{eq:photon-wf-l}
	\Psi^{\gamma^*\to q\bar q}_{ss'} = \frac{-e_f e}{2\pi \sqrt{\pi}} Q z(1-z) K_0(\epsilon r_T) \delta_{s,-s'}.
\end{equation}
Here $z$ is the longitudinal momentum fraction of the photon carried by the quark and $\rt$ is the transverse separation of the quarks. The spins of the quark and antiquark are denoted by $s$ and $s'$, $e_f$ is the fractional charge of the quark, $\varepsilon$ is the polarization vector and $\epsilon=\sqrt{Q^2z(1-z)+m_f^2}$ with $m_f$ being the quark mass. This wave function is needed when the total virtual photon-proton cross section is calculated in \se\ref{sec:dis-dipole} when considering deep inelastic scattering. Detailed derivations for these results are shown e.g. in \res \cite{Kovchegov:2012mbw,gradu}.

\section{CGC as an effective field theory}
\label{sec:cgc-intro}
The Color Glass Condensate (CGC) is an effective field theory that describes the QCD at the high-energy limit where gluon densities are assumed to be so large that they correspond to strong classical color fields. For a review of the CGC, see e.g. \res \cite{Iancu:2003xm,Gelis:2010nm,Albacete:2014fwa}. 
As the color fields are strong, of the order of $A \sim 1/g$, terms of the order $g A$ must be resummed to all orders. 
The evolution of the color fields as a function of energy (or Bjorken-$x$) is obtained by calculating the quantum corrections to the classical fields via non-linear CGC evolution equations, known as the JIMWLK and BK equations, that include all $(gA)^n$ corrections. These equations are discussed in more detail in \se\ref{sec:evolution}. When considering scattering processes, the $(gA)^n$ contributions to the scattering amplitude are summed using the Wilson lines as discussed shortly.

\begin{figure}[tb]
\begin{center}
\includegraphics[width=0.5\textwidth]{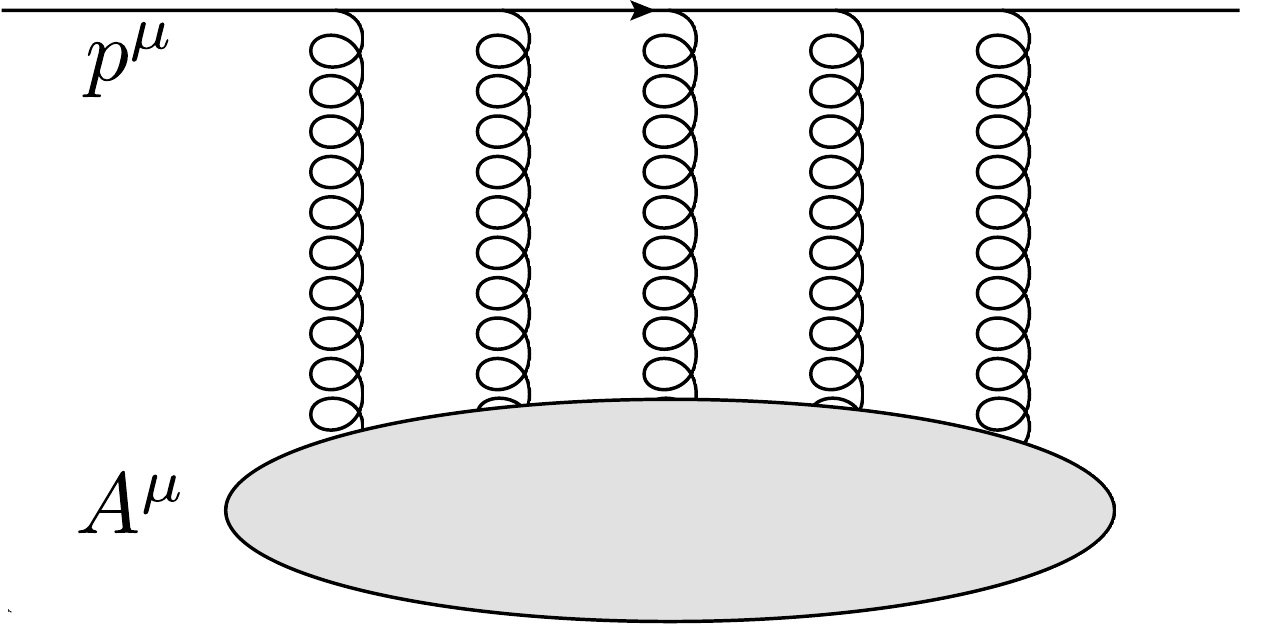}
\caption{A quark with momentum $p^\mu$ scattering multiple times off the color field $A^\mu$. }
\label{fig:multiple_scattering}
\end{center}
\end{figure}

In the CGC picture the most convenient degrees of freedom are not quarks and gluons. To understand this, let us consider a quark moving in positive $z$ direction at high energy with momentum $p$ and scattering off a target consisting of a CGC (that is, off a strong color field, for example a heavy nucleus) as shown in \fig\ref{fig:multiple_scattering}. The quark scatters multiple times when propagating through the target, but at high energy the transverse position of the quark can be considered to be fixed. This can be seen by noticing that the change of the quark transverse position during the interaction is $\Delta \xt \sim R k_T/E$, where $k_T$ is the transverse momentum obtained by the quark during the interactions with the target, $E$ is the energy of the quark in the target rest frame and $R$ is the longitudinal size of the target. Thus, the change of the quark position in the transverse plane is suppressed by the large energy $E$. This is known as an eikonal approximation.

Now assuming that the quark momentum $p$ has very large plus component, the quark-gluon vertex is proportional to $g \bar u(p) \gamma^\mu u(p-q)\approx  g \bar u(p) \gamma^\mu u(p)$. This is a Lorentz 4-vector, and as the only available vector in the problem is $p^\mu$, we must have the quark-gluon vertex proportional to $p^\mu$ in this limit (the same result can be obtained more formally by using the Gordon identity). When calculating the quark scattering off the strong color field $A$, the vertex is contracted with the field $A^\mu$. Thus, the scattering amplitude is proportional to $p^\mu A_\mu$. Moreover, as $p^\mu$ has only one large component, namely $p^+$, we get $p^\mu A_\mu = p^+ A^-$, and we expect that only the minus component of the color field is required to describe the scattering. From now on we work in the light one gauge where $A^+=0$.

Let us then solve $A^-$ from the Dirac equation
\begin{equation}
\label{eq:dirac}
	(i \slashed \partial - g \slashed A(x))\psi(x)=0
\end{equation}
using an ansatz that the solution can be written as a free particle solution multiplied by a factor ($\nc \times \nc$ matrix) $V(x)$ as
\begin{equation}
	\label{eq:ansatz-de-wilson}
	\psi(x) = V(x) e^{-i p \cdot x} u(p) .
\end{equation}
Substituting this ansatz into the equation \eqref{eq:dirac} we obtain
\begin{equation}
	\gamma_\mu \left[i \partial^\mu V(x) + p^\mu V(x) - g A^\mu(x) V(x)\right] u(p)=0. 
\end{equation}
The term in the brackets must vanish with all $\mu$. To see this, note that if $a^\mu \gamma_\mu=0$, then we can write $a^\mu \gamma_\nu \gamma_\mu =0$, and using the anticommutator of the gamma matrices we get $2 a_\nu - a^\mu \gamma_\mu \gamma_\nu=0$, where the second term vanishes giving $a_\nu=0$. Especially we get
\begin{equation}
\label{eq:wilson-line-evolution}
	\partial_+ V(x) = - ig A^-(x) V(x),
\end{equation}
where we used the fact that $\partial^- = \partial_+$ and that $p^- \approx 0$.

The solution to this differential equation is an exponential function. As $V(x)$ is a matrix, we notice that the equation is solved with a path ordered expansion
\begin{multline}
\label{eq:wilson-line-expanded}
	V(x^+,x^-,x_T) = 1 - ig \int^{x^+}_{-\infty} \der z^+ A^-(z^+,x^-,x_T) \\
	 +  \frac{(-ig)^2}{2!} \int^{x^+}_{-\infty} \der z^+ \int^{x^+}_{-\infty} \der z'^+ \PO \left[A^-(z'^+, x^-, x_T) A^-(z^+, x^-, x_T) \right] + \dots \\
 \end{multline}
The path ordering $\PO$ is defined such that the non-commuting fields $A^-$ (that are $\nc \times \nc$ matrices) are arranged according to their $x^+$ component. In order to see more clearly that \eqref{eq:wilson-line-expanded} solves \eq\eqref{eq:wilson-line-evolution} we explicitly calculate the second order term:
\begin{multline}
\label{eq:po-lasku}
	\frac{(-ig)^2}{2!}\int^{x^+}_{-\infty} \der z^+ \int^{x^+}_{-\infty} \der z'^+ \PO \left[A^-(z'^+, x^-, x_T) A^-(z^+, x^-, x_T) \right] \\
	= \frac{(-ig)^2}{2} \int_{-\infty}^{x^+} \der z^+ \int_{-\infty}^{z^+} \der z'^+ A^-(z^+,x^-,\xt)A^-(z'^+,x^-,\xt)  \\
	+ \frac{(-ig)^2}{2}\int_{-\infty}^{x^+} \der z'^+ \int_{-\infty}^{z'^+} \der z^+ A^-(z'^+,x^-,\xt)A^-(z^+,x^-,\xt) \\
	=  (-ig)^2\int_{-\infty}^{x^+} \der z^+ \int_{-\infty}^{z^+} \der z'^+ A^-(z^+,x^-,\xt)A^-(z'^+,x^-,\xt),
\end{multline}
where we have changed variables $z\leftrightarrow z'$ in the second integral. Differentiating \eq\eqref{eq:po-lasku} with respect to $x^+$ gives the previous term in \eq\eqref{eq:wilson-line-expanded} multiplied by $-igA^-(x)$. This procedure generalizes to higher order terms, and the solution \eqref{eq:wilson-line-expanded} becomes
\begin{equation}
	V(x^+,x^-,\xt) = \PE \left[ -ig \int^{x^+}_{-\infty} \der z^+ A^-(z^+, x^-, x_T) \right]
\end{equation}

Consider then a scattering problem where the incoming quark comes from negative infinity and is measured far away from the target where $x^+=\infty$. From \eq \eqref{eq:ansatz-de-wilson} we observe that as the particle propagates through the target, it acquires a phase
\begin{equation}
\label{eq:wilson-line}
	V(x_T) = \PE  \left[-ig \int_{-\infty}^\infty \der z^+ A^-(z^+, x^-=0, x_T) \right],
\end{equation}
called \emph{Wilson line}. Here we have set $x^-=0$, as at high energy the particle approximatively propagates along the $x^+$ light cone, or equivalently, the wave function oscillates as $e^{ip^+x^-}$, and as $p^+$ is large, the wave function averages to zero if $x^-$ is not approximately zero. For a similar discussion in case of QED, see \re\cite{Bjorken:1970ah}. 
Note that the Wilson line sums all powers of $gA^-$, which can be interpreted as allowing the quark to scatter any number of times off the target, and the Wilson line resums these multiple scatterings. 

Let us then consider a process where a color neutral quark-antiquark dipole (with quark color $i$ and transverse coordinates $\xt$ and $\yt$) scatters off the target (strong color field). The incoming state averaged over colors is 
\begin{equation}
	|\text{in}\rangle = \frac{1}{\nc} |q_i(\xt) \bar q_i(\yt)\rangle,
\end{equation}
and when the quarks have picked up the phase factors while propagating through the target, the outgoing state is
\begin{equation}
	|\text{out}\rangle = \frac{1}{\nc} V_{ij} V^\dagger_{j'i} |q_{j}(\xt) \bar q_{j'}(\yt)\rangle.
\end{equation}
Note that we are implicitly summing over repeated indices. Let us then compute the forward elastic scattering amplitude by counting the number of color neutral dipoles in the outgoing state:
\begin{equation}
	S = \langle q_k(\xt) \bar q_k(\yt)|\text{out}\rangle = \frac{1}{\nc} V_{ij}(\xt) V^\dagger_{j'i}(\yt) \delta_{jk} \delta_{j'k} = \frac{1}{\nc} \tr V(\xt) V^\dagger(\yt).
\end{equation}
Note that, using the optical theorem, the forward elastic scattering matrix $S$ can be used to calculate the total cross section. As the scattering matrix also contains the situation when nothing happens, it is useful to define the dipole amplitude
\begin{equation}
\label{eq:nqq}
	N_{q\bar q} = 1 - \frac{1}{\nc} \tr V(\xt) V^\dagger(\yt),
\end{equation}
which now includes all information about the interactions with the target. In practice the color field of the target off which the dipole is scattering is not known, so the trace of the two Wilson lines must be averaged over the possible color field configurations:
\begin{equation}
\label{eq:n-def}
	N(\xt,\yt) = 1 - \frac{1}{\nc} \langle \tr V(\xt) V^\dagger(\yt) \rangle.
\end{equation}
Note that the Wilson lines, and thus the dipole amplitude $N$, implicitly depend on the target color field and on the kinematics of the scattering process, especially on at which $x$ the target is probed.
As it is shown later in this thesis, many cross sections can be expressed in terms of the dipole amplitude. Thus, the relevant degrees of freedom in the Color Glass Condensate picture are actually the Wilson lines.
In the limit of very large nucleus, one can derive the so called McLerran-Venugopalan model for the dipole amplitude~\cite{McLerran:1994ni}
\label{sec:mv-discussion}
\begin{equation}
	\label{eq:mv}
	N(r=|\xt-\yt|) = 1 - \exp\left[ -\frac{r^2 \qso^2}{4} \ln \left( \frac{1}{r\lqcd} + e \right) \right],
\end{equation}
where $\qso$ parametrizes the characteristic transverse momentum scale of the gluons in the nucleus.

To demonstrate the convenience of the Wilson lines as degrees of freedom let us briefly discuss the gluon distribution of a large nucleus. In collinear factorization it is assumed that the partons in the proton (or nucleus) carry zero transverse momentum.  This approximation is not exactly valid at small-$x$, because the small-$x$ partons are created in emissions of gluons with smaller and smaller longitudinal momentum fraction, a sequence of processes that can create a finite amount of transverse momentum.
%This approximation is not valid at small-$x$. To see this, we consider a nucleon moving in the positive $z$ direction along the light cone such that $P^+$ is very large, and in the nucleon wave function there is a quantum fluctuation (a quark or  a gluon) which carries a factor $x$ of the longitudinal momentum and transverse momentum $\kt$.  The fluctuation is now localized in $x^-$, as the uncertainty principle states that $\Delta x^- \sim 1/(xP^+)$. On the other hand, assuming that all the $p^-$ of the fluctuation is of the same order as $P^-$, and requiring that the nucleon is on a mass shell we get $\Delta x^+ \sim 2xP^+/\kt^2$. This states that the small-$x$ fluctuations are delocalized in $x^+$,
%than larger-$x$ parent partons that appear to be frozen. 
At small enough $x$ the transverse momenta of the fluctuations can eventually be of the same order as the momenta of the produced hadrons in the scattering processes, see especially discussion in  \ch\ref{ch:dihad} where this phenomenon manifests itself in the production of two semihard hadrons.

\begin{figure}[tb]
\begin{center}
\includegraphics[width=0.7\textwidth]{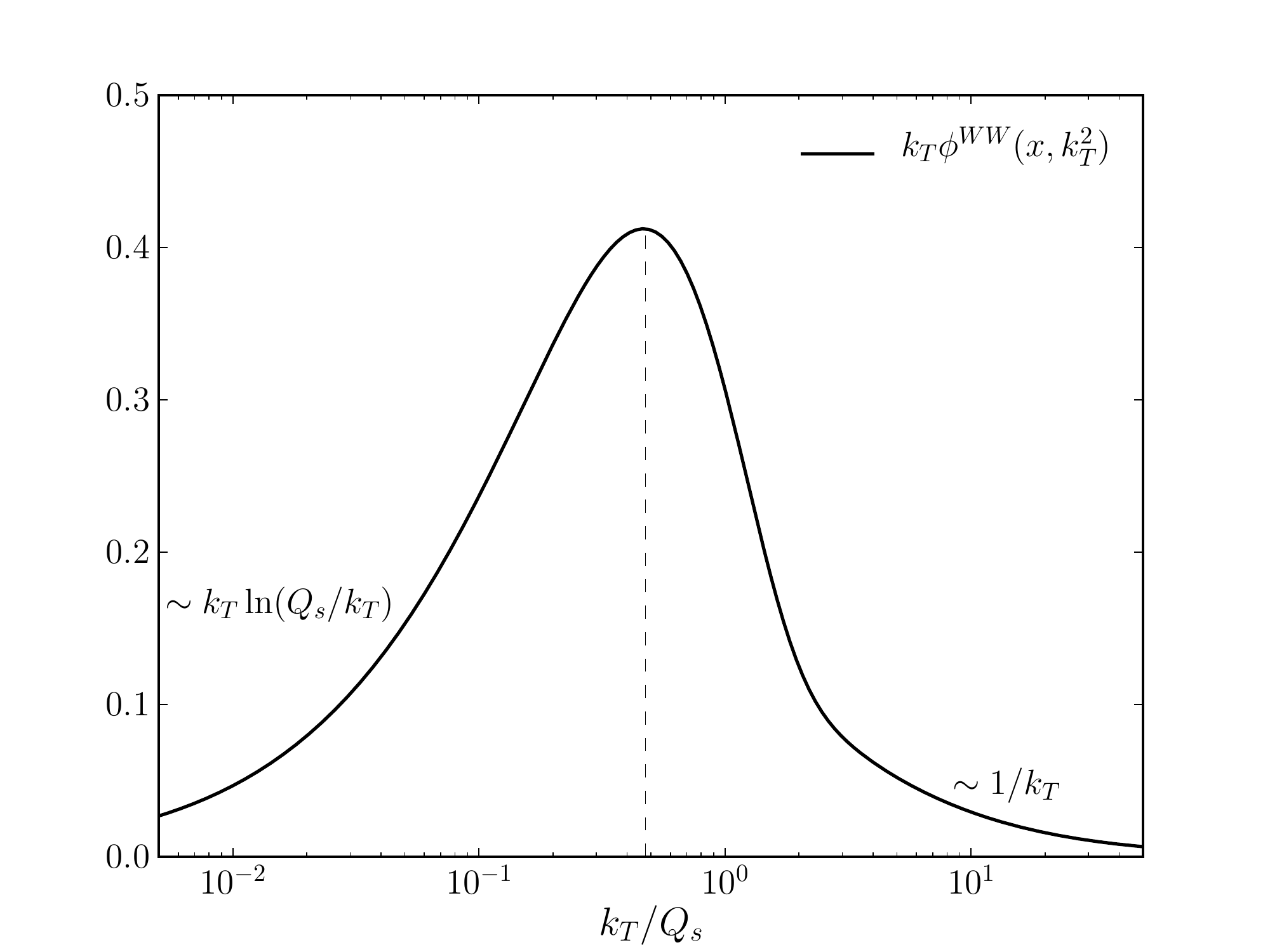}
\caption{Distribution of gluons in transverse momentum space computed from the MV model \eqref{eq:mv} in arbitrary units. Most of the gluons have transverse momenta of the order of $\qs$, which is shown as a dashed line.}
\label{fig:ww_ugd}
\end{center}
\end{figure}

The transverse momentum dependence of the nucleus (of nucleon) gluon distribution can be calculated by evaluating the expectation value of the gluon number density operator $a^\dagger(\kt, k^+)a(\kt,k^+)$ in the nuclear state. This gives the so called \emph{Weizsäcker-Williams distribution} $\phi^{WW}$, which is the unintegrated gluon distribution function of the nucleus~\cite{Dominguez:2011wm} (for a more pedagogical discussion, see \re\cite{Kovchegov:2012mbw}). It is directly related to the collinear factorization gluon distribution function $xg$ via
\begin{equation}
	\phi^{WW}(x,Q^2) = \frac{\partial xg(x,Q^2)}{\partial Q^2},
\end{equation}
and can be computed from the dipole amplitude $N$:
\begin{equation}
	\phi^{WW}(x,\kt) = \frac{S_T}{\pi^2 \as} \frac{\nc^2-1}{\nc} \int \frac{\der^2 \rt}{(2\pi)^2} \frac{e^{-i \kt \cdot \rt}}{\rt^2} N(x,\rt).
\end{equation}
Here $S_T$ is the transverse area of the nucleus, and we have neglected the impact parameter dependence of the dipole amplitude.

The second gluon distribution function is called the \emph{dipole gluon distribution} $\varphi^\text{dipole}$, and is related to the Fourier transform of the dipole amplitude as
\begin{equation}
\label{eq:dipole-ugd}
	\varphi^\text{dipole}(x,\kt) = S_T \frac{\kt^2 \nc}{2\pi^2 \as} \int \frac{\der^2 \rt}{(2\pi)^2} e^{-i \kt \cdot \rt} S(x,\rt).
\end{equation}
The dipole gluon distribution has no number density interpretation, as it contains both initial and final state interactions~\cite{Dominguez:2011wm}. As we will discuss later, particle production cross sections in hadronic collisions and in lepton-proton deep inelastic scattering (DIS) are proportional to this distribution. On the other hand, the WW gluon distribution can be probed for example in dijet production that we discuss in \ch\ref{ch:dihad}, see also discussion in \re\cite{Dominguez:2011wm}. 

%TODO slitys miksi menee 1/kt?
The Weizsäcker-Williams unintegrated gluon distribution computed from the MV model and multiplied by the two-dimensional phase space factor $\kt$, which gives the number of gluons with momentum $\kt$, is shown in \fig \ref{fig:ww_ugd}. At large transverse momenta the number of gluon density drops like $1/\kt$. At low momenta the distribution behaves like $\kt \ln (\qs/\kt)$, and most of the gluons have transverse momenta of the order of the saturation scale $\qs$ (see e.g. \re\cite{Kovchegov:2012mbw}), defined as $N(\rt^2=2/\qs^2)=1-e^{-1/2}$. If this scale is much larger than $\lqcd$, strong coupling constant is small and perturbation theory can be used to describe the wave function of the nucleus or of the proton. This  definition of $\qs^2$ is not unique, as it could as well be defined by using a different constant value, or e.g. from the shape of the Weizsäcker-Williams distribution. The corresponding high-energy behaviors, for example the fact that the characteristic gluon transverse momentum is proportional to the saturation scale, are identical.

\section{High energy evolution equations}
\label{sec:evolution}

\subsection{The BK equation}
\label{sec:bk}

Let us find the energy, or equivalently Bjorken-$x$, dependence of the dipole scattering amplitude. A more detailed discussion and derivation can be found from my MSc thesis~\cite{gradu}.

\begin{figure}[tb]
\begin{center}
\includegraphics[width=0.7\textwidth]{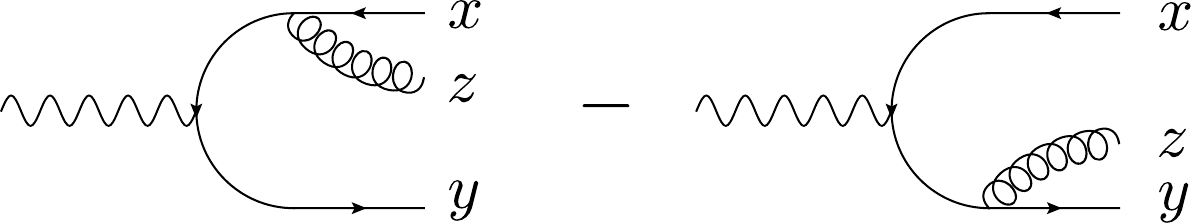}
\caption{Virtual photon splitting to quark-antiquark pair which subsequently emits a gluon. The quark and antiquark transverse coordinates are $x$ and $y$, and $z$ is the gluon position.}
\label{fig:gamma-qq-gluon}
\end{center}
\end{figure}

Consider a virtual photon-target scattering in the dipole frame where the high-energy photon fluctuates to a quark-antiquark dipole which scatters off the low-energy target.  The forward elastic scattering amplitude for this process is called dipole amplitude and denoted by $N$. To obtain the energy evolution of the amplitude we consider what happens when the dipole is boosted to higher rapidity. The boost opens a larger phase space for the quark or the antiquark to emit a gluon, and after the gluon emission the full $q\bar q g $ system can interact with the target. 

The gluon emission from a quark can be computed using light cone perturbation theory, see discussion in \se\ref{sec:lightcone} and e.g. \re \cite{Brodsky:1997de} or a detailed calculation in my MSc thesis~\cite{gradu}. The gluon can be emitted either from the quark or from the antiquark as shown in \fig \ref{fig:gamma-qq-gluon}. Let us choose to label the coordinates of the quarks by $\xt$ and $\yt$, and the gluon transverse coordinate is $\zt$. Recall that the we use the eikonal approximation and assume that the transverse positions are fixed during the interaction.
%The fact that the relative sign between the diagrams is minus is easy to justify using the color neutrality: when quark and gluon are close to each other, the gluon can not see a color charge and gluon emission is not possible. {\bf jos ei lasketa tarkasti kannattaako puhua merkistä}

\begin{figure}[tb]
\begin{center}
\includegraphics[width=0.7\textwidth]{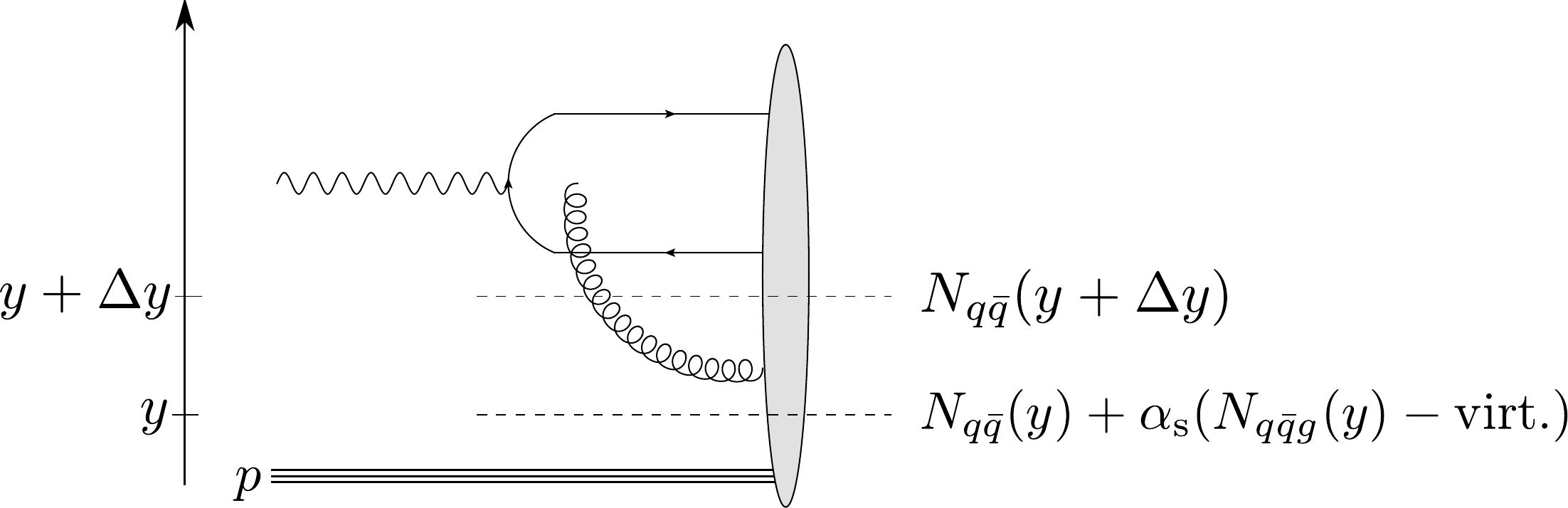}
\caption{The emitted gluon can be seen as a part of the target wave function (rapidity cut at the lower dashed line) or as a part of the dipole wave function (rapidity cut at  the upper dashedline).}
\label{fig:renormalization}
\end{center}
\end{figure}

The emitted gluon can be counted as being part of the dipole wave function, in which case we have $q\bar q g$ system scattering off the target. On the other hand, the gluon can also be counted as being part of the target, in which case we have $q\bar q$ system scattering off a target with a larger gluon density, having a larger rapidity difference between the systems. This is illustrated in \fig \ref{fig:renormalization}, where we also note that in addition to real correction originating from a gluon emission also a virtual correction coming from the wave function normalization requirement. The physical observables can not depend on this arbitrary separation of scales (whether the gluon is chosen to be a part of the dipole or of the target), thus equating the scattering amplitudes in both cases gives an evolution equation 
\begin{multline}
\label{eq:bk-valitulos}
	N_{q\bar q}(y+\Delta y, \rt) = N_{q\bar q}(y,\rt) + \frac{\as \nc}{2\pi^2} \Delta y \int \der^2 \rt' \frac{\rt^2}{\rt'^2 (\rt-\rt')^2} \\
	\times \left[ N_{q\bar q g}(y, \rt, \rt') - N_{q\bar q}(y,\rt) \right],
\end{multline}
where $N_{q\bar q}$ is the scattering amplitude for the $q\bar q$ dipole, and $N_{q\bar q g}$ is the same amplitude for quark-antiquark-gluon system. The dipole amplitude is evaluated at rapidity $y$, which corresponds to Bjorken-$x$ obtainable from the relation $y=\ln 1/x$.

Let us then work in the large-$\nc$ limit where the emitted gluon can be seen as a quark-antiquark dipole. This is because the color structure of the gluon is approximately the same as that of the dipole, as the number of gluon color states is $\nc^2-1 \approx \nc^2$. Thus, instead of having a $q\bar q$ dipole with transverse separation $\rt=\xt-\yt$ and a gluon at point $\zt$, we have two $q\bar q$-dipoles with transverse separations $\rt'=\xt-\zt$ and $\rt-\rt'=\zt-\yt$. 
In the mean field limit the expectation value for the product of two scattering matrices factorizes and the probability for this system not to scatter is obtained as a product of probabilities for the two dipoles to not to scatter:
\begin{equation}
	\langle S_{q\bar q g}(\rt,\rt')\rangle \approx \langle S_{q\bar q}(\rt') \rangle \langle S_{q\bar q}(\rt-\rt') \rangle.
\end{equation}
Noticing that $S_{q\bar q g}=1-N_{q\bar q g}$ and writing the \eq \eqref{eq:bk-valitulos} as a differential equation we get
\begin{multline}
\label{eq:bk}
	\partial_y N(\rt) = \frac{\as \nc}{2\pi^2} \int \der^2 \rt' \frac{\rt^2}{\rt'^2(\rt-\rt')^2} \\
	\times \left[ N(\rt') + N(\rt-\rt') - N(\rt) - N(\rt')N(\rt-\rt') \right]
\end{multline}
which is the Balitsky-Kovchegov evolution equation first derived in \res \cite{Balitsky:1995ub,Kovchegov:1999yj}. Note that we use the notation $N=\langle N_{q\bar q} \rangle$, see \eqs. \eqref{eq:nqq} and \eqref{eq:n-def}. The interpretation of \eq\eqref{eq:bk} is clear: to get the dipole amplitude at higher rapidity one has to consider the scattering of two daughter dipoles generated by a gluon emission, and the original dipole must be removed. The subtraction of the non-linear term corresponds to removing double counting in the case where both daughter dipoles scatter off the target. 

In the small scattering amplitude limit, the non-linear term in \eq\eqref{eq:bk} can be neglected, and one obtains the so called BFKL (Balitsky-Fadin-Kuraev-Lipatov) equation~\cite{Kuraev:1977fs,Balitsky:1978ic}, whose derivation actually predates the derivation of the BK equation. The main feature of the BFKL evolution is that it makes the dipole amplitude, and the unintegrated gluon distribution function, to grow exponentially, which eventually violates unitarity of the theory and the Froissant bound. The non-linear contribution included in the BK evolution tames the growth at larger distances and restores unitarity. For more details, we refer the reader e.g. to \re\cite{Kovchegov:2012mbw}.

The BK equation is a leading order evolution equation where the strong coupling constant $\as$ is fixed. In QCD, next to leading order corrections are known to be significant in many processes. 
Thus the inclusion of NLO corrections to the BK equation is an important task. The full NLO BK equation is available (derived in \re \cite{Balitsky:2008zza}), and we shall discuss it more in \se \ref{sec:nlobk}. In phenomenological applications, however, the first step towards the full NLO evolution is to incorporate the running coupling corrections into the leading order BK equation and assume, that this takes into account most of the NLO corrections. 

The most widely used running coupling prescription is derived by Balitsky in \re \cite{Balitsky:2006wa}. In addition to that, there exist also other prescriptions to include the running coupling effects into the BK equation. The reason for having different possible running coupling corrections is that there is no unique way to determine which NLO terms are counted as being part of the running of $\as$, as it is a scheme-dependent choice. 
As an example, we mention the prescription derived by Kovchegov and Weigert  in \re \cite{Kovchegov:2006vj} that is shown in \re\cite{Albacete:2007yr} to agree with the Balitsky prescription when the different scheme choice is taken into account. Before these perturbative calculations became available, the running coupling effects were estimated by e.g. evaluating the strong coupling constant at the scale set by the parent dipole, see for example \re\cite{Albacete:2004gw}.

In phenomenological applications the Balitsky prescription is usually used, as it gives a slower evolution speed which is consistent with the experimental data (see discussion in \se \ref{sec:fits}). In the Balitsky prescription the BK kernel is replaced by
\begin{multline}
\label{eq:bk-balitsky}
	\frac{\as \nc}{2\pi^2}  \frac{\rt^2}{\rt'^2(\rt-\rt')^2 } \to \frac{\as(\rt^2)\nc}{2\pi^2} \left[ \frac{\rt^2}{\rt'^2(\rt-\rt')^2} + \frac{1}{\rt'^2}\left(\frac{\as(\rt')}{\as(\rt-\rt')} -1\right) \right. \\
	\left. + \frac{1}{(\rt-\rt')^2} \left( \frac{\as(\rt-\rt')}{\as(\rt')}-1\right) \right].
\end{multline}
One can check that the dominant scale is the smallest of the dipole sizes $|\rt|$, $|\rt'|$ and $|\rt-\rt'|$.

When a running coupling kernel is used one has to evaluate the strong coupling constant $\as$ as a function of transverse separation. For this, we use an expression
\begin{equation}
\label{eq:alphas-r-csqr}
	\as(\rt^2) = \frac{12\pi}{(11\nc-2\nf)\ln \left( \frac{4C^2}{\rt^2 \lqcd^2}\right)}.
\end{equation}
The uncertainty in the Fourier transform of the expression of $\as$ from the momentum space to the coordinate space is parametrized by introducing a factor $C^2$. Note that one can argue that the scale should be chosen as $C^2=e^{-2\gamma_E} \approx 0.3152$, as suggested in \re \cite{Kovchegov:2006vj} (see also discussion in \paper \cite{Lappi:2012vw} and in \se\ref{sec:jimwlk}). However, as we will discuss in \se \ref{sec:fits}, in order to get an evolution speed comparable with the experimental data $C^2$ must be adjusted properly.

\begin{figure}[tb]
\begin{center}
\includegraphics[width=0.6\textwidth]{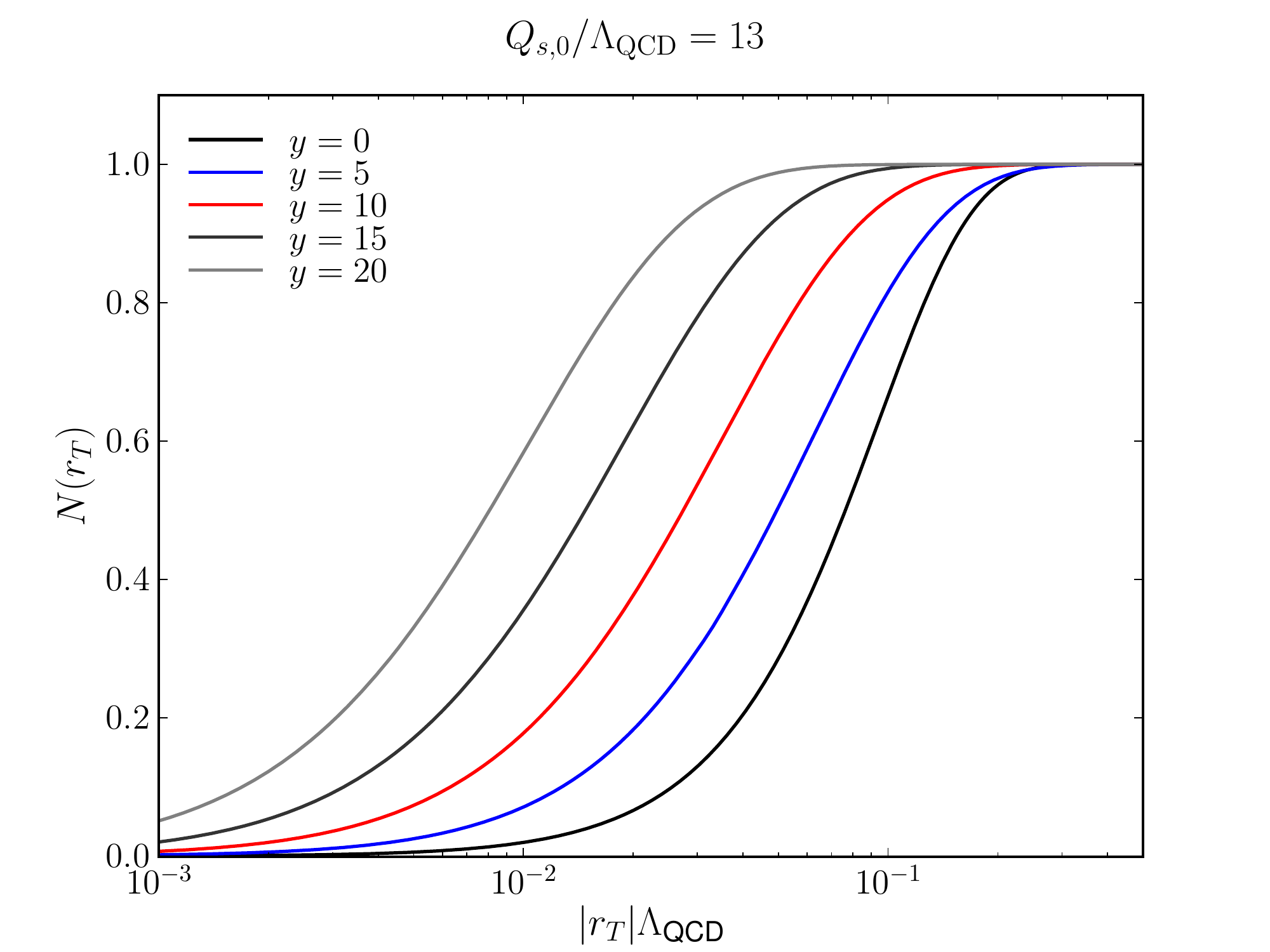}
\caption{Dipole amplitude from the BK equation at different rapidities using an MV model initial condition.}
\label{fig:amplitude}
\end{center}
\end{figure}

To demonstrate the effect of the BK evolution the dipole amplitude $N(\rt)$ is shown in \fig\ref{fig:amplitude} at the initial condition (MV model \eqref{eq:mv} with $\qso/\lqcd=19$) and at higher rapidities. The Balitsky running coupling \eqref{eq:bk-balitsky} is used. The general trend of the evolution can be seen from the figure: near the initial condition the solution approaches the asymptotic shape which changes very little after a few rapidity steps. Later in the evolution the solution propagates to smaller values of $|\rt|$, which can be interpreted as having an increasing gluon density at smaller distance scales. Physically, this is a result of gluon splitting $g\to gg$. The dipole amplitude saturates to unity at larger dipoles, which can be seen to be a consequence of gluon recombination $gg \to g$ processes that balances the gluon splitting. This interpretation in terms of gluon densities is cleaner if the BK equation is written in transverse momentum space as done in \re\cite{Marquet:2005zf} (see also \re\cite{gradu}).

The discussion above is valid for infinitely large targets with rotational invariance where the impact parameter can be neglected. In principle one could include the impact parameter dependence in the BK equation to describe e.g. the fact that if the dipole size is much larger than the impact parameter and the size of the target, the dipole will not scatter as the quarks are far away from the target. The impact parameter dependent BK equation is, however, known to develop unphysical Coulomb tails that should be regulated by confinement scale physics (see e.g. \res \cite{GolecBiernat:2003ym,Berger:2010sh,Berger:2011ew,
Berger:2012wx}). As it can not be considered ready for phenomenological applications, only the impact parameter independent BK equation is used in this work.

\subsection{The JIMWLK equation}
\label{sec:jimwlk}

The JIMWLK equation~\cite{JalilianMarian:1996xn,JalilianMarian:1997jx, JalilianMarian:1997gr,Iancu:2001md, Ferreiro:2001qy, Iancu:2001ad, Iancu:2000hn}, named after Jalilian-Marian, Iancu, McLerran, Weigert, Leonidov and Kovner, gives the rapidity (or Bjorken-$x$) evolution of the probability distribution of the Wilson lines $W_Y[V]$. As the energy is increased, large-$x$ color sources emit new gluons that become sources for further emissions and modify the probability distribution. 
The equation can be used to calculate the rapidity dependence of different correlators of Wilson lines. In this sense the JIMWLK equation is more general than the BK equation discussed in \se\ref{sec:bk} which only gives the rapidity evolution of a dipole (correlator of two Wilson lines). Although the BK equation was derived before JIMWLK, it can be obtained from the JIWMLK equation in the mean field and large-$\nc$ limit, see for example \paper\cite{Lappi:2012vw}.

%To be precise, the JIMWLK equation describes the rapidity evoluiton of the probability distribution for Wilson lines that are random SU($\nc$) matrices. 

As we will discuss in \chs\ref{ch:dis} and \ref{ch:sinc}, for example the deep inelastic scattering and single inclusive particle production cross sections can be expressed in terms of the dipole operators only. However, when considering multiparticle production, one generally needs more complicated operators made of more than two Wilson lines. In \ch\ref{ch:dihad} the two-particle production cross section is shown to require knowledge of the so called quadrupole operator which is a trace of four Wilson lines. The evolution for these higher-point functions can be obtained from the JIMWLK equation.

The running coupling corrections to the BK equation are known to be very large~\cite{Albacete:2007yr} and have an important effect on phenomenological calculations. For the JIMWLK equation, the full NLO equation is known~\cite{Balitsky:2013fea} but a numerical solution is still lacking. On the other hand, the leading order JIMWLK equation has been solved numerically (see e.g. \re\cite{Rummukainen:2003ns}). For phenomenological applications it would be useful to have a way to include the running coupling corrections in the JIMWLK equation similarly as in the BK equation.  
%by using the Balitsky running coupling as shown in \eq\eqref{eq:bk-balitsky}. 
What is proposed in \paper \cite{Lappi:2012vw} is a running coupling prescription which keeps the functional form of the JIMWLK equation intact and has the same limiting behaviors as the BK equation with the Balitsky running coupling prescription shown in \eq\eqref{eq:bk-balitsky}.

For numerical calculations the JIMWLK equation is written as a Langevin equation for a single Wilson line.
Following  \re\cite{Blaizot:2002np} the equation can be written as
%, the JIMWLK equation in the Langevin form is
\begin{equation}\label{eq:langevin1}
\frac{\der}{\der y} V_y(\xt) = i t^a V_y(\xt)  \left[
\int_\zt
\varepsilon_y(\xt,\zt)^{ab,i} \; \xi_y(\zt)^b_i  + \sigma(\xt)^a
\right] ,
\end{equation}
where $V_y$ is the Wilson line at rapidity $y$ as defined in \eq\eqref{eq:wilson-line} and $i=1,2$ is a transverse spatial index. 
This corresponds to a random walk in the space of SU($\nc$) matrices, described by a stochastic noise term $\xi_y$ and a deterministic part $\sigma(\xt)^a$. Averaging over the noise corresponds to averaging with the probability distribution $W_Y[V]$.

The coefficient of the stochastic term $\xi_y$ is 
\begin{equation}
 \varepsilon_y(\xt,\zt)^{ab,i} = \left(\frac{\as}{\pi}\right)^{1/2}
K(\xt-\zt)^i
\left[1-U_y^\dagger(\xt)  U_y(\zt)\right]^{ab},
\end{equation}
where $U_y$ is the Wilson line in the adjoint representation. The kernel $K(\xt)^i$ is the gluon emission light cone wave function which, in the continuum limit, reads $K(\xt)^i = \xt^i/\xt^2$. 
%When the JIMWLK equation is solved on a periodic lattice, the kernel must be modified to satisfy the  boundary conditions. 
The noise $\xi$ is a random variable which is taken to be Gaussian and local in transverse coordinate and rapidity with zero expectation value and
\begin{equation}
\label{eq:noice-correlator}
	\langle \xi_y(\xt)^{a,i} \xi_{y'}(\yt)^{b,j} \rangle=  \delta^{ab} \delta^{ij} \delta^{(2)}(\xt-\yt) \delta(y-y').
\end{equation}
The second term in the evolution equation \eqref{eq:langevin1} is a deterministic term 
%\begin{equation}
%	\sigma_\xt^a = -i \frac{\as} {2\pi^2}
%\int_{\zt} S_{\xt -\zt}
% \widetilde{\tr} \left[ T^a U_\xt^\dag U_\zt\right].
%\end{equation}
which is numerically demanding to compute as it would require one to reconstruct the adjoint representation Wilson line. However, as discussed in \paper \cite{Lappi:2012vw} and in \re\cite{Mueller:2001uk}, the
deterministic term is needed if one wants to write the JIWMLK equation in the Langevin form, \eq \eqref{eq:langevin1}, as a multiplication of $V$ from only one side.
If the JIMLWK equation is written in a form where the Wilson line is multiplied from both right and left, the deterministic term is not needed and one obtains
\begin{multline}\label{eq:jtimestepsymm}
V_{y+\der y}(\xt) = 
\exp\left\{-i\frac{\sqrt{\as \der y }}{\pi}\int_\zt 
  K(\xt-\zt) \cdot ( V_y(\zt) \xi_y(\zt) V_y^\dagger(\zt)) \right\}
\\
\times
V_y(\xt)
\exp\left\{i\frac{\sqrt{\as \der y }}{\pi}\int_\zt 
  K(\xt-\zt) \cdot \xi_y(\zt) \right\},
\end{multline}
which is equivalent to \eq\eqref{eq:langevin1} up to order $\der y$. In the numerical calculations presented in \paper\cite{Lappi:2012vw} equation \eqref{eq:jtimestepsymm} (with running coupling modifications discussed next) is solved.

Let us now consider the evolution of an operator consisting of Wilson lines. In order to calculate the operator at rapidity $y+\der y$ one has to expand all Wilson lines using \eq\eqref{eq:jtimestepsymm} up to order $\der y$, or equivalently $\xi^2$, and take the expectation value over the noise term $\xi$. Only the contribution $\sim \as$ remains in the evolution equation, which physically corresponds to the fact that in the leading order JIMWLK equation the evolution is obtained by calculating an emission of a single gluon. 
%The correlator between two noice terms $\xi$ can be interpreted as a gluon emission in both amplitude and in the complex conjugate. 
Consider now an evolution step where two noise terms $\xi(\ut)$ and $\xi(\vt)$ are contracted. The delta function in the correlator \eqref{eq:noice-correlator} can be now interpreted as follows. First, we note that if $\xi(\ut)$  corresponds to the gluon emission at coordinate $\ut$ in the amplitude, then $\xi(\vt)$ is the absorption of a gluon at $\vt$ in the complex conjugate amplitude. Let us denote the momentum of the emitted gluon by $\kt$. Now, the amplitude is proportional to $e^{i\kt \cdot \ut}$, and the complex conjugate amplitude to $e^{-i \kt \cdot \vt}$. When all possible gluon momenta are integrated over, the delta function $\delta^{(2)}(\ut-\vt)$ is obtained, which is  part of the correlator \eqref{eq:noice-correlator}.

To include the running coupling we refer to a general result in gauge theories that the beta function can be computed by considering higher order corrections to the gluon propagator. Note that this is exactly what is done e.g. in \re\cite{Balitsky:2006wa} when the Balitsky running coupling prescription for the BK equation is derived by taking into account the quark loop corrections to the gluon propagator. Now the only scale available for the running $\as$ is the transverse momentum of the emitted gluon $\kt$, so we propose to replace the fixed coupling correlator
\begin{equation}
\as \left\langle \xi(\xt)^{a,i} \xi(\yt)^{b,j} \right\rangle = \as \delta^{ab}
\delta^{ij} \int \frac{\der^2 \kt}{(2\pi)^2}
e^{i \kt\cdot(\xt-\yt)}
\end{equation}
by 
\begin{equation}\label{eq:newcorr}
\left\langle \eta(\xt)^{a,i} \eta(\yt)^{b,j} \right\rangle = \delta^{ab}
\delta^{ij} \int \frac{\der^2 \kt}{(2\pi)^2}
e^{i \kt\cdot(\xt-\yt)} \as(\kt).
\end{equation}
It is shown in \paper \cite{Lappi:2012vw} that if the BK equation is derived from the JIMWLK equation with the correlator \eqref{eq:newcorr}, one obtains an equation that has the same kernel as the BK equation with the Balitsky running coupling in the limit where the parent dipole or one of the daughter dipoles is very small.

Let us then present our numerical results for the JIMWLK equation with running coupling. The equation is solved using the algorithm presented in \re\cite{Rummukainen:2003ns}. We compare our running coupling results with the JIMWLK equation solved using the ``square root'' running coupling. The square root coupling is the simplest modification to the JIMWLK equation which includes the running of $\as$. In this prescription the coupling $\sqrt{\as}$ is evaluated at the scale which is the argument of the kernel $K(\xt)$. For comparison the BK equation is also solved using the Balitsky running coupling. See \paper\cite{Lappi:2012vw} for details. 

The strong coupling constant in the transverse coordinate space is evaluated as 
\begin{equation}
\label{eq:as-r-jimwlk-nlobk}
  \as(\rt) = \frac{4\pi}{
\beta \ln\left\{\left[
       \left(\frac{\mu_0^2}{\lqcd^2}\right)^{\frac{1}{c}}
      +\left(\frac{4e^{-2 \gamma_\mathrm{E}}}{\rt^2\lqcd^2}\right)^{\frac{1}{c}} \right]^{c}
\right\},
}
\end{equation}
with $c=0.2$, $\mu_0^2/\lqcd^2=2.5$ and $\beta=\frac{11}{3}\nc - \frac{2}{3}\nf$. The difference to the expression \eqref{eq:alphas-r-csqr} used in the previous section is that \eq\eqref{eq:as-r-jimwlk-nlobk} has a smoother infrared cutoff parametrized by the constant $c$. 
%We use \eq\eqref{eq:as-r-jimwlk-nlobk} when studying the JIMWLK equation, and also when solving the NLO BK equation in \se\ref{sec:nlobk}.
%This expression is different than \eq\eqref{eq:alphas-r-csqr} which is used in phenomenological applications in \chs \ref{ch:dis}--\ref{ch:dihad}. 
Also the scale at which the coupling is evaluated is not a fit parameter. Instead, it is given by $4e^{-2\gamma_e}/\rt^2$ which is taken from the explicit Fourier transform of the kernel calculated e.g. in \re\cite{Balitsky:2006wa}. As we will discuss in \ch\ref{ch:dis}, the BK fits to experimental data suggest the scale to be $\sim 20/\rt^2$ for $\lqcd\sim 200 \mev$, unless the evolution is slowed down by other effects (see discussion of the next to leading order BK equation in \se\ref{sec:nlobk}).

\begin{figure}[tb]
\centering
	\begin{minipage}[t]{0.5\textwidth}
	\includegraphics[width=\textwidth]{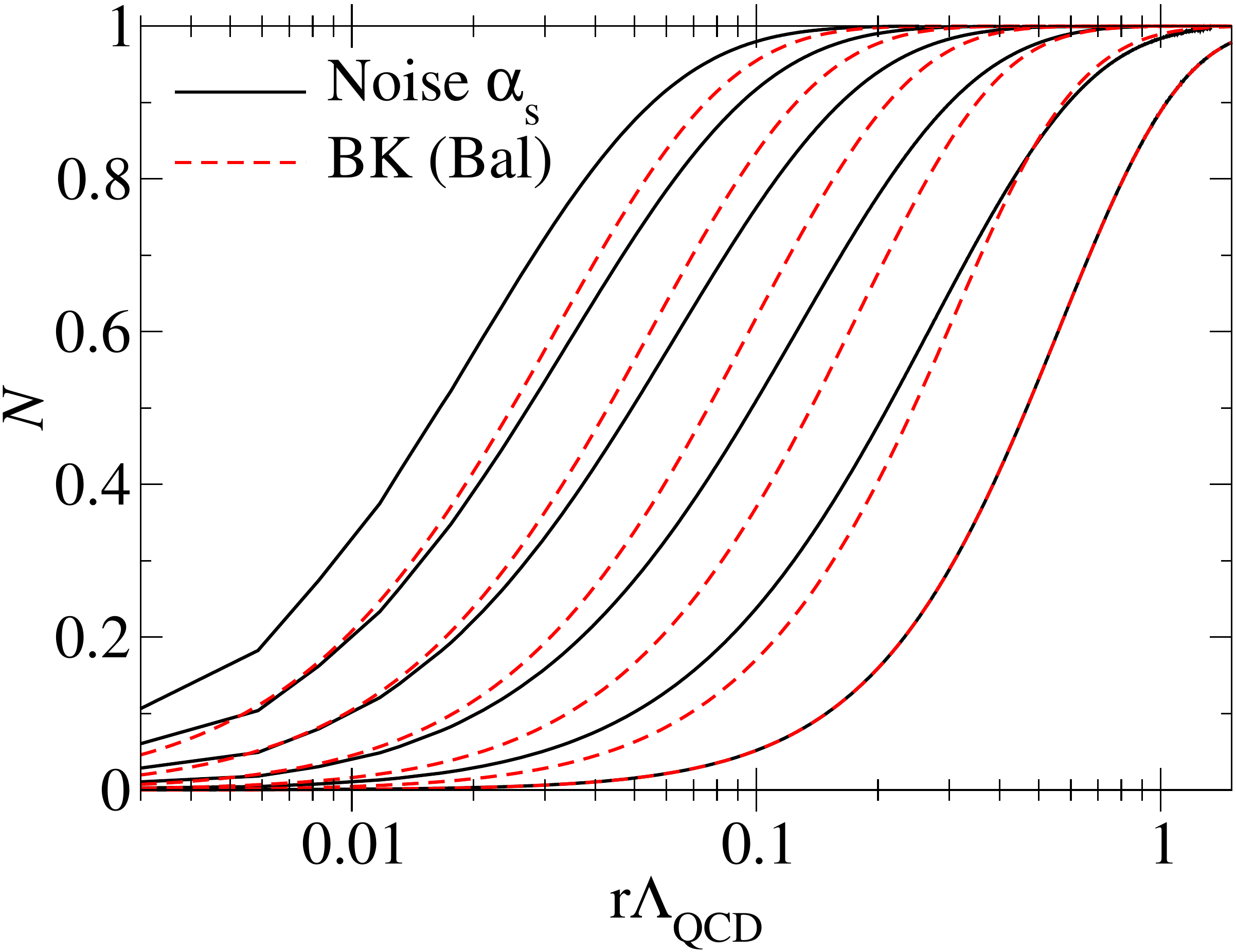}
	\caption{Evolution of the dipole amplitude $N$ from the JIMWLK equation with running coupling compared to the BK equation. The lines show the amplitude at intervals of 2 units in rapidity. Figure from \paper\cite{Lappi:2012vw}.}
	\label{fig:jimwlk-noise-bk}
	\end{minipage}%
	~
	\begin{minipage}[t]{0.5\textwidth}
	\includegraphics[width=1.0\textwidth]{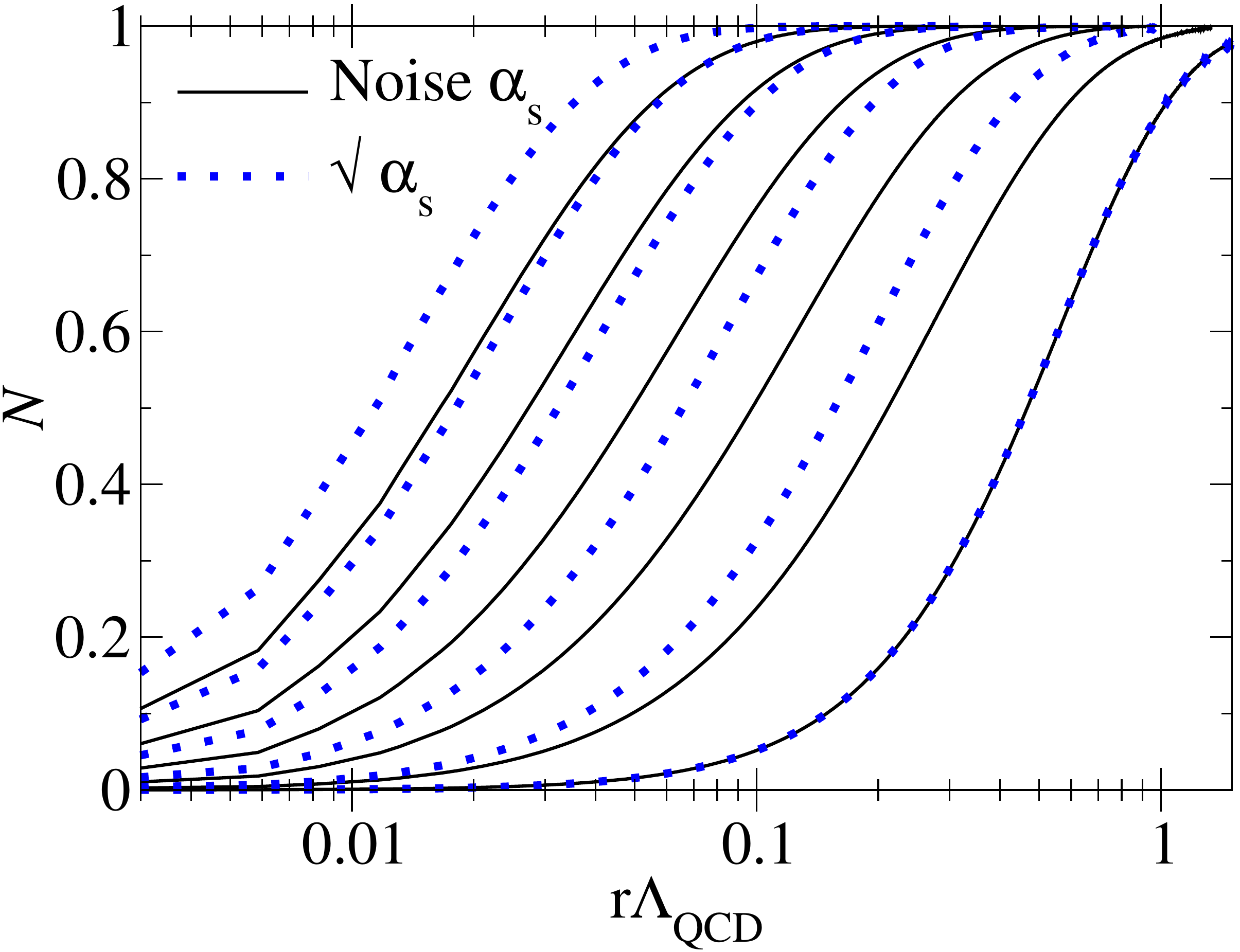}
	\caption{Evolution of the dipole amplitude using the proposed ``noise'' running coupling compared with the results using the square root prescription. The lines show the amplitude at intervals of 4 units in rapidity. Figure from \paper\cite{Lappi:2012vw}.}
	\label{fig:jimwlk-noice-sqrt}
	\end{minipage}	
\end{figure}

\begin{figure}[tb]
\centering
	\begin{minipage}[t]{0.5\textwidth}
	\includegraphics[width=\textwidth]{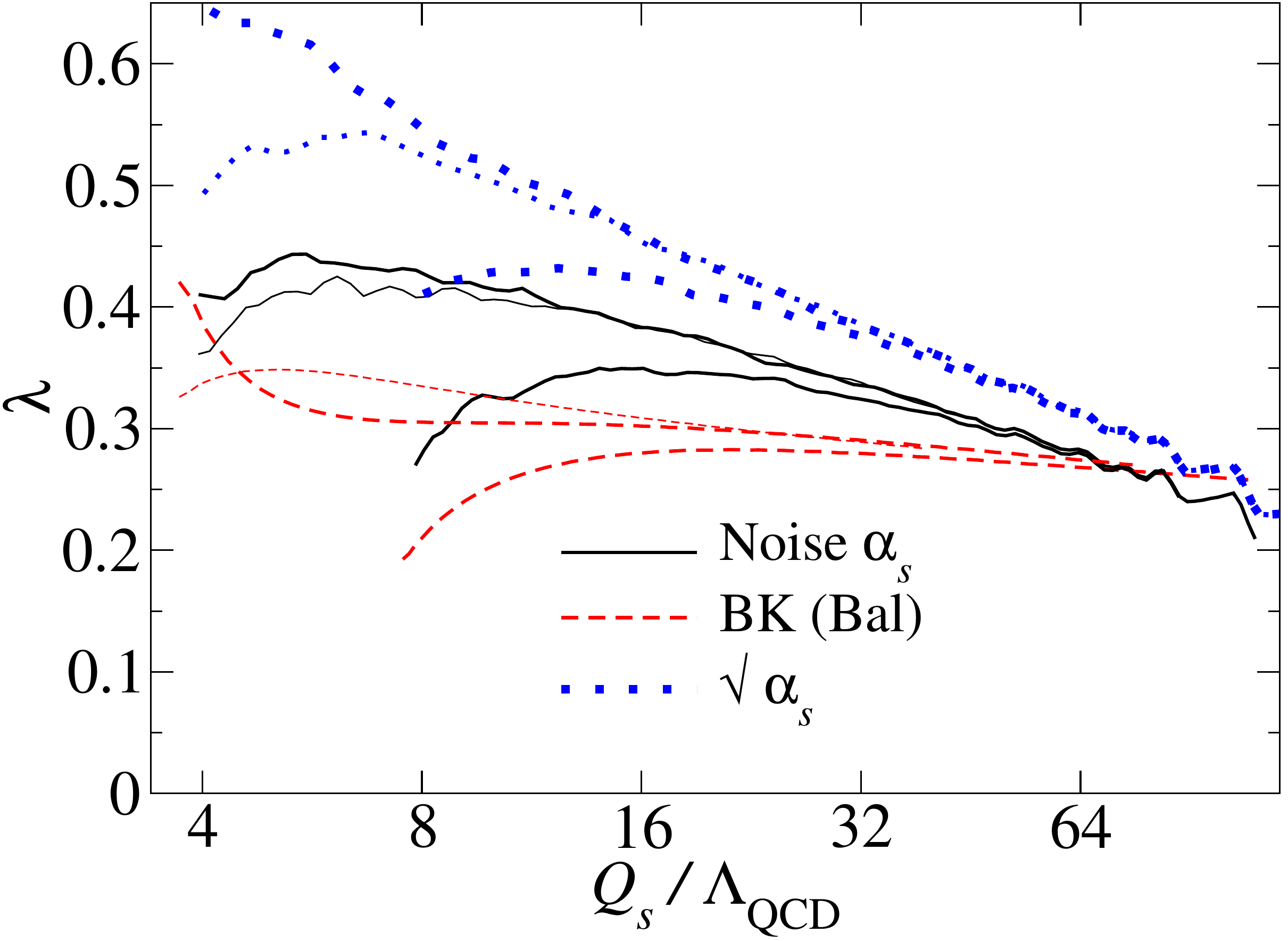}
	\caption{Evolution speed of the saturation scale as a function of $\qs$ calculated using the noise and square root prescriptions and compared to the BK results. The thin lines are obtained by using a smoother infrared freezeing  with $c=1.5$ instead of $c=0.2$. Figure from \paper\cite{Lappi:2012vw}.}
	\label{fig:jimwlk-lambda}
	\end{minipage}%
	~
	\begin{minipage}[t]{0.5\textwidth}
	\includegraphics[width=1.0\textwidth]{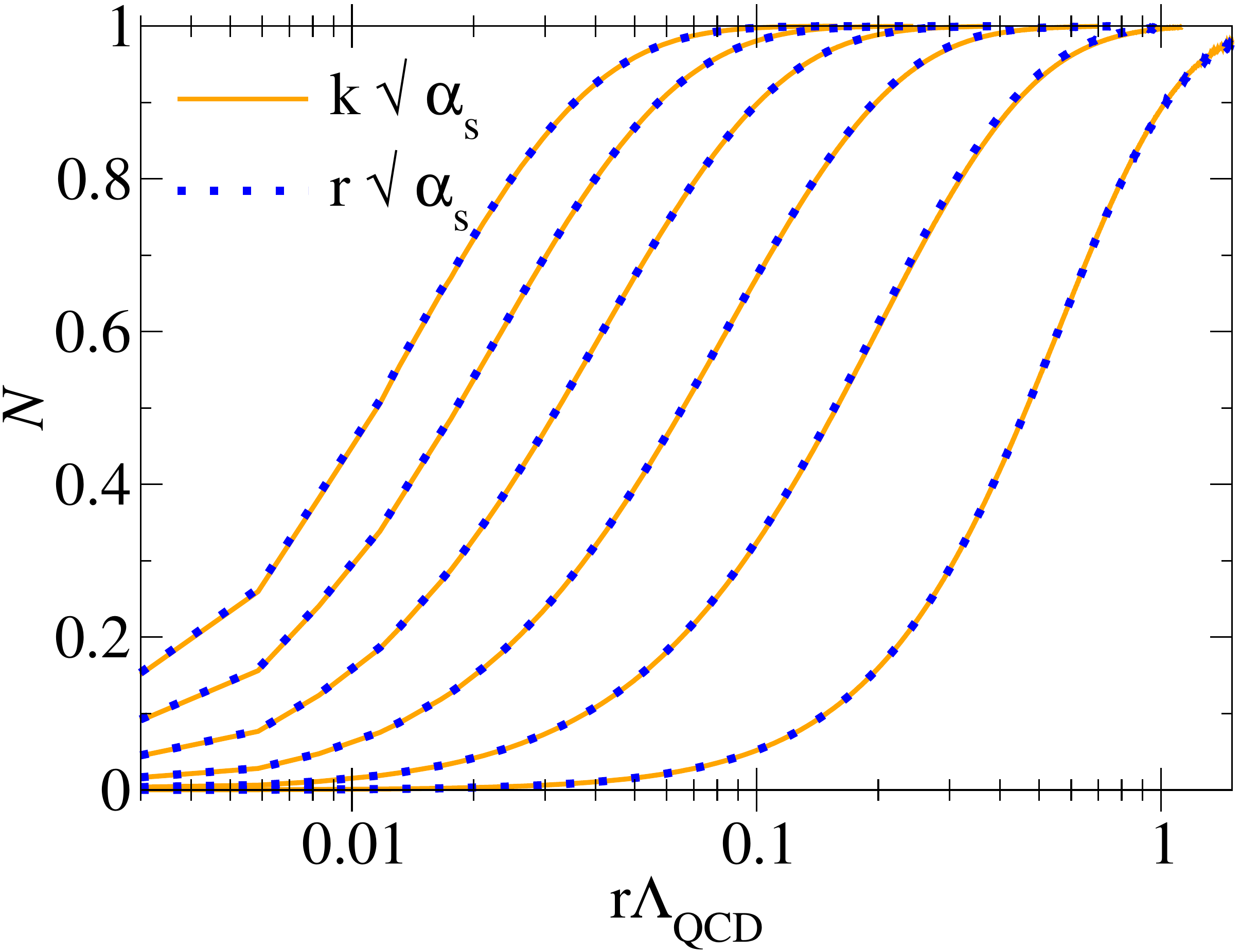}
	\caption{Evolution of the scattering amplitude with the square root running coupling evaluated in coordinate space at scale $4e^{-2\gamma_E}/\rt^2$ and in momentum space at scale $\kt^2$. Figure from \paper\cite{Lappi:2012vw}.
	}
	\label{fig:jimwlk-k-r}
	\end{minipage}	
\end{figure}

Figure \ref{fig:jimwlk-noise-bk} shows dipole amplitude $N$ (which is a correlator of two Wilson lines) obtained by solving the JIMWLK equation with the proposed noise running coupling, \eq\eqref{eq:newcorr}. The result is compared with the solution to the BK equation with the Balitsky running coupling. We find that the BK equation leads to a slower evolution speed and that the shapes of the solutions are slightly different.
To better see the effect of the noise running coupling on the JIMWLK equation
we show in \fig\ref{fig:jimwlk-noice-sqrt} the JIMWLK solutions obtained using our noise running coupling and the square root coupling prescriptions. Changing the running coupling prescription to the noise coupling reduces the evolution speed but leaves the shape of the solution roughly the same.

To characterize the evolution speed we show in \fig\ref{fig:jimwlk-lambda} the evolution of the saturation scale $\qs$ defined as
\begin{equation}
	\lambda = \frac{\der \ln Q_s^2}{\der y},
\end{equation}
where $N(\rt^2=2/\qs^2)=1-e^{-1/2}$. The calculation is done with two different values for the infrared freezing parameter $c$ in \eq\eqref{eq:as-r-jimwlk-nlobk}. The dependence on this constant is quite different between the BK and JIMWLK calculations at small $\qs$. The BK equation is shown to give the smallest evolution speed, and the Noise running coupling prescription slows down the JIMWLK evolution at all saturation scales. At larger $\qs$ the evolution speed from the JIMWLK equation drops below that of the BK, which is an artifact from $\qs$ becoming of the same order as the lattice ultraviolet cutoff.

Finally, in order to demonstrate why the strong coupling is evaluated at scale set by $4e^{-2\gamma_E}/\rt^2$ we solve the JIMWLK equation using two different running coupling prescriptions. First is the same square root prescription used before, where the evolution kernels are in the form $\sqrt{\as(\rt)}K(\rt)$. This is denoted by $r\sqrt{\as}$. We compare this with the momentum space square root prescription, denoted by $k\sqrt{\as}$. This is obtained by noticing that in the JIMWLK equation of a dipole one only has dot products of kernels $K$. These dot products can be Fourier transformed into the momentum space, and the coupling constant can then be evaluated as a function of transverse momentum. We write the dot product as
\begin{equation}
	\sqrt{\as} K(\xt) \cdot v = -i \int \frac{\der^2 \kt}{2\pi} \sqrt{\as(\kt)} e^{-i \kt \cdot \xt} \frac{\kt \cdot \vt}{\kt^2},
\end{equation}
where $\vt$ in practice is the kernel $K$ with the same or different argument.
%where the Fourier transform of $K(\xt)$ is multiplied by the square root of the coupling and the kernels are of the form
%\begin{equation}
%	i\int \der^2 \kt \sqrt{\as(\kt)} \frac{e^{-i \kt \cdot \xt}}{\kt^2}.
%\end{equation}
When evaluating the strong coupling in momentum space $4e^{-2\gamma_E}/(\rt^2 \lqcd^2)$ is replaced by $\kt^2/\lqcd^2$ in \eq\eqref{eq:as-r-jimwlk-nlobk}. The obtained dipole amplitudes are shown in \fig\ref{fig:jimwlk-k-r}, and the agreement between the results obtained with different running couplings are remarkably similar. This justifies the usage of the constant $4e^{-2\gamma_E}$ when evaluating the transverse coordinate space strong coupling constant.

\section{The BK equation at next to leading order}
\label{sec:nlobk}
The BK equation describes the energy evolution of the dipole amplitude at leading order accuracy. As the dipole amplitude encodes the relevant degrees of freedom to describe a scattering process, the BK evolution also describes the energy dependence of many observables such as single inclusive particle production and deep inelastic scattering. In \paper \cite{Lappi:2015fma} we studied the BK equation at next to leading order (NLO) accuracy.

The leading order calculations can give a good physical description of a scattering process, but the next to leading order corrections to the perturbative calculations of cross sections can be numerically large. Thus it is important to perform CGC calculations at next to leading order accuracy in $\as$ and compare the results quantitatively to experimental data. For example, the single inclusive cross section is already known at NLO accuracy~\cite{Chirilli:2011km,Chirilli:2012jd,Stasto:2013cha,Altinoluk:2014eka},
and similar calculations for the deep inelastic scattering cross section also exist~\cite{Balitsky:2010ze,Beuf:2011xd}. A crucial ingredient in phenomenological NLO calculations is an NLO evolved dipole amplitude.

The next to leading order BK equation has been derived in~\cite{Balitsky:2008zza}, but a numerical solution for the equation has not been available previously (for the NLO BFKL equation~\cite{Fadin:1996nw,Fadin:1998py,Ciafaloni:1998gs} a solution exists~\cite{Avsar:2011ds}). The equation can be written for the scattering matrix $S=1-N$ as 
\begin{multline}
\label{eq:nlobk}
 	\partial_y S(r) = \frac{\as \nc}{2\pi^2} K_1 \otimes [S(X)S(Y)-S(r)] \\
		+ \frac{\as^2 \nc^2}{8\pi^4} K_2 \otimes [S(X)S(\zt-\zt')S(Y')-S(X)S(Y)]  \\
		+ \frac{\as^2 \nf \nc}{8\pi^4} K_f \otimes S(Y)[S(X')-S(X)].
 \end{multline} 
Here convolutions $\otimes$ are calculated by integrating over the transverse coordinates of the emitted gluons $\zt$ and $\zt'$, and we use a notation $r=|\xt-\yt|$, $X=|\xt-\zt|$, $Y=|\yt-\zt|$, $X'=|\xt-\zt'|$ and $Y'=|\yt-\zt'|$ . The kernels are
\begin{align}
K_1 &= \frac{r^2}{X^2Y^2} \left[ 1+\frac{\as\nc }{4\pi} \left(  \frac{\beta}{\nc} \ln r^2\mu^2 - \frac{\beta}{\nc} \frac{X^2-Y^2}{r^2} \ln \frac{X^2}{Y^2} \right. \right. \nonumber \\
	& \left.\left. + \frac{67}{9} - \frac{\pi^2}{3} - \frac{10}{9} \frac{\nf}{\nc} - \ln \frac{X^2}{r^2} \ln \frac{Y^2}{r^2} \right) \right] 
\\
K_2 &= -\frac{2}{(z-z')^4} + \left[ \frac{X^2 Y'^2 + X'^2Y^2 - 4r^2(z-z')^2}{(z-z')^4(X^2Y'^2 - X'^2Y^2)} \right. \nonumber  \\
	&\left.  + \frac{r^4}{X^2Y'^2(X^2Y'^2 - X'^2Y^2)} + \frac{r^2}{X^2Y'^2(z-z')^2} \right]   \ln \frac{X^2Y'^2}{X'^2Y^2} 
\\
 K_f &= \frac{2}{(z-z')^4}  
	- \frac{X'^2Y^2 + Y'^2 X^2 - r^2 (z-z')^2}{(z-z')^4(X^2Y'^2 - X'^2Y^2)} \ln \frac{X^2Y'^2}{X'^2Y^2} 
\end{align}
The kernel $K_1$ consists of a LO BK kernel $r^2/(X^2Y^2)$ and an NLO correction. Part of the NLO corrections, especially the term involving the renormalization scale $\mu^2$, should be absorbed into the running of $\as$. What other terms are absorbed into the coupling is a scheme choice, and here we adapt the choice derived in~\cite{Balitsky:2006wa} and replace the terms including the beta function coefficient $\beta = \frac{11}{3}\nc - \frac{2}{3}\nf$ by the Balitsky running coupling given in \eq\eqref{eq:bk-balitsky}. The other $\as^2$ terms we choose to evaluate at the scale set by the parent dipole $r^2$, as it is the only available external scale. Thus, the kernel $K_1$ is written as
\begin{multline}
	\frac{\as \nc}{2\pi^2} K_1 = \frac{\as(r) \nc}{2\pi^2} \left[\frac{r^2}{X^2Y^2} + \frac{1}{X^2} \left(\frac{\as(X)}{\as(Y)}-1\right) + \frac{1}{Y^2} \left(\frac{\as(Y)}{\as(X)}-1\right) \right] \\
		+ \frac{\as(r)^2 \nc^2}{8\pi^3} \frac{r^2}{X^2Y^2} \left[ \frac{67}{9} - \frac{\pi^2}{3} - \frac{10}{9} \frac{\nf}{\nc} - 2\ln \frac{X^2}{r^2} \ln \frac{Y^2}{r^2} \right],
\end{multline}
and the coupling constants multiplying kernels $K_2$ and $K_f$ are replaced by $\as(r)$. For the running coupling we use the same expression as in the case of JIMWLK analysis in \se \ref{sec:jimwlk}, \eq \eqref{eq:as-r-jimwlk-nlobk}.

The equation \eqref{eq:nlobk} is written in the large-$\nc$ limit where the evolution equation can be written in terms of dipole operators only. If finite-$\nc$ corrections were included, the evaluation of the derivative $\partial_y S(r)$ would require one to evaluate correlators of up to six Wilson lines. Without a numerical solution to the NLO JIMWLK equation~\cite{Balitsky:2013fea,Kovner:2013ona}, which is currently not available, a possible way to evaluate the higher point functions would be to use the so called Gaussian approximation which allows one to write any higher point function in terms of the dipole as described in \re \cite{Dominguez:2012ad}. As the finite-$\nc$ corrections to the leading order BK equation are known to be very small~\cite{Kovchegov:2008mk}, the inclusion of the $\nc$ suppressed terms to the NLO BK equation was not attempted in \paper\cite{Lappi:2015fma}.

The Wilson lines are by definition (see \eq \eqref{eq:wilson-line}) conformally invariant, so the evolution equation should not break this invariance in a conformal field theory. In QCD the conformal invariance is broken by the running coupling effects, but if the running of $\as$ is not taken into account, the NLO BK equation should be conformally invariant.
Let us explicitly check the invariance. First, the Wilson lines are clearly invariant under translations and rotations. The only non-trivial symmetry that has to be checked is the invariance under the spatial inversion. Consider a transformation $x^\mu \to x^\mu/x^2$ with respect to a point with $x^-=0$. Now $(x^+,\xt)^2=-\xt^2$. The Wilson line now transforms as
\begin{align}
V(x_T) &= \PE  \left[-ig \int_{-\infty}^\infty \der x^+ A^-(x^+, x_T) \right] \nonumber \\
&\to \PE \left[ -ig\int_{-\infty}^\infty \frac{\der x^+}{-\xt^2} A^-\left(\frac{x^+}{-\xt^2}, \frac{\xt}{-\xt^2}\right) \right] = V\left(\frac{\xt}{-\xt^2}\right),
\end{align}
which shows the invariance.

However, the NLO BK equation \eqref{eq:nlobk} is not conformally invariant. The running coupling part (proportional to $\beta$) is expected to break the invariance, but in addition to that the double logarithmic term $\ln X^2/r^2 \ln Y^2/r^2$ in \eq\eqref{eq:nlobk} is not invariant in conformal transformations. Before proving that, let us first show that the leading order BK equation is indeed invariant. Again, the only non-trivial transformation is inversion. To show this, it is easiest to write the two-dimensional vectors as  complex numbers $x=x_1+ix_2$, $y=y_1+iy_2$ and $z=z_1 + iz_2$. Now, under an inversion all vectors $x,y,z$ transform as
\begin{equation}
	z = z_1 + i z_2 \to \frac{z_1 + i z_2}{z_1^2 + z_2^2} = \frac{z}{z \bar z} = \frac{1}{\bar z},
\end{equation}
where $\bar z = z_1 - i z_2$ is the complex conjugate of $z$. Now the squared distances transform as
\begin{equation}
	|x-y|^2 = (x-y)(\bar x - \bar y) \to \left( \frac{x}{|x|^2} - \frac{y}{|y|^2}\right)\left( \frac{\bar x}{|x|^2} - \frac{\bar y}{|y|^2}\right) = \frac{|x-y|^2}{|x|^2|y|^2}.
\end{equation}
Using this result one can directly transform the kernel:
\begin{equation}
	\frac{|r|^2}{|X|^2 |Y|^2} \to \frac{|r|^2 |z|^4}{|X|^2|Y|^2}.
\end{equation}
As the integration measure transforms as $\der^2 z \to \der^2 z/|z|^4$, the leading order BK equation is found to be invariant under the inversion.

On the other hand, the double logarithmic term transforms as
\begin{equation}
	\ln \frac{|X|^2}{|r|^2} \ln \frac{|Y|^2}{|r|^2} \to \ln \left( \frac{|X|^2}{|r|^2} \frac{|y|^2}{|z|^2} \right) \ln \left( \frac{|X|^2}{|r|^2} \frac{|x|^2}{|z|^2} \right) 
\end{equation}
and is not invariant under the inversion.
Similarly one can check that the other combinations of transverse coordinates in the kernels $K_2$ and $K_f$ are invariant under inversions.

% lähteitä laskuihin: 1102.4040 ja hep-ph/0306279

The conformal symmetry breaking is a consequence of having a cutoff in the longitudinal direction when deriving the NLO BK equation. This cutoff violates the conformal symmetry.
If the NLO BK equation is derived in the fully conformal $N=4$ supersymmetric Yang-Mills (SYM) theory (which is similar to QCD but has no running coupling), the resulting equation still has the double logarithmic term and is not conformally invariant~\cite{Balitsky:2009xg}. This confirms that the conformal breaking is an artifact of the use of a cutoff which does not respect the required symmetries.

To solve this problem it was proposed in \re\cite{Balitsky:2009xg} to write the equation in terms of the conformal dipole $S(r)^\text{conf}$, defined as
\begin{equation}
\label{eq:confdipole}
	S(r)^\text{conf} = S(r) - \frac{\as \nc}{4\pi^2} \int \der^2 z \frac{r^2}{X^2Y^2} \ln \frac{ar^2}{X^2Y^2} \left[S(X)S(Y) - S(r) \right].
\end{equation}
Here $a$ is an arbitrary dimensional constant which will cancel from the evolution equation. When the NLO BK equation is derived for the conformal dipole, the double logarithmic term disappears from the equation, and an additional term
\begin{equation}
	\frac{2r^2}{X^2Y'^2 (z-z')^2} \ln \frac{r^2 (z-z')^2}{X'^2Y^2}
\end{equation}
appears in the kernel $K_2$. The conformal invariance is then restored if running of $\as$ is not taken into account. In $N=4$ SYM the corresponding equation is fully conformal.

The first numerical solution to the NLO BK equation is presented in \paper \cite{Lappi:2015fma}. As an initial condition in this analysis we used a modified McLerran-Venugopalan model \eqref{eq:mv} with an anomalous dimension $\gamma$
\begin{equation}
	N(r,y=0) = 1 - \exp \left[ \frac{(\rt^2 \qso^2)^\gamma}{4} \ln \left(\frac{1}{r\lqcd}+e\right)\right],
\end{equation}
where the anomalous dimension controls the power-like tail of the dipole amplitude for small dipoles. This parametrization is used in leading order (with Balitsky running coupling corrections from \eq\eqref{eq:bk-balitsky}) fits to DIS data e.g. in \re \cite{Albacete:2010sy} and in \paper \cite{Lappi:2013zma}. There is, however, no particular reason why a full NLO fit would prefer the same values for the fit parameters. Thus we only explore the general properties of the evolution equation with different parametrizations for the initial condition.

As we will demonstrate later, the dipole amplitude may turn negative during the evolution and may not satisfy the requirement $N(r)\to 0$ when $r\to 0$, which follows from the definition \eqref{eq:n-def}. In this case also the convolution with $K_1$ would not coverge. To avoid this problem we freeze $N(r)=0$ in the region where it would become negative.

\begin{figure}[tb]
\centering
	\begin{minipage}[t]{0.5\textwidth}
	\includegraphics[width=\textwidth]{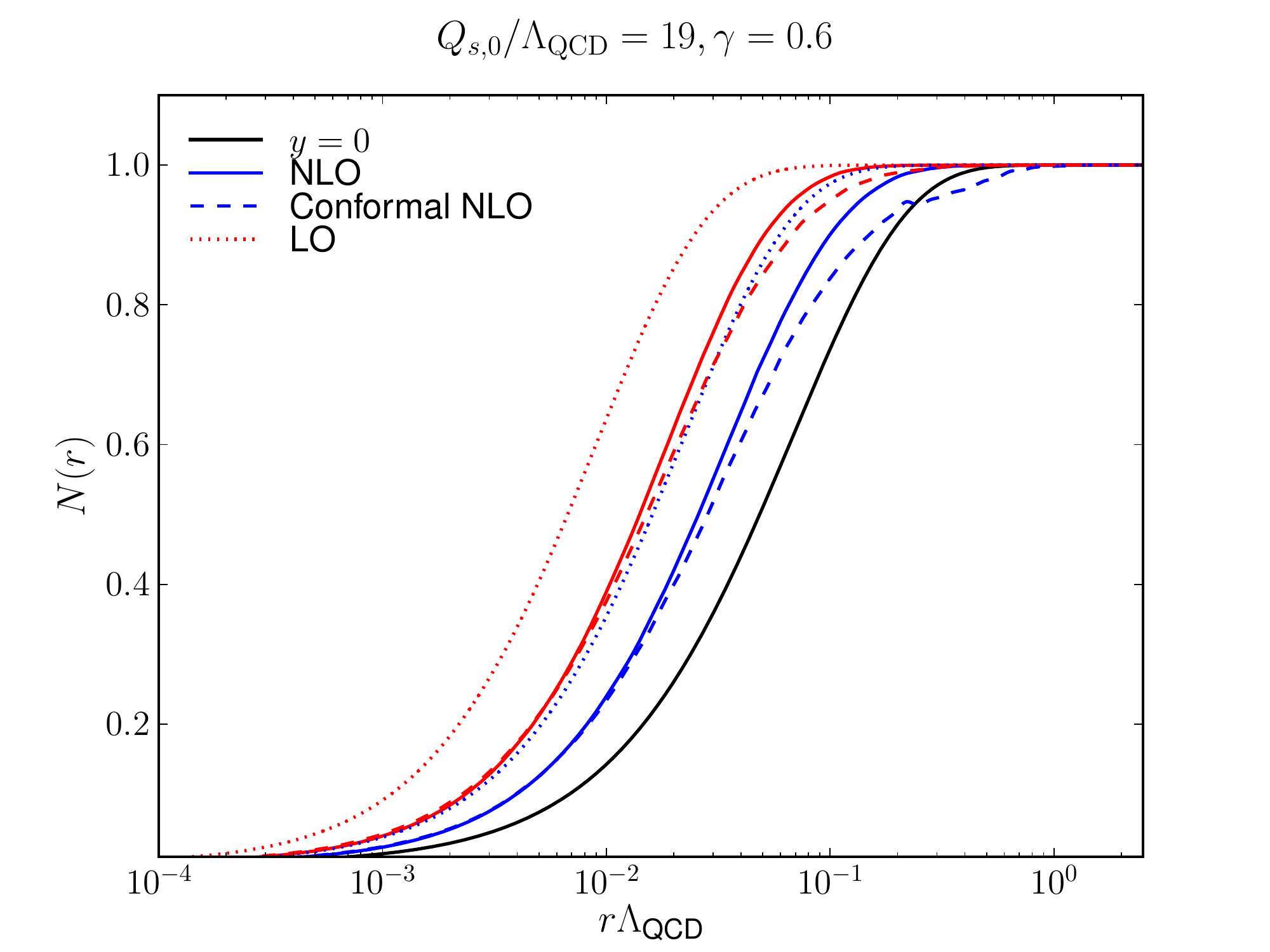}
	\caption{Dipole amplitude and conformal dipole amplitude at initial condition and at rapidities $y=5$ and $y=10$ (from riht to left) compared to leading order BK equation. Figure from \paper \cite{Lappi:2015fma}}
	\label{fig:nlobk_amplitude}
	\end{minipage}%
	~
	\begin{minipage}[t]{0.5\textwidth}
	\includegraphics[width=1.0\textwidth]{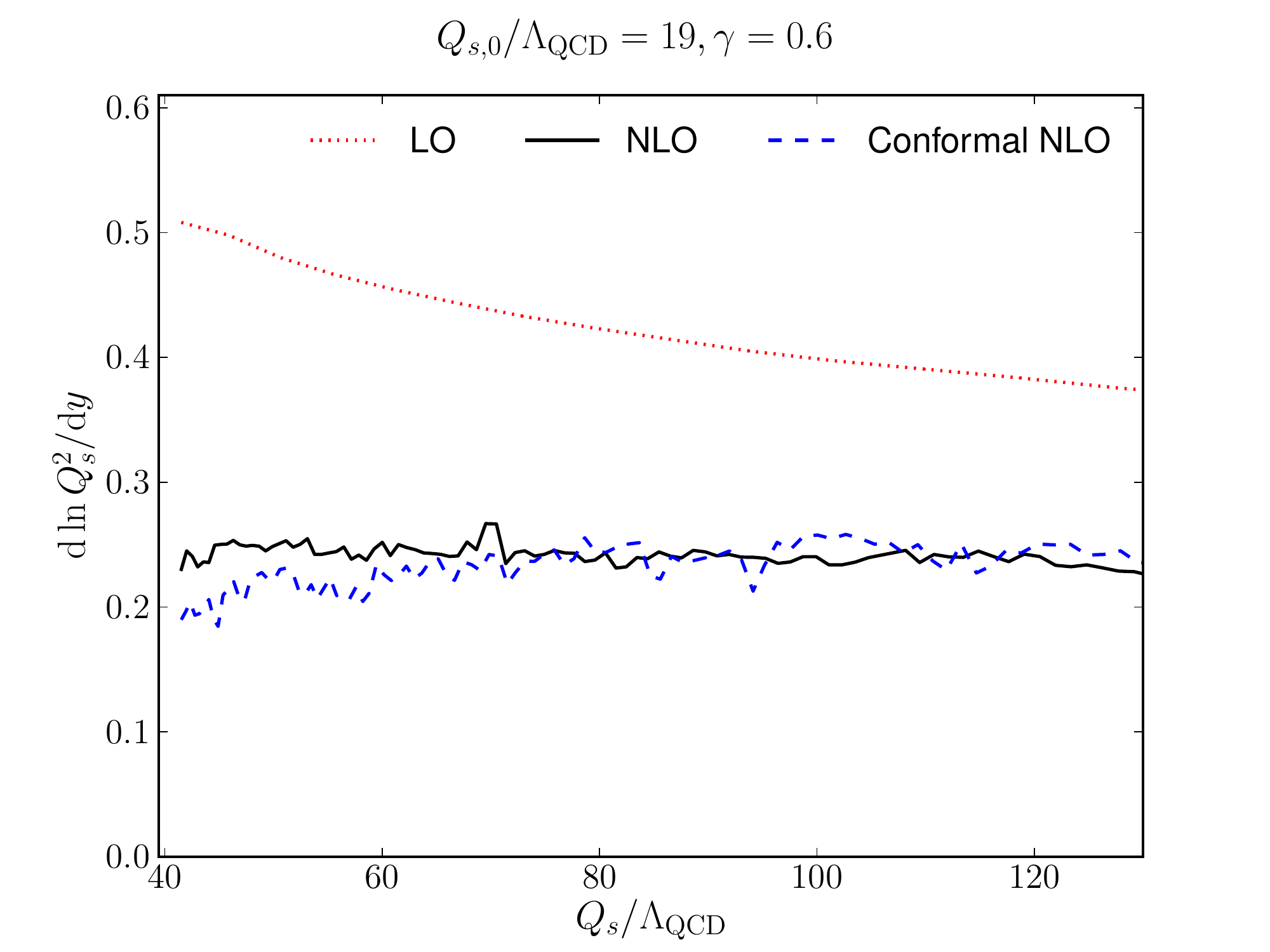}
	\caption{Evolution speed of the saturation scale for conformal and non-conformal dipoles as a function of the saturation scale compared to the LO BK equation. Figure from \paper \cite{Lappi:2015fma}.
	}
	\label{fig:dqs_nlobk}
	\end{minipage}	
\end{figure}

The dipole amplitudes obtained by solving the NLO BK equation for non-conformal and conformal dipoles are shown in \fig\ref{fig:nlobk_amplitude}. The results are compared with the leading order solution, and as an initial condition we use parameters $\qso/\lqcd\sim 19,\gamma=0.6$ chosen such that the dipole amplitude increases at all dipole sizes throughout the evolution considered here. We observe that the NLO corrections slow down the evolution speed of both NLO dipoles equally compared to the leading order evolution. The shape of the dipole amplitude, on the other hand, remains roughly unchanged. The leading order BK equation is solved using the Balitsky running coupling prescription~\cite{Balitsky:2006wa} in order to get comparable results.

To quantify more precisely the effect of the NLO corrections we study the evolution speed of the saturation scale $\lambda$, defined as
\begin{equation}
	\lambda = \frac{\der \ln Q_s^2}{\der y},
\end{equation}
where the saturation scale is defined as
\begin{equation}
	N(r^2=2/Q_s^2) = 1-e^{-1/2}.
\end{equation}
Note that $Q_s$ is different from $\qso$ at the initial condition: with $\qso/\lqcd\sim 19$ we get $Q_s/\lqcd \sim 40$. The results are shown in \fig\ref{fig:dqs_nlobk}. The NLO evolution for conformal and non-conformal dipoles is roughly equally fast, and significantly slower than the evolution speed at leading order.
The slower evolution speed at NLO can somewhat be expected from the leading order fits where the scale at which the transverse coordinate space strong coupling constant is evaluated is fitted to the data. In order to get a slow enough evolution speed comparable with the experimental data this scale, parametrized by $C^2$ in \eq \eqref{eq:alphas-r-csqr}, must be taken to be relatively large. See discussion in  \papers\cite{Lappi:2012vw,Lappi:2013zma}.

\begin{figure}[ptb]
	\subfloat[$\gamma=0.6$]{%
		\includegraphics[width=0.333\textwidth]{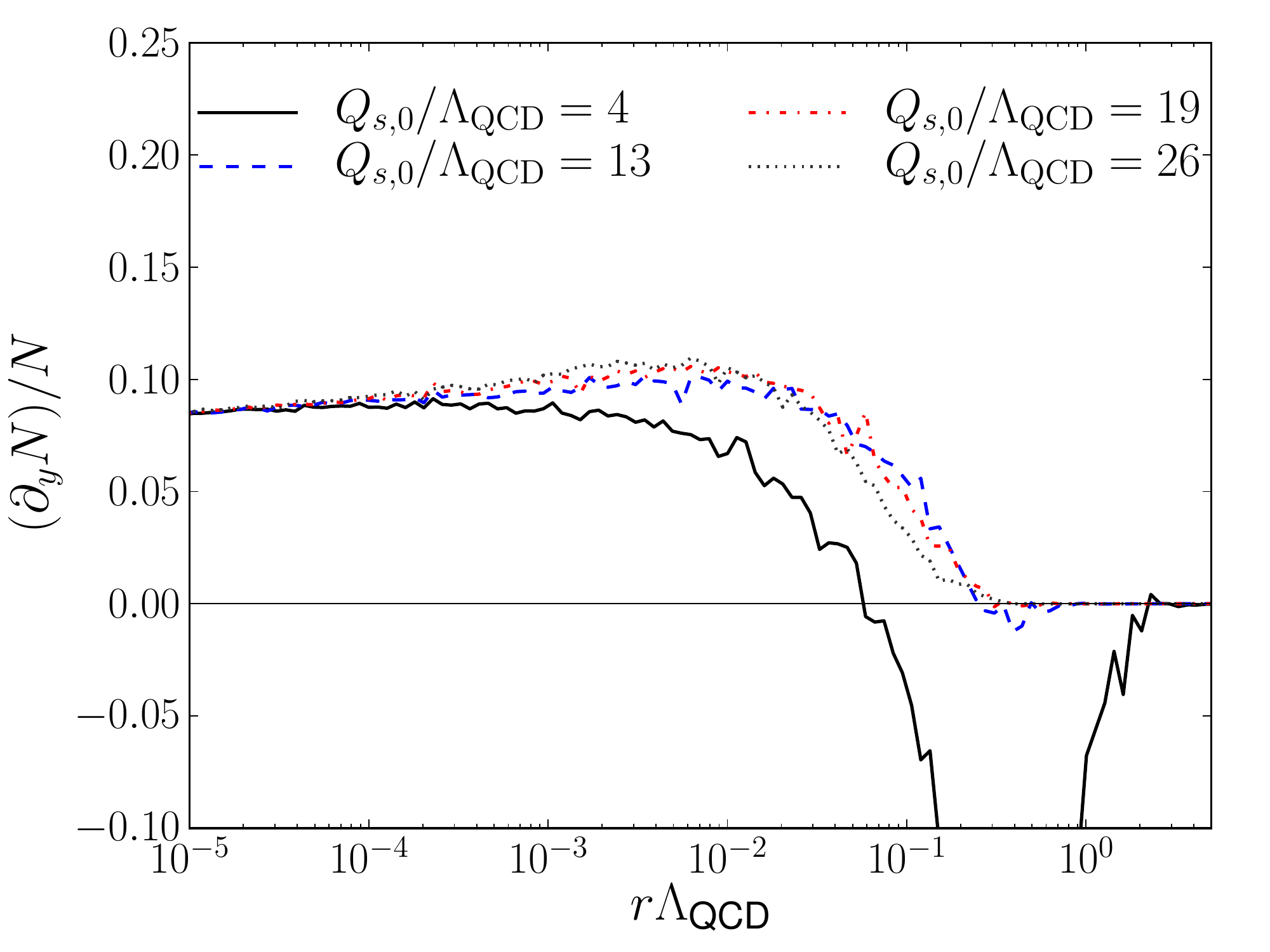}
		}% \hspace{-1.13em} %
	\subfloat[$\gamma=0.8$]{%
		\includegraphics[width=0.333\textwidth]{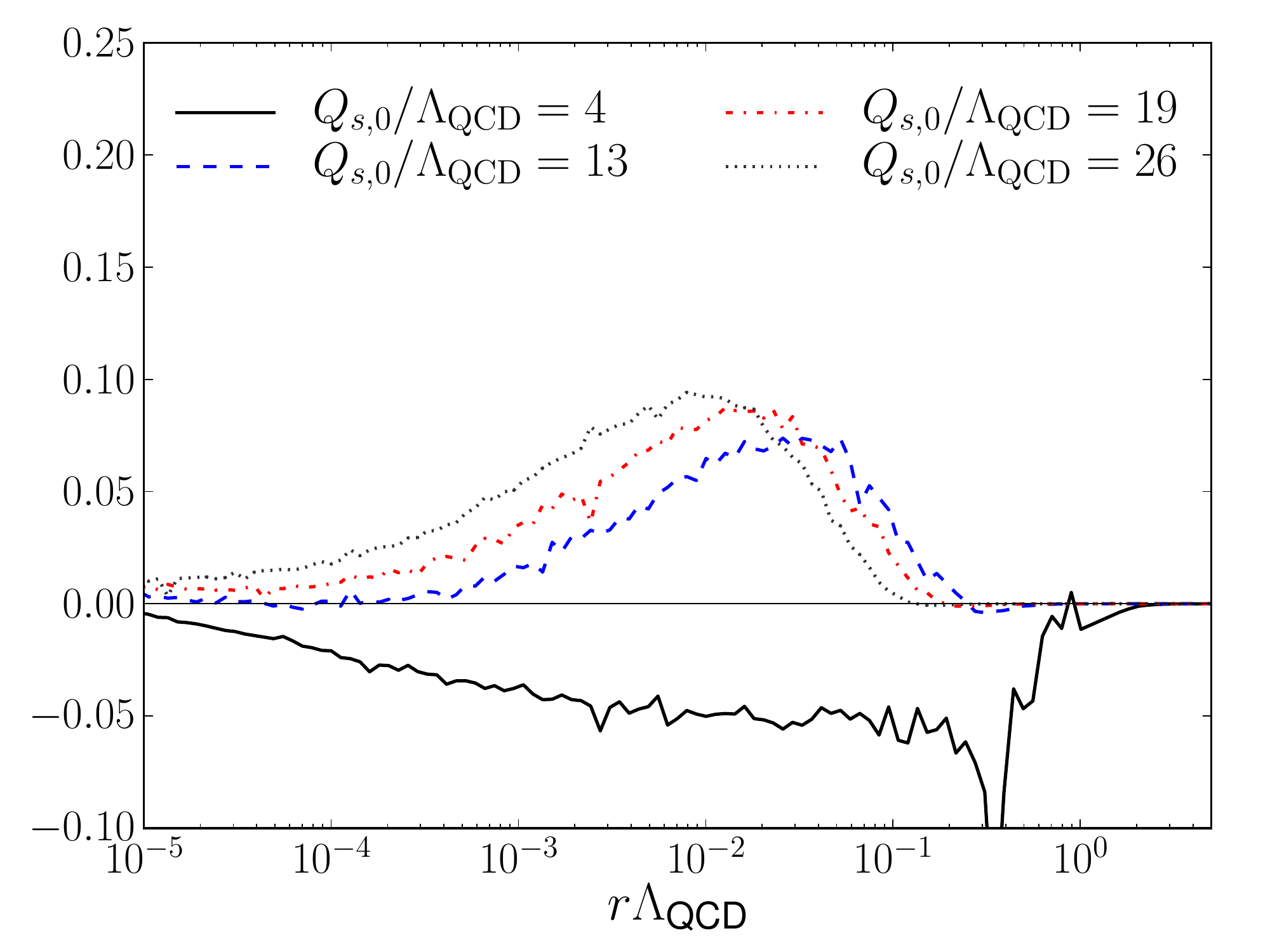}
		}% \hspace{-1em} %
	\subfloat[$\gamma=1.0$]{%
		\includegraphics[width=0.333\textwidth]{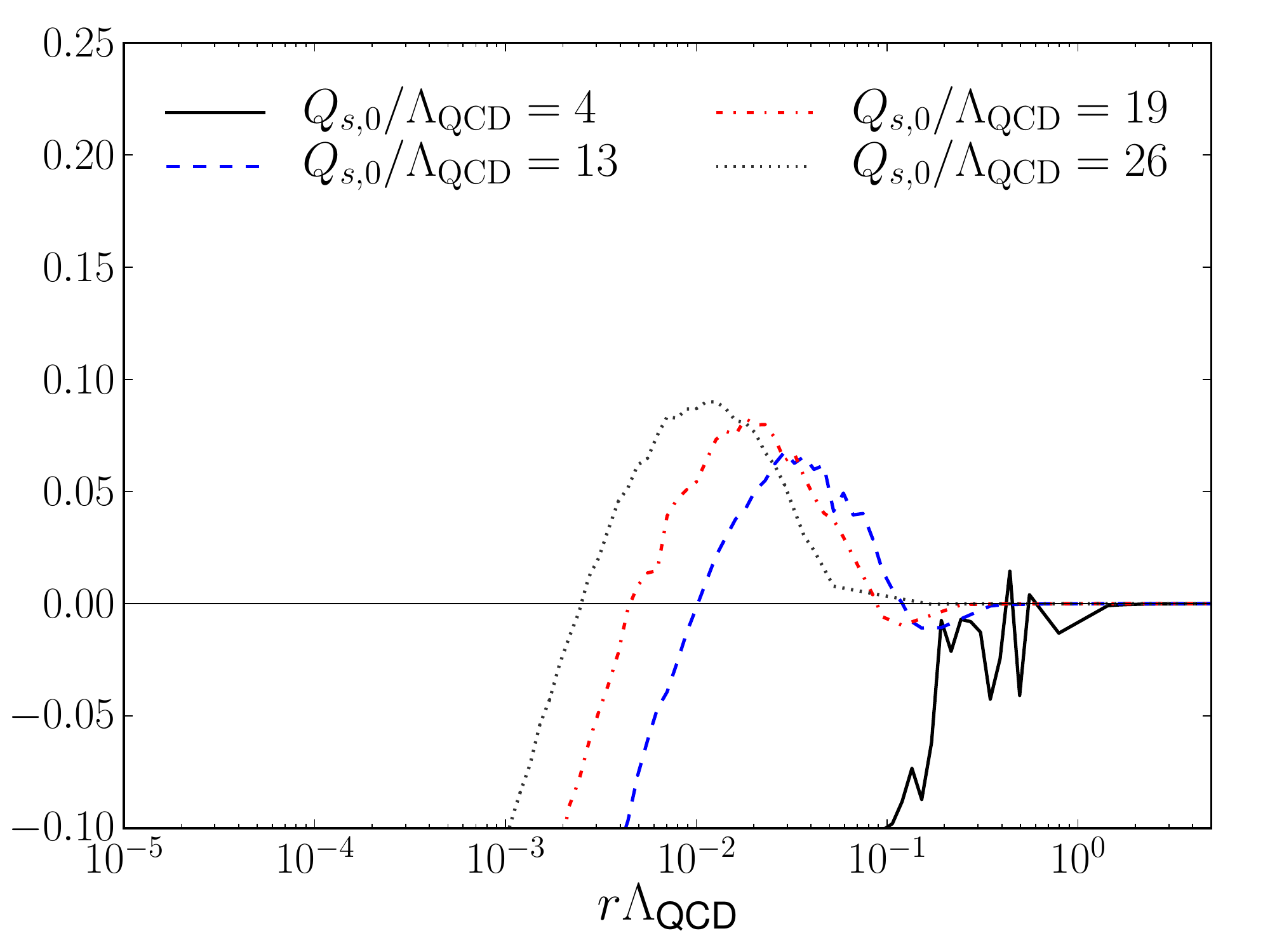}
		}	
	\caption{Logarithmic derivative of the dipole amplitude (evolution speed) at initial condition with different values for the saturation scale and the anomalous dimension. Figure from \paper \cite{Lappi:2015fma}.}
	\label{fig:dndy_dipole_y0}
\end{figure}

To study the evolution of the dipole amplitude as a function of the dipole size we show in \fig\ref{fig:dndy_dipole_y0} the logarithmic derivative of the dipole amplitude $\partial_y N(r)/N(r)$ at the initial condition with different values for the anomalous dimension $\gamma$. Generally one would expect the dipole amplitude to increase with rapidity, corresponding to the physical picture of having more gluons inside the hadron at small $x\sim e^{-y}$. What we find is that for small initial saturation scale $\qso$ the NLO corrections are so large around $r\sim 1/Q_s$ (which is the dominant scale) that the solution propagates in the ``wrong'' direction.

In order to interpret the logarithmic derivative shown in \fig\ref{fig:dndy_dipole_y0} note that if $\partial_y N/N$ has a constant positive value, the amplitude grows exponentially in rapidity and propagates to smaller dipole sizes. This is what happens with initial anomalous dimension $\gamma=0.6$, and marginally also with $\gamma=0.8$. With $\gamma=1.0$, on the other hand, we have $\partial_y N/N\sim \ln r$, which drives the amplitude towards a steeper shape. Eventually this leads to a singularity in the logarithmic derivative and $N(r)$ changing its sign at finite $r$. Similar results were also found in \re\cite{Iancu:2015vea}. As shown in \paper\cite{Lappi:2015fma}, the shape of $\partial_y N/N$ does not change significantly during the evolution. In other words, the evolution is sensitive to the initial condition still after a few rapidity steps. It is also shown that the conformal dipole has a similar evolution, but the amplitude crosses zero at smaller $r$, meaning that with the conformal dipole the ``negativity problem'' is less severe, but it still exists.

\begin{figure}[ptb]
	\subfloat[$y=1$]{%
		\includegraphics[width=0.333\textwidth]{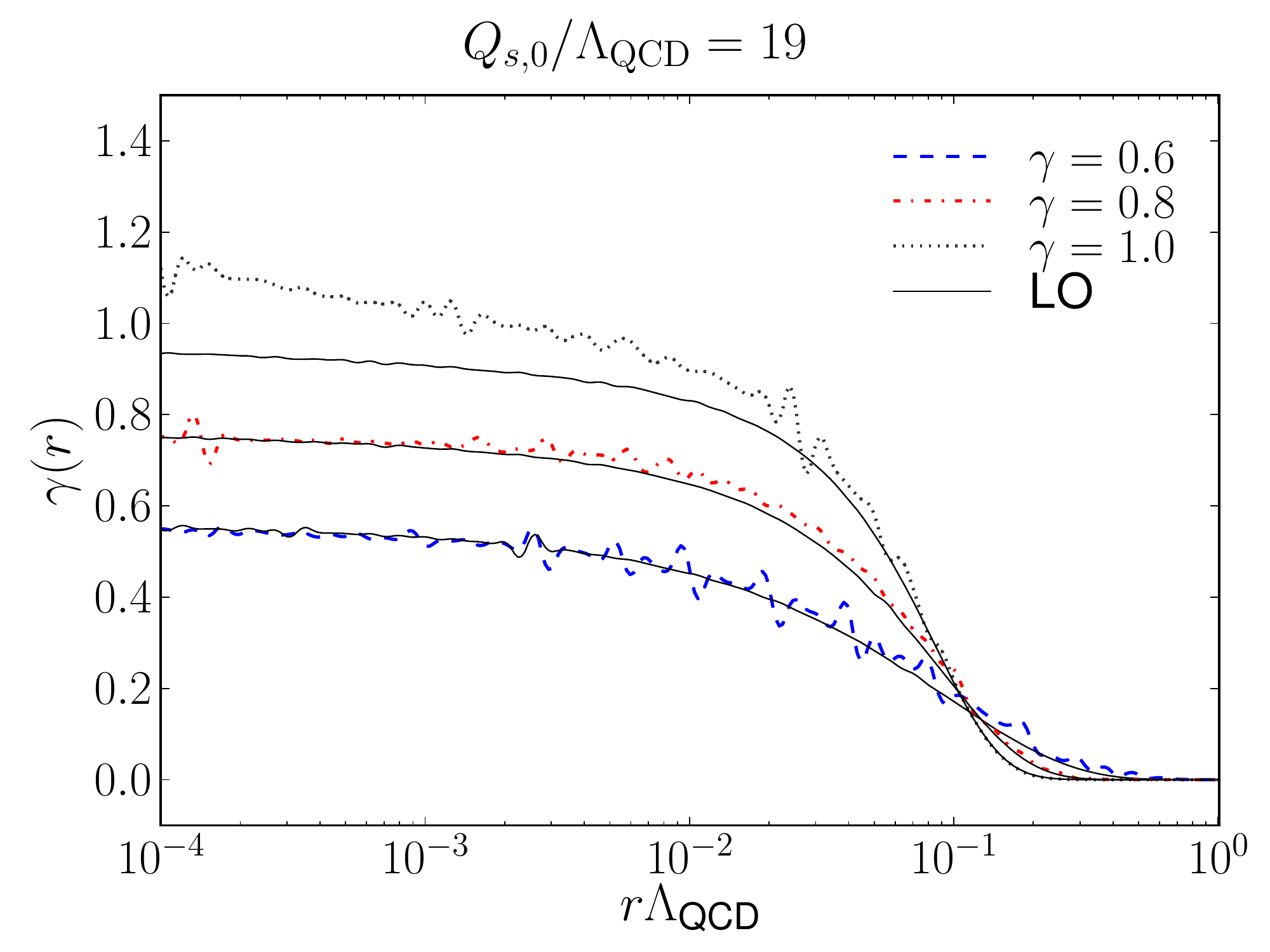}
		}%
	\subfloat[$y=5$]{%
		\includegraphics[width=0.333\textwidth]{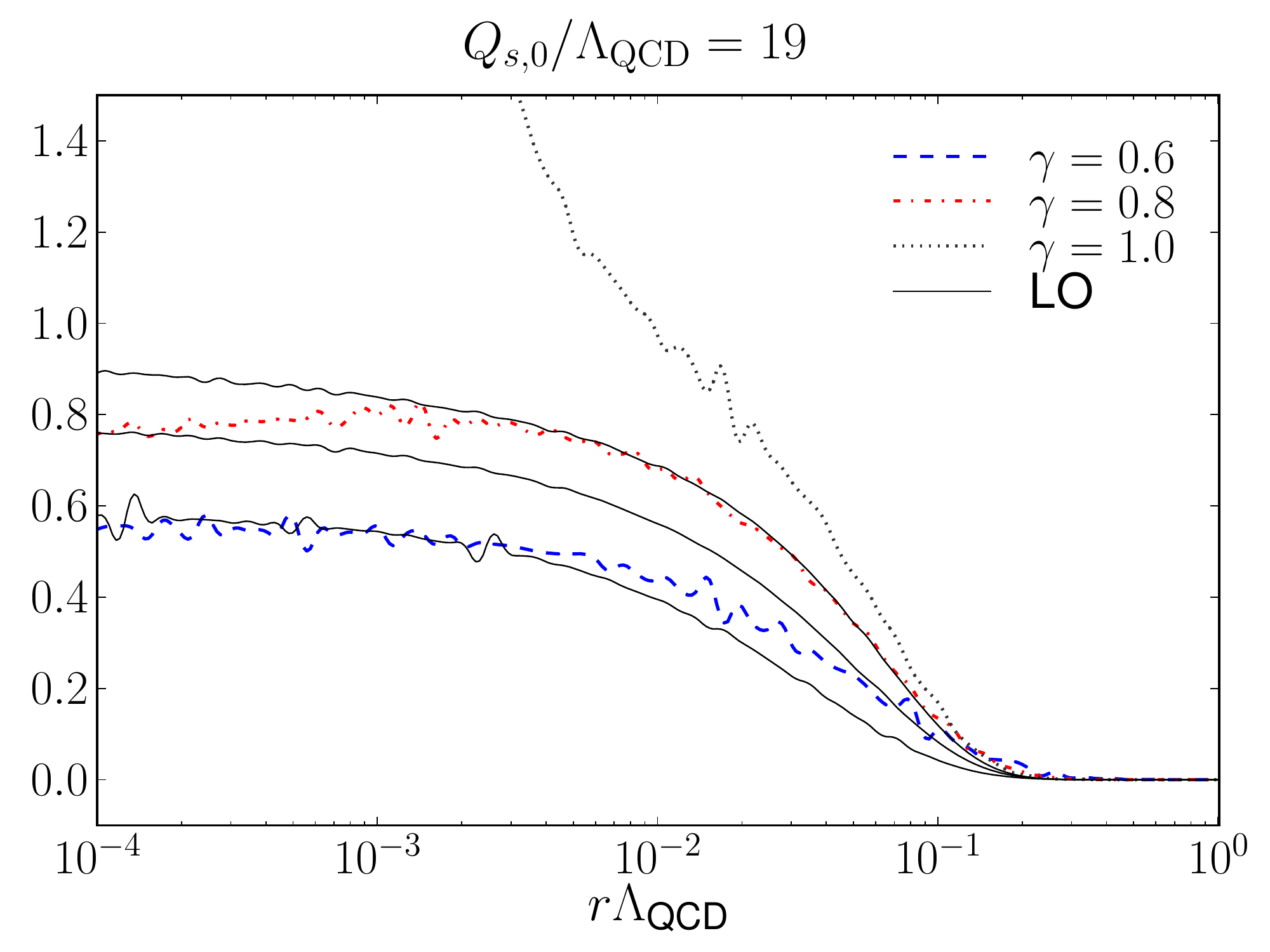}
		}%
	\subfloat[$y=30$]{%
		\includegraphics[width=0.333\textwidth]{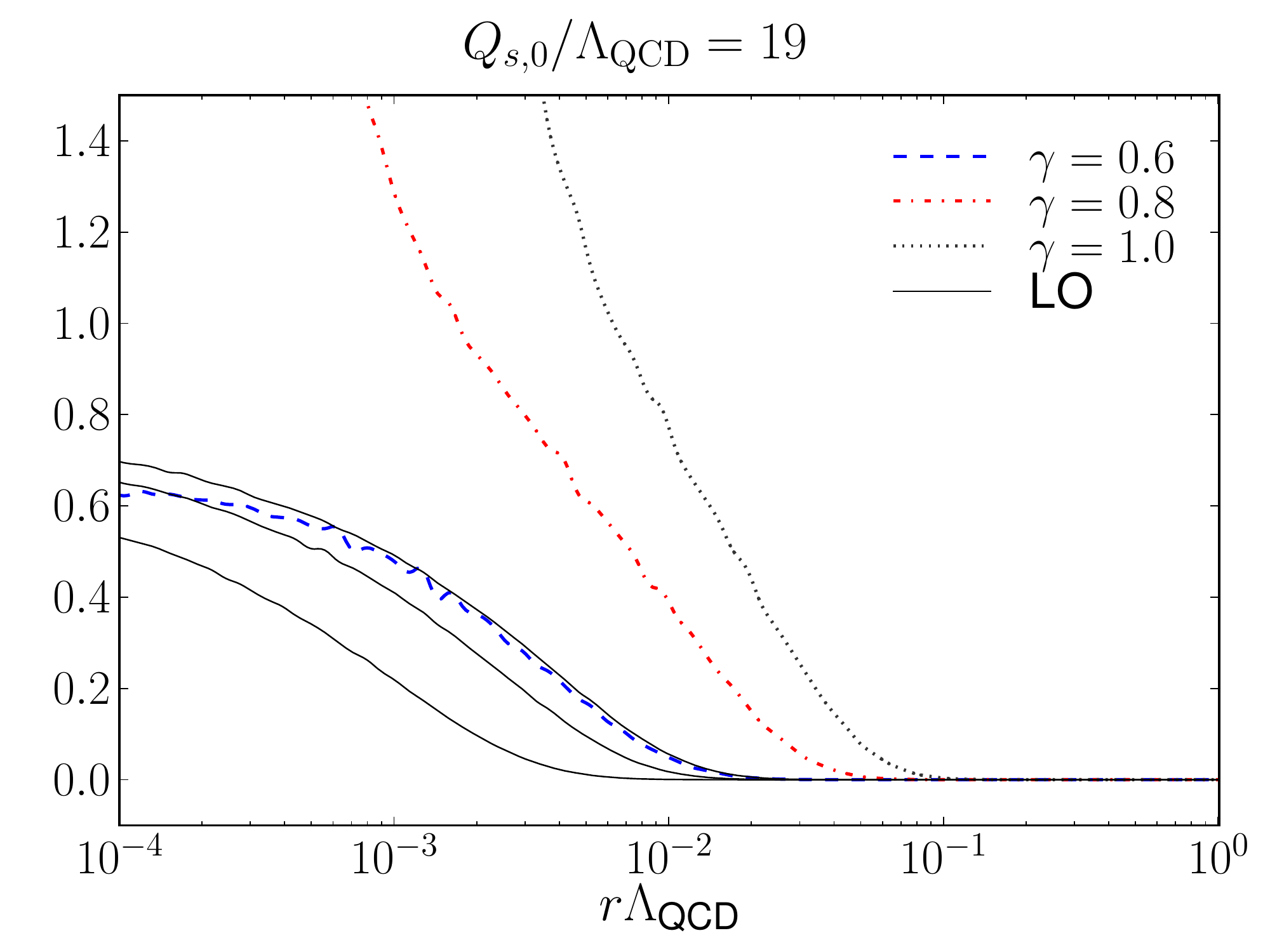}
		}			%
	\caption{Anomalous dimension $\gamma(r)$ as a function of dipole size at rapidities $y=1,5$ and $30$  with different initial anomalous dimensions. The solid lines are the solutions of the LO equation and the different dashed lines correspond to NLO evolution of non-conformal dipole with different initial anomalous dimensions. Figure from \paper \re\cite{Lappi:2015fma}.}
	\label{fig:gammaqcd}
\end{figure}

In order to study the evolution of the dipole amplitude shape we calculate the anomalous dimension as a function dipole size, defined as
\begin{equation}
	\gamma(r) = \frac{\der \ln N(r)}{\der \ln r^2}.
\end{equation}
The results are shown in \fig\ref{fig:gammaqcd} where the anomalous dimension $\gamma(r)$ is evaluated at rapidities $y=1$, $y=5$ and at $y=30$ with different initial anomalous dimensions. The fact that the solution becomes unstable is clearly visible with $\gamma=1$ where the anomalous dimension grows very rapidly already at $y=5$. With $\gamma=0.8$ much longer evolution in rapidity is needed before the unstable region is reached. With small anomalous dimension in the initial condition (here $\gamma=0.6$), the unstable behavior is not observed within the studied evolution range. Note that for $\gamma=1.0$ the solution does not evolve significantly from $y=5$ to $y=30$ as the evolution is dominated by region where the amplitude would be negative we have frozen $N(r)=0$. The conclusions for the conformal dipole are similar.

\begin{figure}[tb]
	\subfloat[$\gamma=0.6$]{%
		\includegraphics[width=0.333\textwidth]{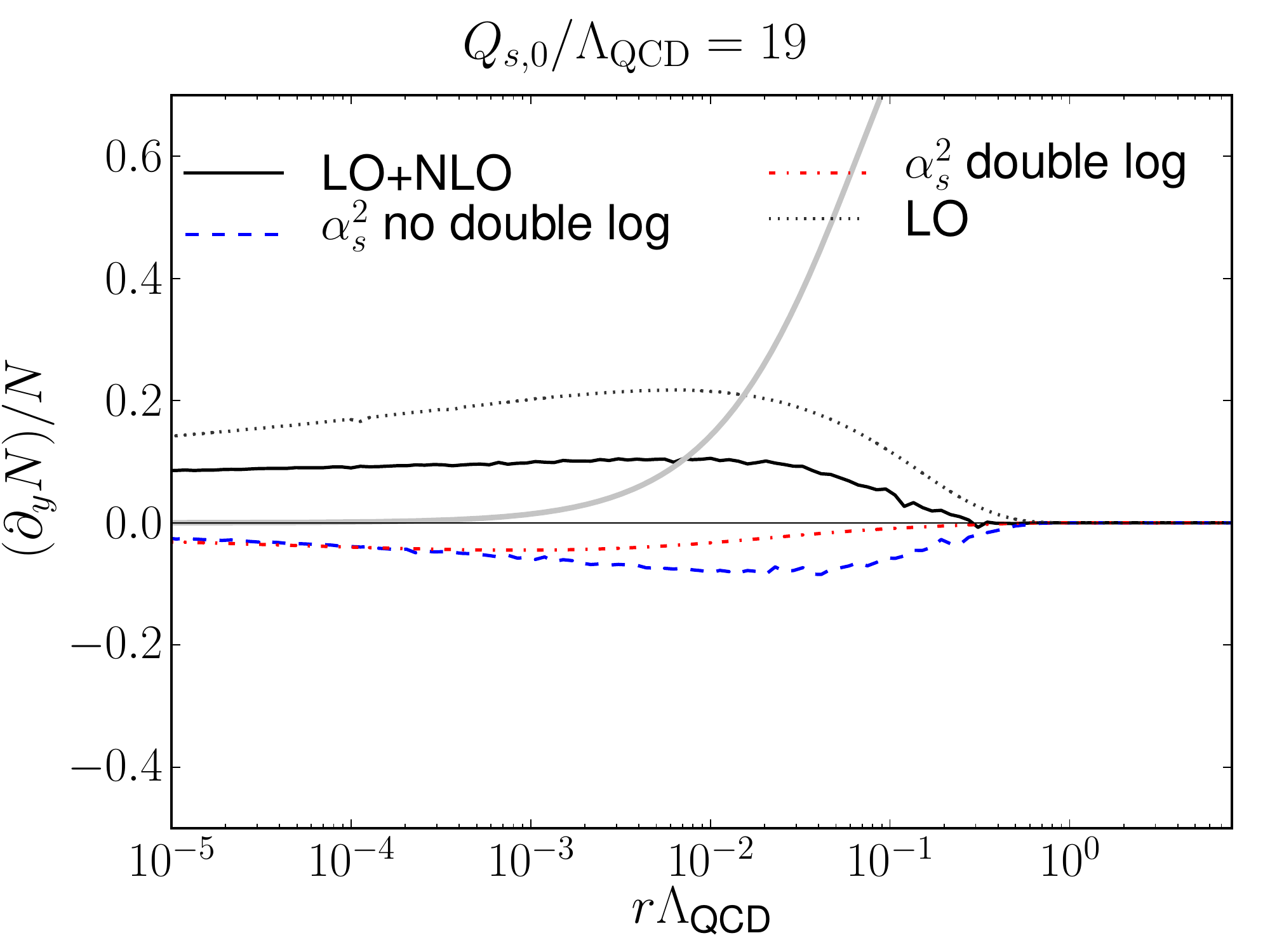}
		}%
	\subfloat[$\gamma=0.8$]{%
		\includegraphics[width=0.333\textwidth]{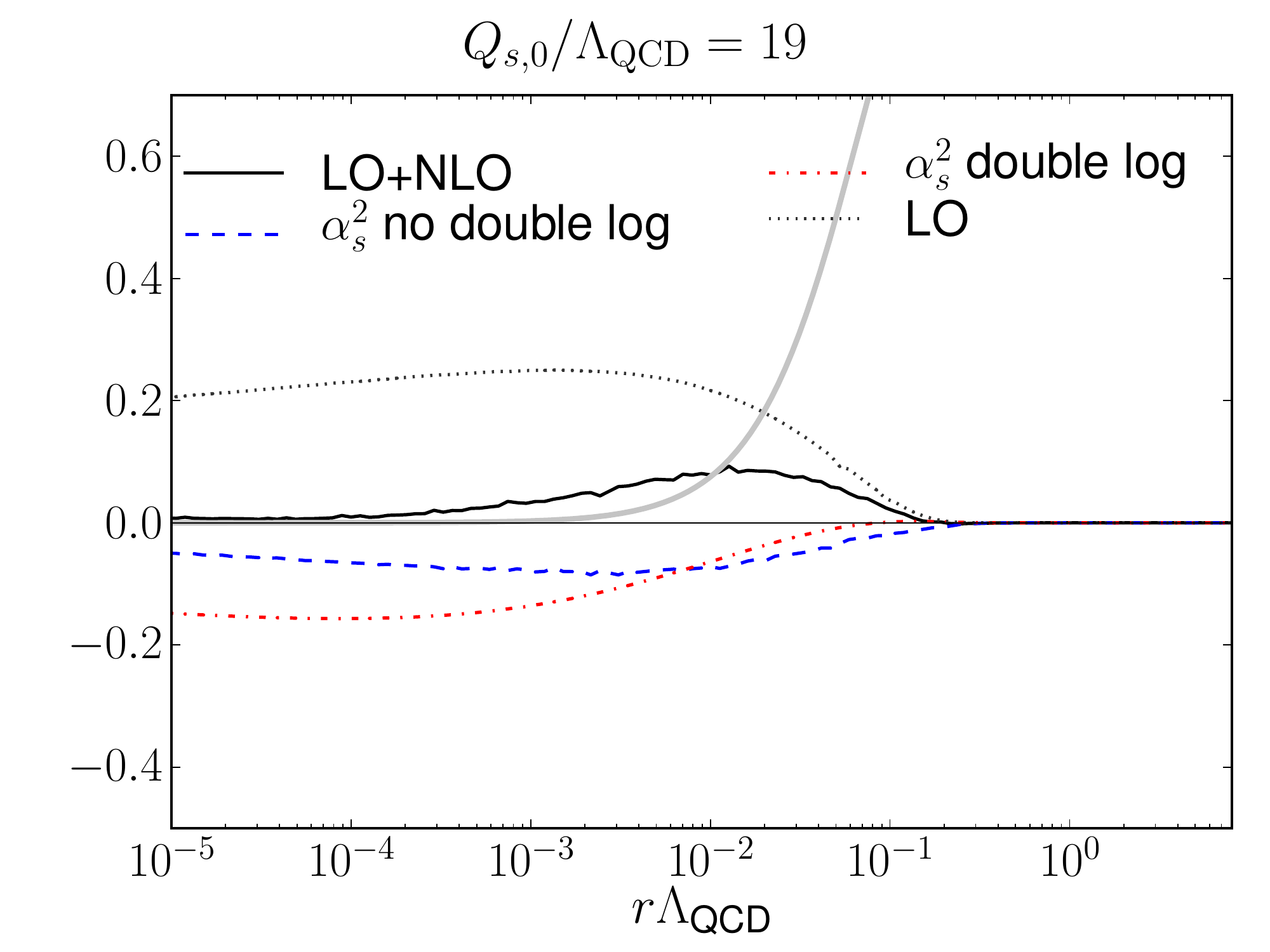}
		}%
	\subfloat[$\gamma=1.0$]{%
		\includegraphics[width=0.333\textwidth]{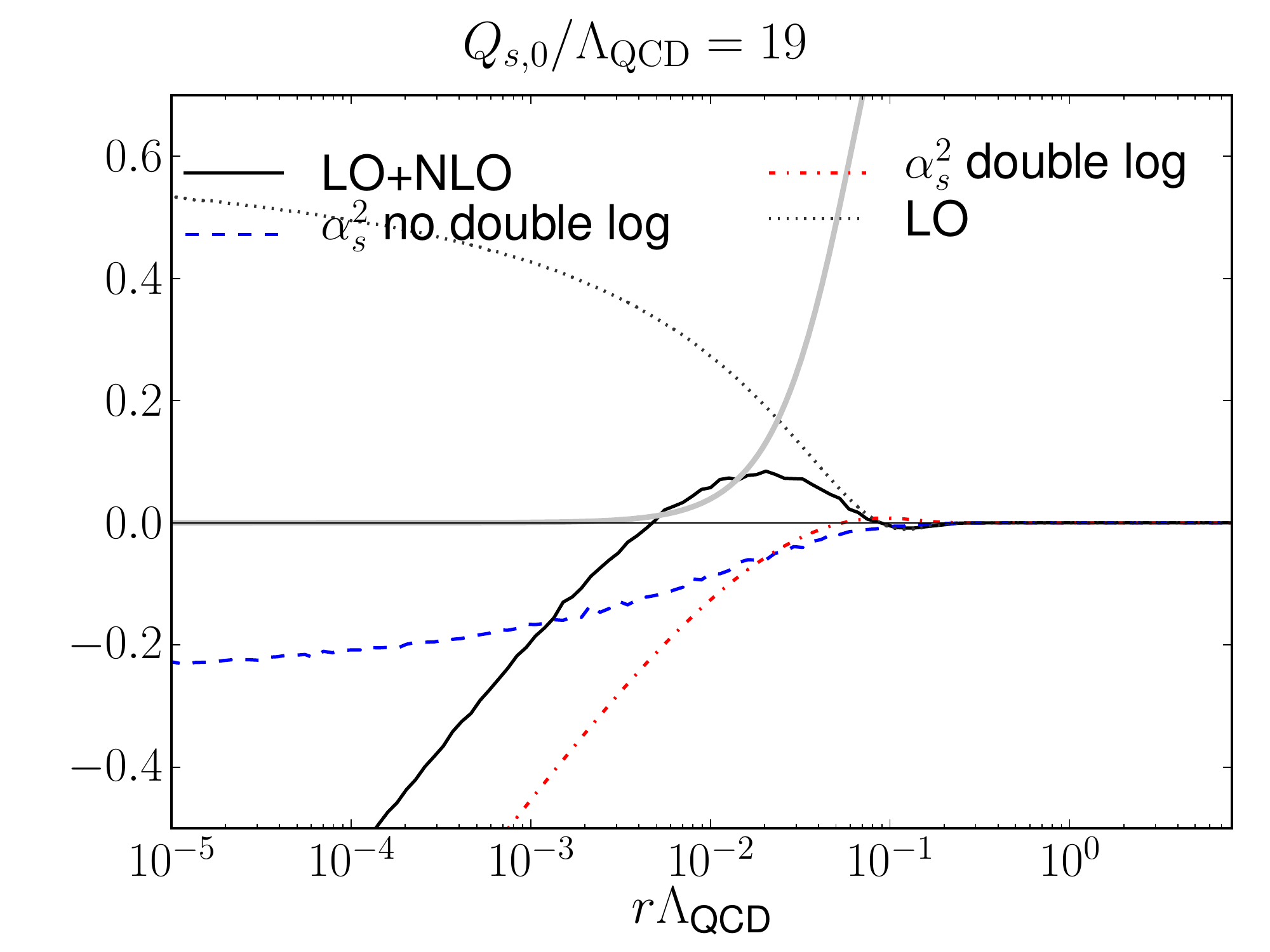}
		}%
	\caption{Evolution speed of the dipole amplitude at the initial condition. Shown are separately the full NLO and LO results and the NLO contributions from the conformal (no double logarithmic) and non-conformal (only double logarithmic) terms. Figure from \paper\cite{Lappi:2015fma}. }
	\label{fig:dndy_qcd}
\end{figure}

To study where the unstable behavior originates from we calculate contributions from different terms in the NLO evolution equation to the logarithmic derivative of the dipole amplitude, $\partial_y N/N$. In \fig\ref{fig:dndy_qcd} we show separately the leading order contribution, the NLO contribution originating from the double logarithmic term (which breaks the conformal invariance) $\sim \as^2 \ln X^2/r^2 \ln Y^2/r^2$ and the other NLO contributions. With large anomalous dimension in the initial condition (especially with $\gamma=1.0$) the double logarithmic term drives the evolution speed and is responsible for turning the dipole amplitude negative. With smaller initial anomalous dimension $\gamma=0.6$ both NLO contributions approach zero in the small dipole limit, and $\gamma=0.8$ is again a marginal case.

\begin{figure}[tb]
	\subfloat[$\gamma=0.6$]{%
		\includegraphics[width=0.333\textwidth]{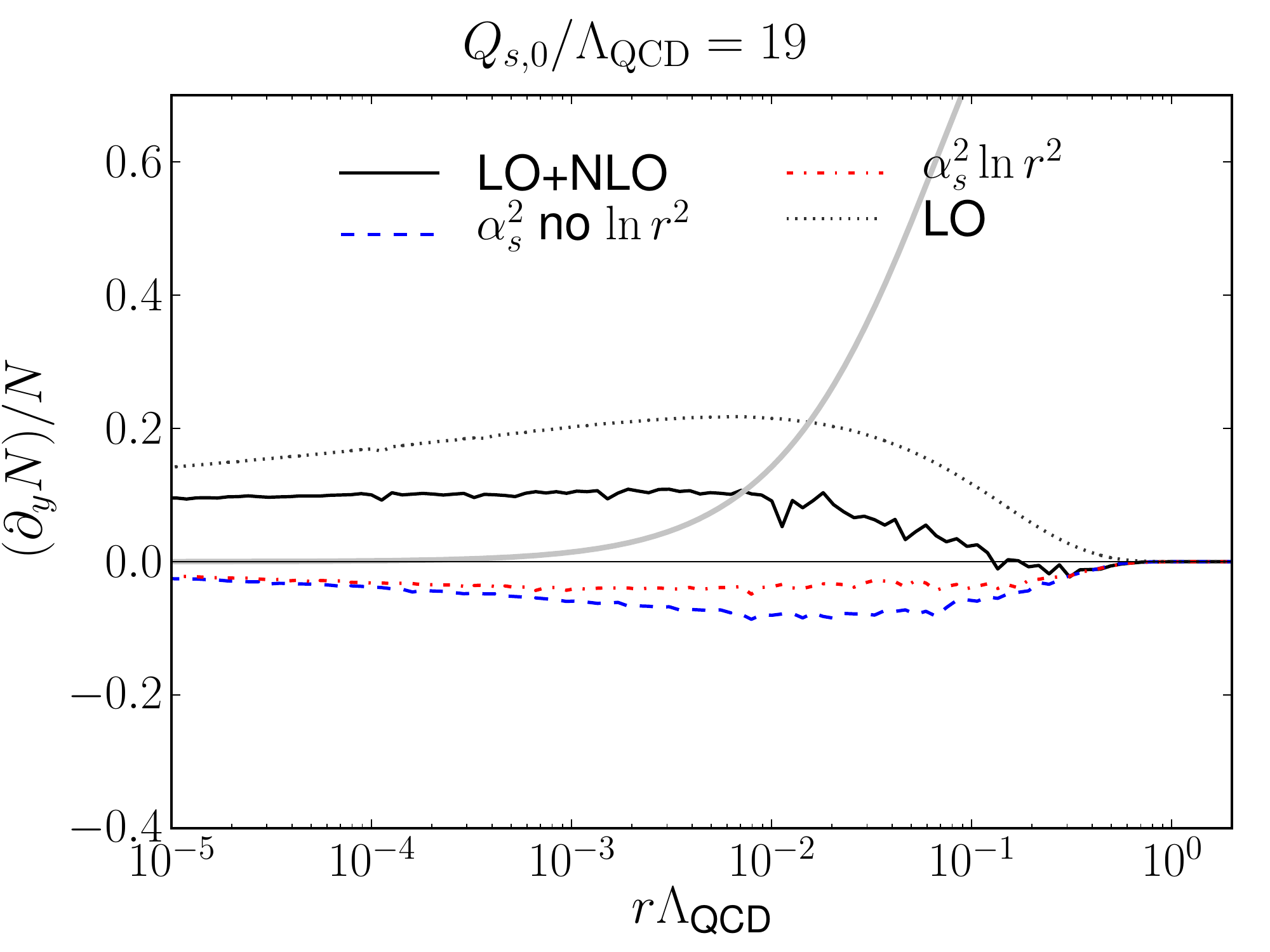}
		}%
	\subfloat[$\gamma=0.8$]{%
	\includegraphics[width=0.333\textwidth]{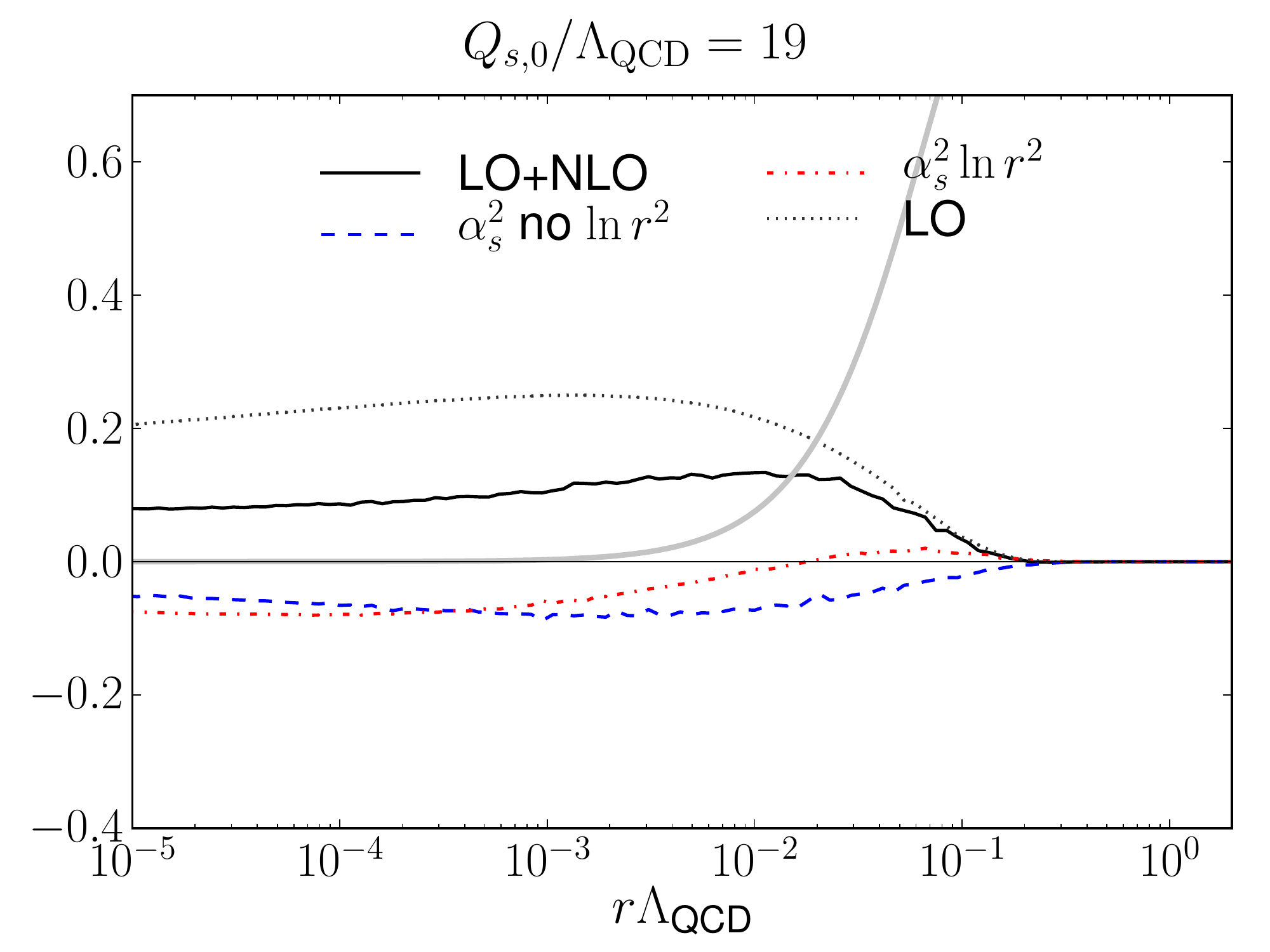}
		}%
	\subfloat[$\gamma=1.0$]{%
		\includegraphics[width=0.333\textwidth]{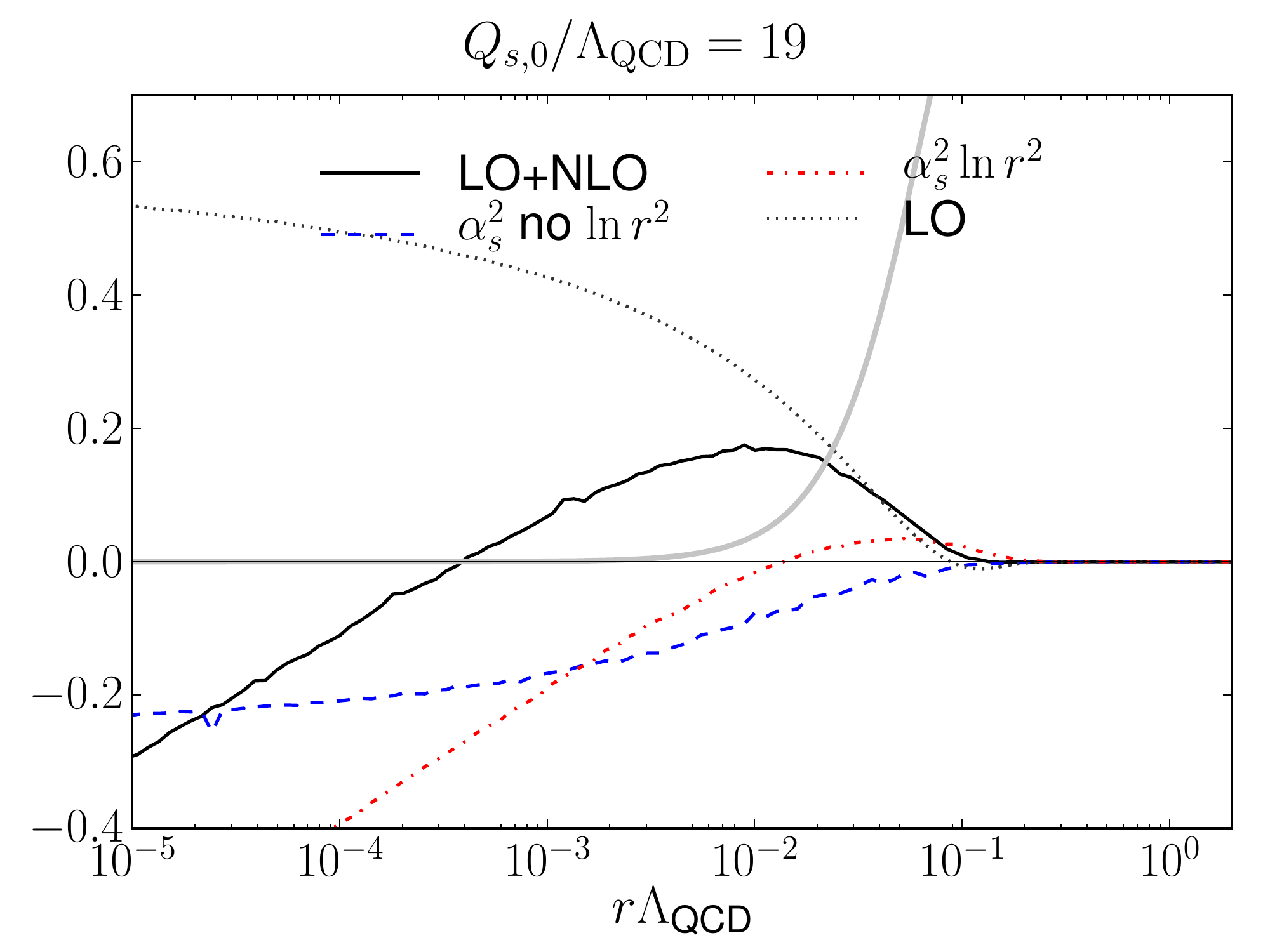}
		}	%
	\caption{Evolution speed of the conformal dipole amplitude at the initial condition. Shown are separately the full NLO and LO results and the contributions from the $\ln r^2$ term and from the other NLO terms. Figure from \paper\cite{Lappi:2015fma}.}
	\label{fig:dndy_confqcd}
\end{figure}

In the evolution equation of the conformal dipole the double logarithmic term is absent, but a new logarithm $\sim \ln r^2$ is present. Note that the only other logarithm is $\sim \ln X^2Y'^2/(X'^2Y^2)$ which explicitly vanishes in the limit of zero parent dipole size. The NLO contributions coming from the $\as^2 \ln r^2$ term and the other $\sim \as^2$ contributions are shown separately in \fig\ref{fig:dndy_confqcd}. In this case the $\as^2 \ln r^2$ is responsible for driving the amplitude negative at small dipoles if the initial anomalous dimension is large. The definition of the conformal dipole \eqref{eq:confdipole} roughly speaking deletes the problematic double logarithmic term from the evolution equation and introduces a new logarithm that makes the evolution unstable in a same way, except that the unstable region is approached only at smaller dipoles.

Based on these results and \paper\cite{Lappi:2015fma} we can conclude that a better theoretical understanding of the NLO BK equation is needed berore it can e.g. be convoluted with the NLO impact factors~\cite{Balitsky:2010ze,Beuf:2011xd} and applied to DIS processes. Possible solutions could be to include something like a kinematical constraint that is suggested for the leading order BK equation in e.g. \re\cite{Beuf:2014uia}. A possible resummation procedure which takes properly into account the time ordering of the subsequent gluon emissions and sums the double logarithmic contributions to all orders has recently been proposed in \re\cite{Iancu:2015vea}. 

The strong dependence on the anomalous dimension at the initial condition is perhaps not surprising.  As discussed e.g. in \res\cite{Munier:2003sj,Marquet:2005ic,Beuf:2008mb}, the leading order BK equation (without running coupling) has a very different behavior depending on the steepness of the initial condition. In particular, as shown in \re\cite{Munier:2003sj}, if the anomalous dimension of the initial condition in the Mellin space is small, then this anomalous dimension sets the evolution speed of the solution. This is known as the \emph{pushed front} solution in the literature. On the other hand, if the anomalous dimension is larger than the critical value, then the critical anomalous dimension sets the evolution speed to the universal value in the so called \emph{pulled front} region.

\chapter{Deep inelastic scattering}
\label{ch:dis}

\section{Probing the proton structure with leptons}

The proton structure can be studied by scattering (virtual) photons off it. As a photon source, a lepton (electron or positron) beam is usually used.

\begin{figure}[tb]
\begin{center}
\includegraphics[width=0.4\textwidth]{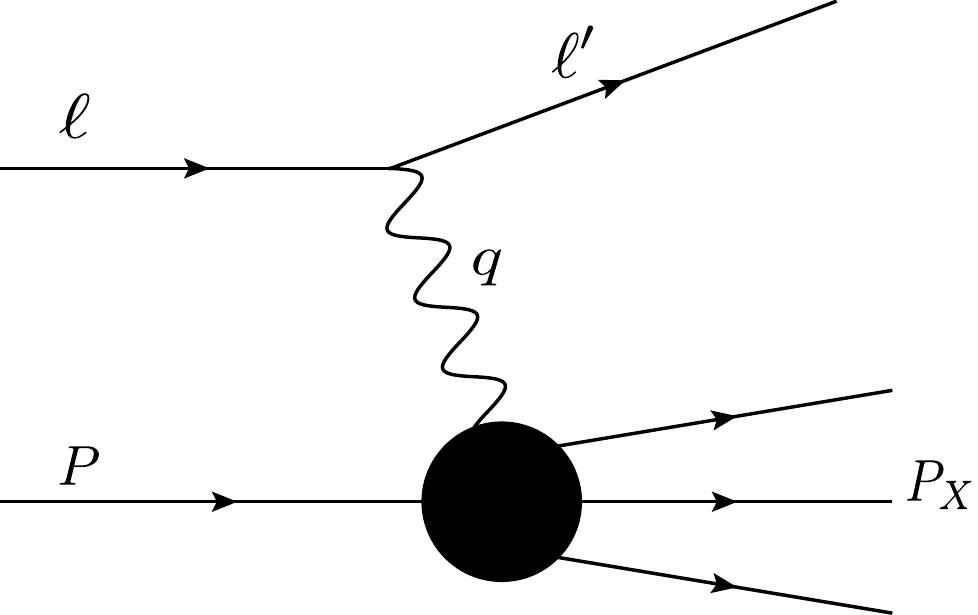}
\caption{Deep inelastic scattering kinematics}
\label{fig:dis}
\end{center}
\end{figure}

Let us consider a process shown in \fig \ref{fig:dis}, where an incoming lepton (momentum $\ell$) interacts with the proton by exchanging a photon with momentum $q$. In order to see the inner structure of the proton, and not just one electric charge with radius $\sim 1$ fm, the photon virtuality $Q^2=-q^2$ must be increased. 
In that case, the momentum transfer becomes very large, and the proton breaks up into fragments denoted by $X$ with an invariant mass $W^2=(P+q)^2$. This process is called  deep inelastic scattering (DIS).

The DIS kinematics, when the target can either be a proton ($A=1$) or a nucleus with mass number $A$, can be expressed using the Lorentz invariant variables
\begin{align}
	Q^2& =-q^2=-(\ell - \ell')^2, \\
	x &= \frac{AQ^2}{2P\cdot q}, \\ 	
	y &= \frac{P\cdot q}{P\cdot \ell}.
\end{align}
In the infinite momentum frame, where the proton momentum is very large, $x$ can be interpreted as the fraction of the proton longitudinal momentum carried by the quark taking part in the scattering process. The virtuality of the photon $Q^2$ sets the distance scale probed in the collision, and the inelasticity $y$ gives the fraction of the electron energy carried by the photon in the target rest frame.

Photon emission from a lepton (which is a point-like particle) is well known and can be computed accurately from Quantum Electrodynamics (QED), but as the proton structure is \emph{a priori} unknown, the photon-proton interaction can not be computed from first principles using perturbative QCD. However, one can parametrize the photon-hadron coupling by introducing the most general form for the hadronic tensor. Taking into account the symmetry requirements, one can eventually write the the invariant cross section for the inelastic electron-proton scattering in terms of two dimensionless proton structure functions $F_1$ and $F_2$, see e.g. \re \cite{quarks-leptons}. The total cross section then becomes
\begin{equation}
	\frac{\der \sigma}{\der x \der Q^2} = \frac{4\pi \alphaem}{Q^4} \frac{1}{x} \left[ xy^2 F_1(x,Q^2) + \left(1-y-\frac{xyM^2}{s-M^2}\right)  F_2(x,Q^2) \right],
\end{equation}
where $M$ is the proton mass and $s=(P+\ell)^2$ is the center-of-mass energy of the proton-lepton system.

The QCD part of the scattering is now included in the structure functions $F_1$ and $F_2$. To leave out the well-known QED part, the experimental measurements  are often reported for the structure functions or for the reduced cross section defined as
\begin{equation}
	\label{eq:reduced-xs}
	\sigma_r(y,x,Q^2) = F_2(x,Q^2) - \frac{y^2}{1+(1-y)^2} F_L(x,Q^2).
\end{equation}
Here the longitudinal structure function is $F_L = F_2 - 2xF_1$. It measures the violation of the Callan-Gross relation which states that $F_L=0$ if the proton consists of only spin-$\frac{1}{2}$ fermions (quarks). Thus, $F_L$ gives in principle a more direct measure of the gluonic structure of the proton. Unfortunately, the precision of the current experimental $F_L$ data from HERA is very limited~\cite{Aaron:2008ad,Chekanov:2009na} compared to measurements of $F_2$ alone~\cite{Aaron:2009aa}.

The structure function $F_2$ can be seen as a measure of parton density, as at leading order one can show that \cite{quarks-leptons}
\begin{equation}
	F_2(x,Q^2)=\sum_i e_i x f_i(x),
\end{equation}
where $e_i$ is the charge of quark type $i$ and $f_i$ is the quark distribution function. Note that at leading order in perturbative QCD the structure function is independent of $Q^2$. When higher order QCD corrections are included, the $Q^2$ dependence of $F_2$ can be obtained by deriving the so called DGLAP evolution equation~\cite{Gribov:1972ri,Gribov:1972rt,Altarelli:1977zs,Dokshitzer:1977sg}.

The journey to measure the internal structure of the proton via DIS began at SLAC experiment in 1960s where the partonic structure of the proton was seen. The theoretical foundations of the parton model were derived around the same time in \res\cite{Feynman:1969ej,Bjorken:1969ja}. 
Currently, the most precise information on the proton structure is obtained from measurements done at DESY-HERA electron/positron-proton collider from 1992 to 2007. The H1 and ZEUS experiments from HERA have published combined results for the proton structure functions and the for the reduced cross section~\cite{Aaron:2009aa}.

\section{DIS in dipole picture}
\label{sec:dis-dipole}
Deep Inelastic Scattering at small $x$ can be described within the Color Glass Condensate framework by looking at the process in the dipole picture. There, in the frame where the proton is at rest the lifetime of the $\gamma^*\to q\bar q$ quantum fluctuations is much larger than the typical timescale of the interaction given that $x\ll 1/(mR)$ where $m$ and $R$ are the target mass and radius (in the target rest frame), respectively. The process then looks such that first the incoming virtual photon splits to quark-antiquark color dipole which  subsequently scatters off the target proton~\cite{Kovchegov:2012mbw}.

The total virtual photon-proton cross section is (see e.g.~\cite{Kowalski:2006hc}, or more pedagogical discussion in \re \cite{Kovchegov:2012mbw}) 
\begin{equation}
	\sigma_{T,L}^{\gamma^*p}(\rt, x) = \sum_f  \int \der^2 \rt \int_0^1 \frac{\der z}{4\pi} [\Psi^*\Psi]_{T,L}^f(\rt,z) \sigma_\text{tot}^{q\bar q}(\rt,x),
\end{equation}
where $\sigma_\text{tot}^{q\bar q}$ is the total dipole-proton cross section. It can be obtained from the dipole amplitude $N$ which is the imaginary part of the dipole-target scattering amplitude by using the optical theorem:
\begin{equation}
	\label{eq:totdipxs_bint}
	\sigma_\text{tot}^{q\bar q}(\rt,x) = 2 \int \der^2 \bt N(\rt, \bt, x).
\end{equation}

The photon wave function squared $[\Psi^*\Psi]_{T,L}^f$ can be interpreted to be proportional to the probability for the $\gamma^*\to q\bar q$ splitting, where $T$ and $L$ refer to transverse and longitudinal polarizations of the virtual photon and $f$ is the quark flavor. The transverse separation of the quarks is $\rt$ and the impact parameter in the collision is $\bt$. The longitudinal momentum fraction of the photon carried by the quark is denoted by $z$.

The photon to dipole splitting can be computed using quantum electrodynamics written on the light cone as discussed in \se\ref{sec:lightcone}. The virtual photon wave functions squared  (summed over spins and polarizations) can be obtained from the light cone calculations presented in \se\ref{sec:lightcone} (see \eqs \eqref{eq:photon-wf} and \eqref{eq:photon-wf-l}) and are~\cite{Kovchegov:2012mbw} (see also my MSc thesis~\cite{gradu} for a detailed derivation) 
\begin{align}
	[\Psi^*\Psi]_T^f &= \frac{2\nc}{\pi} \alphaem e_f^2 \{[z^2 + (1-z)^2] \varepsilon^2 K_1^2(\varepsilon r) + m_f^2 K_0^2(\varepsilon r) \\
	[\Psi^*\Psi]_T^f &= \frac{8\nc}{\pi} \alphaem e_f^2 Q^2 z^2(1-z)^2 K_0^2(\varepsilon r), 
\end{align}
where $\varepsilon^2 = z(1-z)Q^2 + m_f^2$ and $r=|\rt|$.

Note that now all QCD dynamics is included in the virtual photon-proton cross section $\sigma^{\gamma^*p}$, and the ``trivial'' part of the process where the photon is emitted from the lepton is factorized out. The structure functions $F_1$ and $F_2$ can also be related directly to the virtual photon-proton cross sections, and in the high-energy limit we obtain~\cite{Kovchegov:2012mbw}
\begin{align}
	F_2(x,Q^2) &= \frac{Q^2}{4\pi^2 \alphaem} \sigma^{\gamma^*p}_\text{tot} =  \frac{Q^2}{4\pi^2 \alphaem} (\sigma_T^{\gamma^*p} + \sigma_L^{\gamma^*p}) \\
	2x F_1(x,Q^2) &= \frac{Q^2}{4\pi^2 \alphaem} \sigma_T^{\gamma^*p}.
\end{align}
Thus, $F_2$ measures the total cross section. At small-$x$ the photon couples to the sea quarks that originate from gluons, and $F_2$ is related to the total gluonic density of the proton. 
The longitudinal structure function, on the other hand, measures the cross section for the longitudinally polarized virtual photon-proton scattering, as
\begin{equation}
	F_L(x,Q^2) = F_2(x,Q^2) - 2xF_1(x,Q^2) = \frac{Q^2}{4\pi^2 \alphaem} \sigma_L^{\gamma^* p}.
\end{equation}

\section{Extracting initial conditions for BK evolution}
\label{sec:fits}
After the previous discussion we can compute the proton structure functions within the Color Glass Condensate framework given that we know the dipole amplitude $N(\rt, \bt, x)$ for the dipole with size $\rt$ to scatter with impact parameter $\bt$. The $x$ (or energy) evolution of $N$ is given by the Balitsky-Kovchegov equation \eqref{eq:bk} discussed in \se \ref{sec:evolution}. However, the BK equation is a differential equation whose solution requires an initial condition.

The initial condition for the BK equation, the dipole amplitude $N(\rt, \bt, x)$ at some $x=x_0$ is non-perturbative information. It can be obtained by parametrizing the amplitude and fitting the parameters to the experimental data. Note that this is exactly the same procedure that is used to determine the parton distribution functions: there one parametrizes the distributions  at an initial scale $Q_0^2$, and the distributions at higher scale $Q^2>Q_0^2$ are obtained by solving the DGLAP equations.

In current phenomenological literature the initial condition for the BK equation is often taken as an MV-model (see discussion in \se \ref{sec:mv-discussion}) inspired parametrization
\begin{equation}
	\label{eq:modified-mv}
	N(\rt, x=x_0) = 1 - \exp \left[ -\frac{(r^2 Q_{s,0}^2)^\gamma}{4} \ln \left( \frac{1}{r\lqcd} + e_c \cdot e\right) \right].
\end{equation}
The fit parameters are $Q_{s,0}^2$, the anomalous dimension $\gamma$ and an infrared cutoff $e_c$. 
The parameter $Q_{s,0}^2$ can be roughly seen as a measure of the saturation scale, or as a gluon density at initial Bjorken-$x$. The anomalous dimension $\gamma$ controls the behavior of the gluon distributions at large transverse momentum. 
Note that in \eq \eqref{eq:modified-mv} we did not include any impact parameter dependence. In practical calculations it is often assumed that one can replace
\begin{equation}
	\label{eq:b-fact}
	2\int \der^2 \bt N(\rt, \bt, x) \to \sigma_0 N(\rt, x),
\end{equation}
where $\sigma_0$ is twice the transverse area of the target, which must be fitted to the data. 

When the BK equation is solved with the running coupling corrections (see \se \ref{sec:bk}), the strong coupling constant $\as$ must be evaluated at some given distance scale. As discussed in \se \ref{sec:bk}, the uncertainty in the Fourier transform of the expression of $\as$ from the momentum space to the coordinate space is parametrized by factor $C^2$, which is taken to be a fit parameter. See discussion in \se\ref{sec:jimwlk} for a discussion of a possible theoretical value for this parameter.
% We take it to be a fit parameter, and find that the fit prefers
 %significantly larger values for $C^2\sim 5$ than what is argued theoretically in \re \cite{Kovchegov:2006vj} ( $C^2=e^{-2\gamma_E} \approx 0.3152$).

The first dipole amplitude fits with the running coupling BK equation to the HERA $F_2$ data, combined with CERN-SPS and FNAL measurements, were performed by the AAMS collaboration in \re \cite{Albacete:2009fh}. This work was later improved by the AAMQS collaboration~\cite{Albacete:2010sy} where the combined measurements of the reduced cross section (see \eq\eqref{eq:reduced-xs}) from the different HERA experiments~\cite{Aaron:2009aa} (ZEUS and H1) with significantly smaller uncertainties were used. 
It was found in these fits that an anomalous dimension $\gamma>1$ and relatively large $C^2 \sim 5\dots 10$ (compared to theoretical calculation $C^2=e^{-2\gamma_E}\approx 0.3 $ as discussed in \se\ref{sec:jimwlk}) were needed for a good description of the HERA data. The largish value of $C^2$ can be interpreted such that the leading order BK equation tends to give a too fast evolution speed compared to data, and this speed is reduced by a large $C^2$. See also \se \ref{sec:nlobk} for a discussion of NLO effects in the evolution speed of the BK equation.

In \paper \cite{Lappi:2013zma} we studied a possibility to obtain a good fit to HERA data without introducing an anomalous dimension. A motivation for this was the fact that when an unintegrated (transverse momentum dependent) dipole gluon distribution \eq\eqref{eq:dipole-ugd} is computed from the dipole amplitude, one can obtain negative gluon densities using the initial condition obtained in \res \cite{Albacete:2010sy,Albacete:2009fh} before any BK evolution. 
Thus, we allowed the infrared cutoff in the modified MV model \eqref{eq:modified-mv} to vary and parametrized it by introducing a fit parameter $e_c$ and fixing $\gamma=1$. Another difference to the previous works was that we did not include any other experimental data except the combined HERA results, as thanks to the good precision of the HERA $\sigma_r$ data, it anyway dominates the fit. 

We also only considered the light quarks instead of including charm as done in \re \cite{Albacete:2010sy}, even though the charm contribution to the reduced cross section is measured by HERA~\cite{Aaron:2009aa}. The reason for this is that in \re \cite{Albacete:2010sy} it was found that in order to obtain a good fit with heavy quarks one has to introduce additional parameters for the transverse area and initial saturation scale for the charm quarks. As we found this approach inconvenient, 
the heavy quarks were not included in the fit in \paper\cite{Lappi:2013zma}.

We studied three different MV-model inspired parmetrizations for the initial condition. First, we used the standard MV model without modifications by fixing $\gamma=e_c=1$. What is denoted by MV$^\gamma$ is the same parametrization used also in previous works where the anomalous dimension $\gamma$ is a free parameter but we fix $e_c=1$. Similarly in the MV$^e$ parametrization we fixed $\gamma=1$ but took the infrared cutoff parameter $e_c$ to be a fit parameter. 
The fit was performed to the HERA reduced cross section $\sigma_r$ data in kinematical window $x<0.01$ and $Q^2<50$ GeV$^2$. In this kinematics the CGC picture should be valid due to large gluon densities at small $x$, and the upper limit for $Q^2$ is needed as we did not include the $Q^2$ evolution via e.g. DGLAP equations. Note, however, that the results are not independent of $Q^2$, as the virtuality of the photon sets the dominant transverse size $r\sim 1/Q$ for the dipole. 

\begin{table}
\begin{tabular}{|l||r|r|r|r|r|r|r|}
\hline
 & $\chi^2/\text{d.o.f}$ &  $\qso^2$ [GeV$^2$] & $\qs^2$ [GeV$^2$] & $\gamma$ & $C^2$ & $e_c$ & $\frac{\sigma_0}{2}$ [mb] \\
\hline\hline
MV & 2.76 & 0.104 & 0.139  & 1 & 14.5 & 1 & 18.81 \\
MV$^\gamma$ & 1.17 & 0.165 & 0.245 & 1.135 & 6.35 & 1 & 16.45 \\
MV$^e$ & 1.15 & 0.060 & 0.238  & 1 & 7.2 & 18.9 & 16.36 \\
\hline
\end{tabular}
\caption{Parameters from fits to HERA reduced cross section data  for different initial conditions. Also the corresponding initial saturation scales $\qs^2$ defined via \eq \eqref{eq:satscale} are shown. The parameters for the MV$^\gamma$ initial condition are obtained by the AAMQS collaboration \cite{Albacete:2010sy}.
}
\label{tab:fit_params}
\end{table}

The fit results are given in \tab \ref{tab:fit_params}. We observe that some modification to the MV model is needed in order to get a good fit with $\chi^2/\text{d.o.f}\sim 1$. For the MV$^\gamma$ parametrization we obtain a very similar fit result as obtained by the AAMQS collaboration in \re \cite{Albacete:2010sy}, and small deviations can be explained by noticing the use of different data sets and small differences in the numerical setup. In our numerical calculations we will use the AAMQS result for the MV$^\gamma$ parametrization, and we also show only their result in \tab \ref{tab:fit_params}.

The initial saturation scale is controlled by the parameter $\qso$, but the numerical values for it are not directly comparable between the different parametrizations. Thus, we also show in \tab \ref{tab:fit_params} the initial saturation scale, which is a physically relevant quantity, defined via the equation
\begin{equation}
	\label{eq:satscale}
	N\left(r^2 = \frac{2}{\qs^2}\right) = 1-e^{-1/2}.
\end{equation}
The two parametrizations that give equally good fits to the data, MV$^\gamma$ and MV$^e$, also give a consistent result for the proton saturation scale at the initial $x=0.01$: $\qs \approx 0.5 \gev$.

\begin{figure}[tb]
\begin{center}
\includegraphics[width=0.6\textwidth]{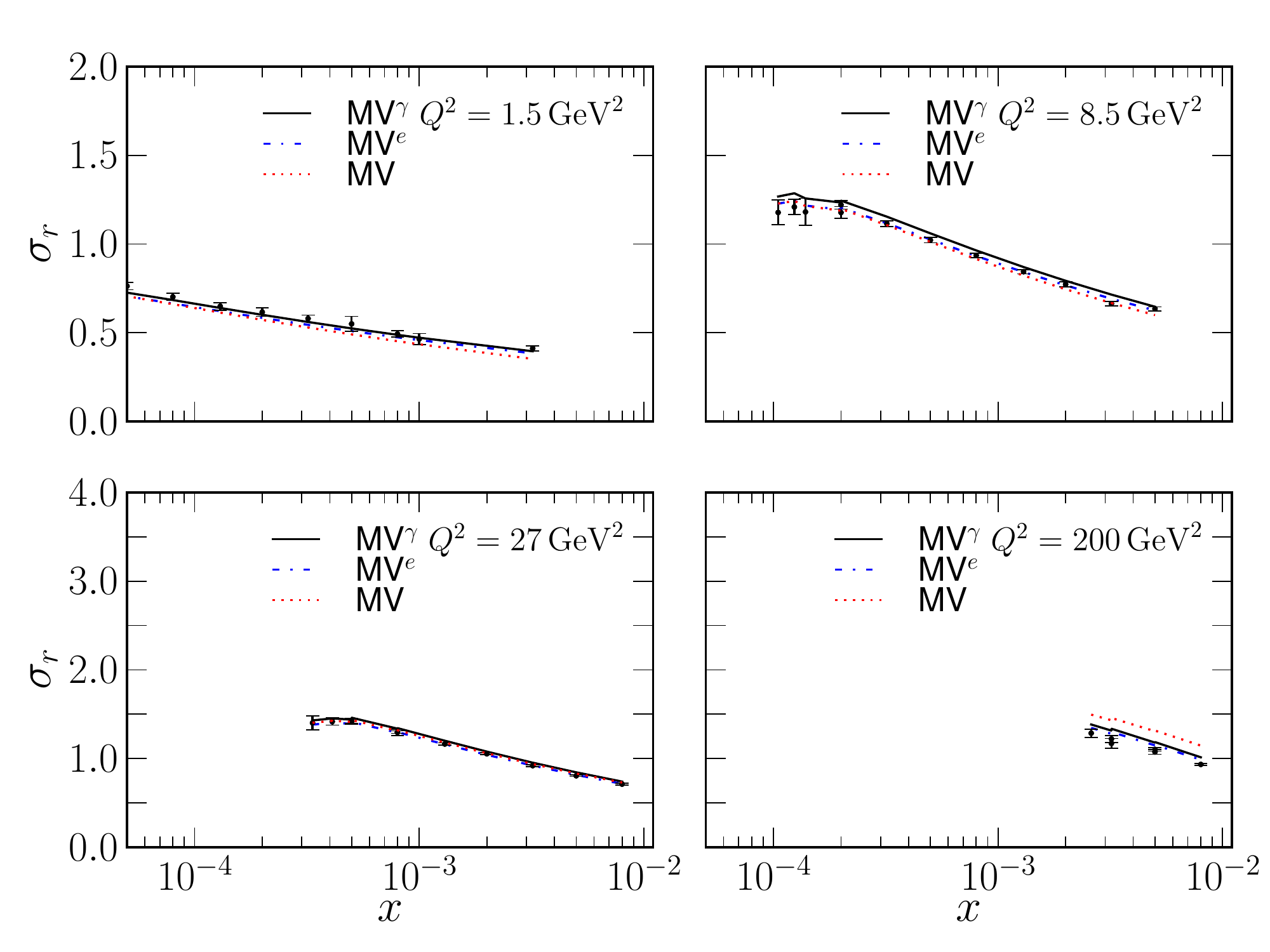}
\caption{Reduced cross section $\sigma_r$ computed using the different fitted parametrizations as an initial condition for the BK evolution compared with the combined HERA data from H1 and ZEUS experiments~\cite{Aaron:2009aa}. Figure from \paper \cite{Lappi:2013zma}.}
\label{fig:sigmar}
\end{center}
\end{figure}

The results obtained using the best fit values are compared with a part of the HERA reduced cross section data in \fig \ref{fig:sigmar} at different values of $Q^2$ (note that $Q^2=200 \gev^2$ data is not included in the fit). Both modified parametrizations seem to give almost identical results that agree very well with the precise HERA data. The standard MV model is not too bad either, but the agreement is clearly worse than with the MV$^\gamma$ or MV$^e$ parametrizations.

In addition to the BK-evolved MV model parametrizations, similar fits have been done with other models for the dipole amplitudes. These are necessary especially if the impact parameter dependence of the dipole amplitude is needed, which is the case when dealing with e.g. diffractive scattering which is the topic of \ch\ref{ch:ddis}. As an example we mention the IPsat model, which has an eikonlaized DGLAP-evolved gluon distribution function~\cite{Kowalski:2003hm} fit in \res\cite{Kowalski:2006hc,Rezaeian:2012ji} and the IIM model~\cite{Iancu:2003ge} that parametrizes the asymptotic properties of
the BK equation, and has been fitted in \res\cite{Soyez:2007kg,Rezaeian:2013tka}. A disadvantage of these parametrizations is that they do not include the full Bjorken-$x$ evolution given by the BK equation.

\section{Dipole amplitude for a heavy nucleus}
\label{sec:dipole-amplitude-nuke}
Currently the DIS experiments with nuclei have only been done with fixed target experiments at SLAC~\cite{Gomez:1993ri} and by the  NMC collaboration~\cite{Amaudruz:1995tq,Arneodo:1995cs}. The kinematical range probed in these measurements is very limited for the small-$x$ part of the phase space. This prevents us from performing a similar fit for the nuclei as what is done for the protons. 

In the coming decades there are plans to build next-generation colliders that could study lepton-nucleus scattering at high energy. In the United States, the plans are to build an Electron-Ion Collider (EIC)~\cite{Accardi:2012qut} by adding an electron beam to  the BNL-RHIC accelerator, which currently collides protons and different nuclei, or adding a nuclear beam to the JLAB-CEBAF facility. At CERN, the possibility to add an electron beam to the Large Hadron Collider, which can accelerate protons and lead nuclei, is considered as a way to build an LHeC collider~\cite{AbelleiraFernandez:2012cc}. 

Before experimental small-$x$ nuclear DIS data becomes available another approach must be taken in order to obtain the dipole-nucleus scattering amplitude $N^A$. We use the optical Glauber model (for a review, see e.g. \re \cite{Miller:2007ri}) to generalize dipole-proton scattering amplitude to dipole-nucleus scattering to obtain an initial condition for the BK evolution. We presented this method in \paper \cite{Lappi:2013zma}.

Let us first write the total dipole-proton cross section for the dipole with a transverse separation $r$ as
\begin{equation}
	\sigma_{q\bar q}^p = \sigma_0 N(r),
\end{equation}
where $\sigma_0$ is the result of the impact parameter integral in \eq \eqref{eq:totdipxs_bint}, and was fitted to HERA data in \se \ref{sec:fits}. In the dilute limit where the dipole is very small, the dipole-nucleus cross section should be just $A$ times dipole-proton cross section (where $A$ is the mass number of the nucleus). On the other hand, for large dipoles we should have
\begin{equation}
	\frac{\der \sigma_{q\bar q}^A}{\der^2 \bt} = 2N(\rt, \bt) \to  2.
\end{equation}
These requirements are satisfied with an exponentiated dipole-nucleus scattering amplitude
\begin{equation}
	\label{eq:dipole-nucleus-1}
N^A(\rt,\bt) = 1-\exp \left( -\frac{A T_A(\bt)}{2} \sigma_{q\bar q}^p \right),
\end{equation}
where $T_A$ is the transverse density of the nucleus at impact parameter $\bt$ which, in practice, is obtained by integrating the nuclear Woods-Saxon~\cite{PhysRev.95.577} density distribution over the longitudinal direction. This form corresponds to an average of the dipole-nucleus scattering amplitude over the fluctuating positions of the nucleons in the nucleus, see e.g. \res \cite{Kowalski:2003hm,Kowalski:2007rw}. 

The scattering matrix $S^A=1-N^A$ calculated from \eq \eqref{eq:dipole-nucleus-1} approaches a limiting value 
\begin{equation}
	\exp \left( -\frac{A T_A(\bt) \sigma_0}{2} \right)  \sim \exp \left( -A^{-1/3}\right),
\end{equation}
instead of zero, in the large dipole limit. This will cause unphysical oscillations to the gluon distribution computed from the dipole amplitude (see discussion in \se\ref{sec:cgc-intro}). Therefore we expand the dipole-proton cross section $\sigma_{q\bar q}^p$ as
\begin{equation}
	\sigma_{q\bar q}^p \approx \sigma_0 \frac{(\rt^2 \qso^2)^\gamma}{4} \ln \left( \frac{1}{|\rt| \lqcd} + e_c \cdot e\right).
\end{equation}
Combining this with \eq \eqref{eq:dipole-nucleus-1} we obtain the dipole-nuclues amplitude at $x=x_0$:
\begin{equation}\label{eq:dipole-nucleus}
	N^A(\rt,\bt) = 1 - \exp\left[ -A T_A(\bt) \frac{\sigma_0}{2} \frac{(\rt^2 \qso^2)^\gamma}{4}  \ln \left(\frac{1}{|\rt|\lqcd}+e_c \cdot e\right) \right]
\end{equation}
which is our result, from \paper \cite{Lappi:2013zma}, for the dipole-nucleus amplitude $N^A$. One advantage of the MV$^e$ parametrization, where $\gamma=1$, can now been seen from \eq\eqref{eq:dipole-nucleus}: 
in this case there is no ambiguity whether the initial saturation scale should scale as $T_A$ or $T_A^{1/\gamma}$. 

To obtain the energy (or Bjorken-$x$) evolution of the dipole-nucleus amplitude we use the obtained fit parameters from lepton-proton DIS quoted in \tab \ref{tab:fit_params} and solve the BK equation separately at different impact parameters using \eq\eqref{eq:dipole-nucleus} as an initial condition. Note that in principle one should evolve the initial condition using an impact parameter dependent BK equation which is not currently possible in practice, as discussed in \se\ref{sec:bk}.

\begin{figure}[tb]
\centering
	\begin{minipage}[t]{0.5\textwidth}
	\includegraphics[width=\textwidth]{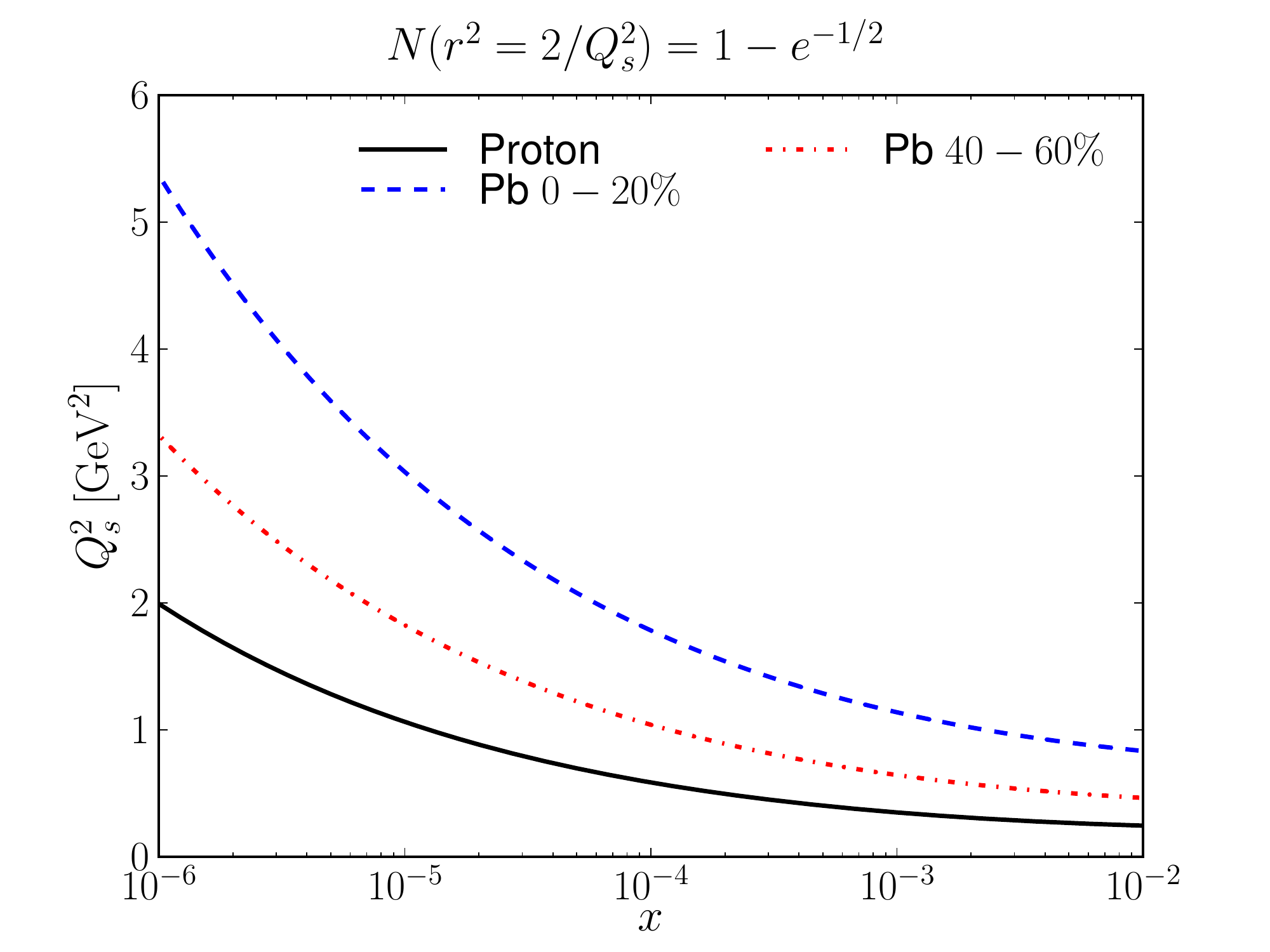}
	\caption{Saturation scale for the proton and the lead nucleus at central and semi-central collisions as a function of $x$ calculated using the MV$^\gamma$ parametrization. }
	\label{fig:qs_p_pb}
	\end{minipage}%
	~
	\begin{minipage}[t]{0.5\textwidth}
	\includegraphics[width=1.0\textwidth]{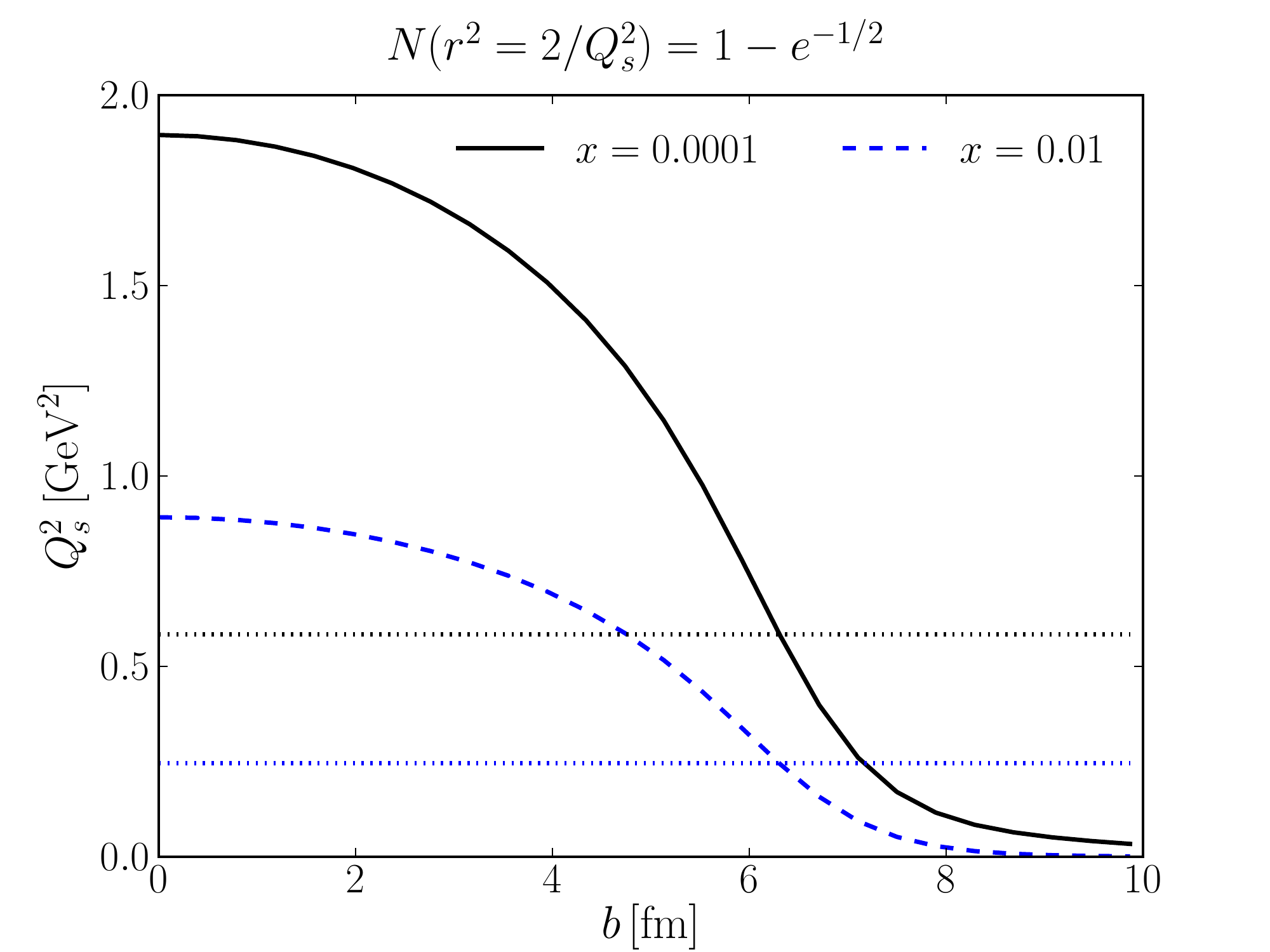}
	\caption{Impact parameter dependence of the saturation scale of the lead nucleus at two different values of $x$. 
	The dashed lines correspond to the proton saturation scales. Figure from \paper \cite{Lappi:2013zma}.
	}
	\label{fig:qs_b}
	\end{minipage}	
\end{figure}

The obtained saturation scales for the proton and the lead nucleus are shown in \figs \ref{fig:qs_p_pb} and \ref{fig:qs_b}. For the lead the saturation scale is shown for the central and mid-central impact parameters (corresponding to $0-20\%$ and $40-60\%$ most central collisions, defined from the Optical Glauber picture~\cite{Miller:2007ri}). The saturation scale of the nucleus is significantly larger than that of the proton still in mid-central centrality classes, and falls below the proton saturation scale only at $b \gtrsim 6$ fm, which corresponds to centrality classes $\gtrsim 70\%$. In that region our parametrization is not exactly valid any more, as the BK evolution would make the nucleus to grow rapidly from the dilute edges. In phenomenological applications we choose to use the parametrization \eqref{eq:dipole-nucleus} only in the region where the saturation scale in the nucleus is larger than in the proton.

\chapter{Exclusive vector meson production}

\label{ch:ddis}

\section{Diffraction in scattering experiments}

\begin{figure}[tb]
\begin{center}
\includegraphics[width=0.4\textwidth]{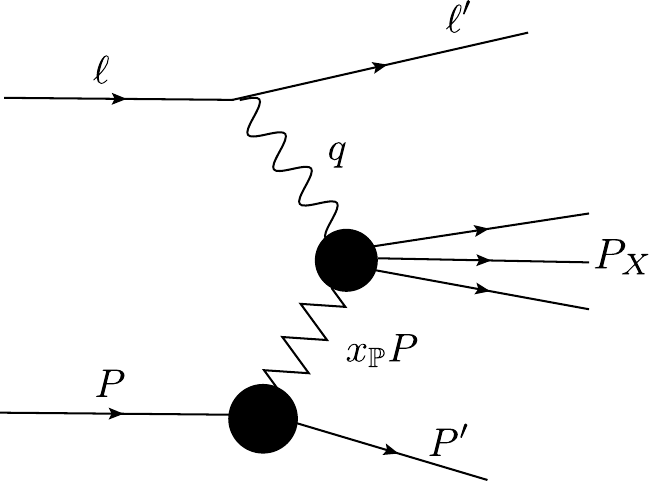}
\caption{Diffractive scattering.}
\label{fig:ddis}
\end{center}
\end{figure}

In high-energy scattering experiments diffractive events are experimentally defined such that a large rapidity gap is present in the event. A rapidity gap means that there is an interval in rapidity (a few units wide) where there are no particles produced. Elastic scattering is a simple example of a diffractive event, but it is also possible to have other events where, for example, a single particle is produced at central rapidity and the scattered particles continue at very forward/backward rapidities.

An example of a diffractive scattering process is shown schematically in \fig \ref{fig:ddis}, where in a lepton-proton scattering the final state particles are the scattered lepton $\ell'$, scattered proton $P'$ and the produced system $X$. We focus on exclusive vector meson production, where the produced system $X$ is for example a $J/\Psi$ meson. The scattered proton can either remain intact or break up, but there can not be exchange of color charge between the proton and the scattered system, or otherwise the breakup of the color strings would fill the rapidity space between the proton and the system $X$. In addition to lepton-proton scattering, diffractive scattering can be studied also with nuclear targets and in hadronic collisions (see \se\ref{sec:upc}).

The photon-proton interaction can be described in terms of an exchange of a color neutral object, known as the pomeron. It is represented in \fig \ref{fig:ddis} by a zigzag line. The kinematical variable $\xpom$ has the similar interpretation as Bjorken-$x$ in DIS: it describes the fraction of the incoming proton (or nucleon) longitudinal momentum carried by the pomeron:
\begin{equation}
\label{eq:xpom}
	\xpom = \frac{(P-P')\cdot q}{P\cdot q} = \frac{M^2 + Q^2 - t}{W^2+Q^2-m_N^2},
\end{equation}
where $q$ is the four-momentum of the emitted photon. In addition, in order to describe the kinematics of diffractive deep inelastic scattering (DDIS) processes, we define the momentum transfer $t=(P-P')^2$ and use the virtuality of the photon $Q^2=-q^2$ and the center-of-mass energy  of the photon-nucleon system $W^2=(P+q)^2$.

Diffractive scattering processes can be used to probe the target gluon distribution very efficiently. This is because at leading order there is an exchange of two gluons between the dipole and the target in the scattering amplitude, as there can not be net exchange of color charge. Thus, the amplitude is proportional to the leading order gluon distribution function $xg$, and the total cross section behaves like gluon distribution squared, see \re\cite{Brodsky:1994kf}. In addition to gluon densities, exclusive vector meson production can be used to probe the spatial distribution of gluons within the proton or a nucleus, see e.g. \res\cite{Caldwell:2009ke,Munier:2001nr,Toll:2012mb}.

Currently the most important diffractive DIS measurements come from HERA, where diffraction off a proton was studied. For nuclear targets, diffractive DIS has so far been  measured only in fixed target experiments such as E665~\cite{PhysRevLett.74.1525} and NMC~\cite{Arneodo1994195} at relatively low center-of-mass energies. In the future, if an electron-ion collider (EIC)~\cite{Accardi:2012qut} or  LHeC~\cite{AbelleiraFernandez:2012cc} is realized, it would open a new era for studies of diffractive DIS off nuclei.

\section{Exclusive vector meson production in the dipole picture}

Let us consider diffractive vector meson production in high-energy deep inelastic scattering, where the target can be either a proton or a nucleus.

\begin{figure}[tb]
\begin{center}
\includegraphics[width=0.6\textwidth]{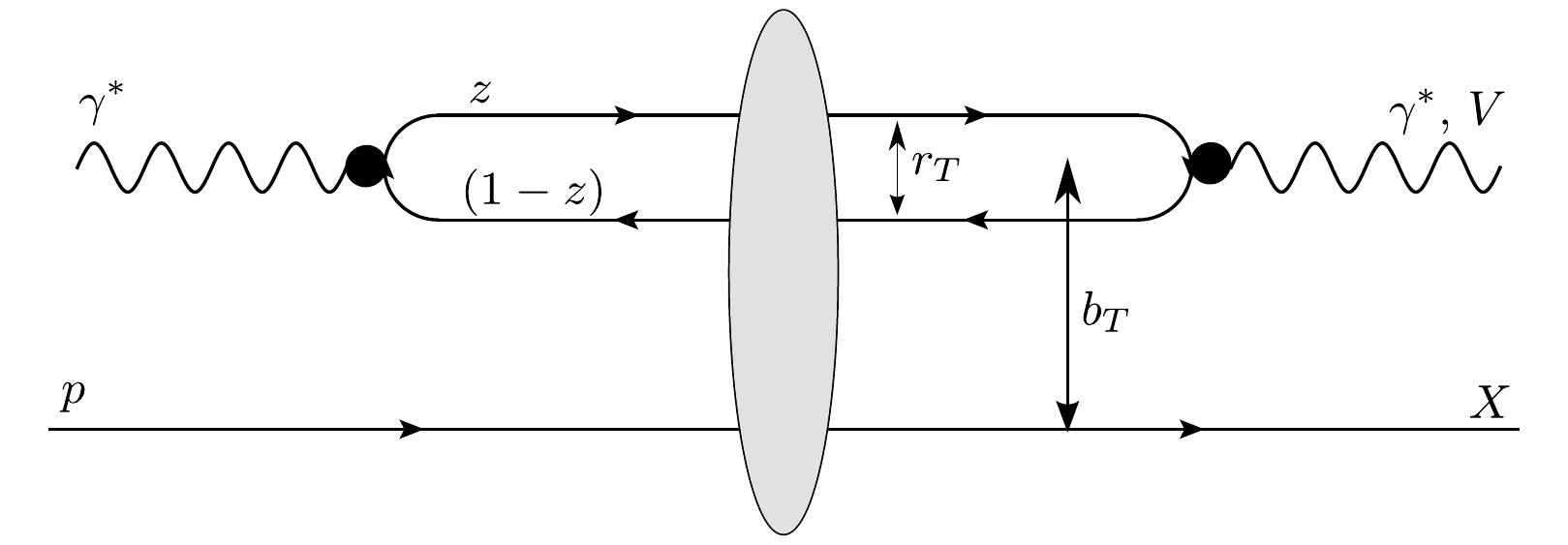}
\caption{Diffractive scattering in dipole picture.}
\label{fig:ddis_dipole}
\end{center}
\end{figure}

The theoretical framework for description of diffractive scattering events, the so called Good-Walker picture, was developed in \re\cite{PhysRev.120.1857}. 
In this framework one has to find the states that diagonalize the imaginary part of the scattering $T$-matrix, which at high energy are the ones where the virtual photon fluctuates into a quark-antiquark dipole long before the dipole interacts with the target. 
The dipole interacts elastically with the target and finally forms a vector meson ($\rho,\omega,J/\Psi,\Upsilon,\dots$). This is shown schematically in \fig \ref{fig:ddis_dipole}. Note that as we consider here high-energy processes where the eikonal approximation is valid, the transverse positions of the quarks can be taken to be fixed during the interaction.

In order to calculate the cross section one has to compute virtual photon splitting to a quark-antiquark dipole (the virtual photon wave function $\Psi^{\gamma^* \to q\bar q}$), which was already discussed in \se \ref{sec:dis-dipole}. In addition, the formation of the vector meson requires knowledge of the vector meson wave function $\Psi_V$, which is in general non-perturbative. In practice for heavy mesons a large contribution to the cross section will come from distance scales set by the meson mass $|\rt| \sim 1/m$. The elastic dipole-target scattering is described by the dipole amplitude $N$.

The cross section for the diffractive vector meson production is derived in \res\cite{Brodsky:1994kf,Ryskin:1992ui} (see also \res\cite{Kovchegov:2012mbw,Kowalski:2006hc}). The scattering amplitude is 
\begin{equation}
\label{eq:ddis-amp}
	\A^{\gamma^*A \to VA} = \int \der^2 \rt \int_0^1 \frac{\der z}{4\pi} [\Psi^{\gamma^*\to q\bar q}\Psi_V^*](\rt,z)  \int \der^2 \bt e^{-i \bt \cdot \Delta} 2N(\xt,\bt,\xpom) ,
\end{equation}
and the differential cross section can be computed as
\begin{equation}	
	\frac{\der \sigma^{\gamma^*A \to VA}}{\der t} = \frac{1}{16\pi} \left| \A^{\gamma^*A \to VA} \right|^2.
\end{equation}
Here $z$ is the longitudinal momentum fraction of the virtual photon carried by the quark, $\bt$ is the impact parameter and $\Delta = \sqrt{-t}$ is the transverse momentum transfer. 

The scattering amplitude \eqref{eq:ddis-amp} has a clear physical interpretation: first the virtual photon splits into quark-antiquark dipole with probability amplitude $\Psi^{\gamma^*\to q\bar q}$, and the dipole then scatters elastically off the target color field described by the dipole amplitude $N$. Finally, the quark-antiquark pair forms the vector meson with probability amplitude $\Psi_V^*$.
The momentum transfer dependence of the scattering amplitude originates from the Fourier transform of the dipole amplitude. It is then necessary to have a detailed description of the impact parameter profile of the target, in contrast to the DIS cross section calculations where we only had to calculate the integral over the impact parameter space in \ch\ref{ch:dis}.

%Note that the impact parameter dependence can not be factorized from the cross section as was the case when the DIS cross section was computed from the dipole picture in \ch\ref{ch:dis}. Instead, the impact parameter dependence of the dipole amplitude is needed in order to calculate the Fourier transform.

As the vector meson wave function can not be computed from first principles using perturbative techniques, different models for it exist in the literature. The wave function is obtained by assuming that the vector meson is predominantly a quark-antiquark state that has the same spin and polarization structure as the virtual photon. The model parameters are fixed by requiring the wave function to reproduce the measured decay width  to the  electron channel (which turns out to be proportional to the wave function at origin) and that it is correctly normalized. 
In this work we study the dependence of our results on the model uncertainties by using two different models for the vector meson wave function, called ``Gaus-LC'' and ``boosted Gaussian''. These wave functions are described in more detail and fitted to data in \re\cite{Kowalski:2006hc}. 

In order to calculate the evolution of the dipole amplitude with impact parameter dependence a solution to the impact parameter dependent BK equation would be needed. However, as discussed in \se \ref{sec:dipole-amplitude-nuke}, it is not currently available. Instead of a BK evolved dipole amplitude we will use different parametrizations for $N(\xt,\bt,\xpom)$  which include impact parameter dependence.

We use mostly the so called IPsat model~\cite{Kowalski:2003hm}, where the dipole amplitude is obtained from an eikonalized DGLAP-evolved gluon distribution function $xg$ whose initial condition is fitted to the HERA deep inelastic scattering data in \re\cite{Kowalski:2006hc}. The transverse profile of the proton is assumed to be Gaussian, and the width of the distribution is a fit parameter.

The second dipole amplitude that we use here to study the dependence of our results on the details of the dipole amplitude is the so called IIM model~\cite{Iancu:2003ge}, which is a parametrization including the most important features o the BK evolution. The model parameters are again fitted to the HERA data\footnote{After the completion of \paper~\cite{Lappi:2010dd}, where the parametrizations were used, newer fits  for both IIM and IPsat models to the much more precise combined HERA data have been published~\cite{Rezaeian:2012ji,Rezaeian:2013tka}} in \re \cite{Soyez:2007kg}. The calculated diffractive $J/\Psi$ production cross section in electron-proton scattering is shown in \fig\ref{fig:hera-diffraction-comparison} as a function of $Q^2$, where the results are compared with the H1~\cite{Aktas:2005xu} and ZEUS~\cite{Chekanov:2004mw} data. The ``Factorized IPsat'' refers to a parametrization where the impact parameter dependence is factorized from the IPsat model, see discussion in \se\ref{sec:diffraction-nucleus}.

\section{Diffraction off a nucleus}
\label{sec:diffraction-nucleus}

When diffractive scattering with nuclear targets is considered, the events can be divided into two classes. In \emph{coherent diffraction} the target nucleus remains fully intact, whereas in \emph{incoherent diffraction} the $\pt$ kick given to the nucleus is large enough to break it up still preserving the rapidity gap.

As shown in \re\cite{Caldwell:2009ke}, the coherent cross section corresponds to performing the average over the nucleon configurations (or the nuclear wave function) at the scattering amplitude level, and squaring the amplitude gives the cross section. Similarly, averaging the cross section instead of the amplitude over the nucleon positions gives the sum of incoherent and coherent cross section, called \emph{quasielastic} cross section.

Coherent diffraction dominates at small momentum transfer $t\sim -1/R_A^2$ (where $R_A$ is the nuclear radius), where the dipole scatters coherently off the whole nucleus. At larger momentum transfer $t\sim -1/R_p$ (where $R_p$ is the proton radius) the probed objects are individual nucleons instead of the nucleus. Experimentally, the $t$ dependence of the coherent cross section is a challenging measurement due to the difficulties of measuring small momentum transfer and the intactness of the nucleus.

The cross section for the quasielastic vector meson production can be written as (see e.g. \re\cite{Kowalski:2006hc})
\begin{equation}
	\label{eq:quasiel-xs}
	\frac{\der \sigma^{\gamma^*A \to VA}}{\der t} = \frac{R_g^2 (1+\beta^2)}{16\pi} \langle |\A(\xpom, Q^2, \Delta_T|^2\rangle_N,
\end{equation}
where the average over the nucleon configurations is denoted by
\begin{equation}
	\langle \mathcal{O}(\{ b_{T,i} \}) \rangle \equiv \int \prod_{i=1}^{A} \left[ \der^2 b_{T,i} T_A(b_{T,i}) \right] 
\mathcal{O}(\{ b_{T,i} \}) .
\end{equation}
The nucleon positions $b_{T,i}$ are assumed to be independent, and as a nuclear thickness function $T_A$ we use the Woods-Saxon distribution~\cite{PhysRev.95.577}. The coherent cross section is obtained by averaging the amplitude before squaring it, $|\langle \A\rangle_N|^2$, and the incoherent cross section is variance $\langle |\A|^2\rangle_N - |\langle \A \rangle_N |^2$ that measures the fluctuations of the gluon density inside the nucleus. The averaged amplitude $\langle \A\rangle_N$ is a smooth function of the impact parameter $\bt$, and its Fourier transfer vanishes rapidly at $\Delta \gtrsim 1/R_A$. Therefore, at large momentum transfer $\Delta$, the quasielastic cross section \eqref{eq:quasiel-xs} is almost purely incoherent.

The factor $1+\beta^2$ takes into account the correction from the real part of the scattering amplitude and $R_g$ corrects for the so called \emph{skewedness effect}, i.e. that the gluons in the target are probed at different $x$, see \re\cite{Shuvaev:1999ce}. The corrections are calculated following the prescription of \re\cite{Watt:2007nr}.
In order to derive the cross section for the incoherent cross section we modify the IPsat model and use a factorized approximation for the scattering matrix $S_p(\rt,\bt,\xpom)=1-N_p(\rt,\bt,\xpom)$ as
\begin{equation}
	S_p(\rt,\bt,\xpom) = 1-T_p(\bt)N_p(\rt,\xpom).
\end{equation}
In \paper \cite{Lappi:2010dd} it was shown that this approximation changes the cross section very little when calculating diffractive vector meson production with proton targets.

The cross section for the coherent diffractive vector meson production is derived in \re\cite{Kowalski:2003hm}, the result being
\begin{multline}
	\langle \A(\xpom,Q^2,\Delta_T) \rangle_N = \int \frac{\der z}{4\pi} \der^2 \rt \der^2 \bt e^{-i \bt \cdot \Delta_T} [\Psi^*_V \Psi^{\gamma^*\to q\bar q}](\rt,Q^2,z) \\
	\times 2\left[ 1-\exp \left\{ -2\pi B_p  A T_A(\bt) N(\rt,\xpom) \right\} \right].
\end{multline}
Here $B_p$ is the width of the Gaussian density distribution of the proton.
The quasielastic (which is almost purely incoherent at large $|t|$) cross section was derived in \paper \cite{Lappi:2010dd} to be
\begin{multline}
\label{eq:incohxs}
	\langle \left| \A(\xpom,Q^2,\Delta_T) \right|^2 \rangle_N  
=
16 \pi B_p A \int \der^2 \bt \der^2 \rt \der^2 \rt' \frac{\der z}{4\pi} \frac{\der z'}{4\pi} \\
\times [\Psi^*_V \Psi^{\gamma^*\to q\bar q}](\rt,Q^2,z) [\Psi^*_V \Psi^{\gamma^*\to q\bar q}](\rt',Q^2,z') \\
\times e^{-B_p \Delta_T^2} e^{-2 \pi B_p A T_A(\bt)
\left[ N(\rt,\xpom) + N(\rt',\xpom) \right] } 
\\ \times
\frac{\pi B_p N(\rt,\xpom) N(\rt',\xpom) T_A(\bt) }
  {1 - 2 \pi B_p T_A(\bt)\left[ N(\rt,\xpom) + N(\rt',\xpom) \right] } .
\end{multline}
The squared amplitude is proportional to $A$ times the squared dipole-proton amplitude, corresponding to independent scattering off the nucleons. This is multiplied by a nuclear attenuation factor
\begin{equation}
\label{eq:attenuation}
	\frac{ e^{-2 \pi B_p A T_A(\bt) \left[ N(\rt,\xpom) + N(\rt',\xpom) \right] }} {1 - 2 \pi B_p T_A(b)\left[ N(\rt,\xpom) + N(\rt',\xpom) \right]} \approx  e^{-2\pi (A-1)T_A(\bt)  \left[ N(\rt,\xpom) + N(\rt',\xpom) \right] },
\end{equation}
which accounts for the requirement that the dipole can not scatter inelastically off the remaining $A-1$ nucleons in order to keep the event diffractive, see \paper \cite{Lappi:2010dd}  for details.

\begin{figure}[tb]
\centering
	\begin{minipage}[t]{0.5\textwidth}
	\includegraphics[width=1.0\textwidth]{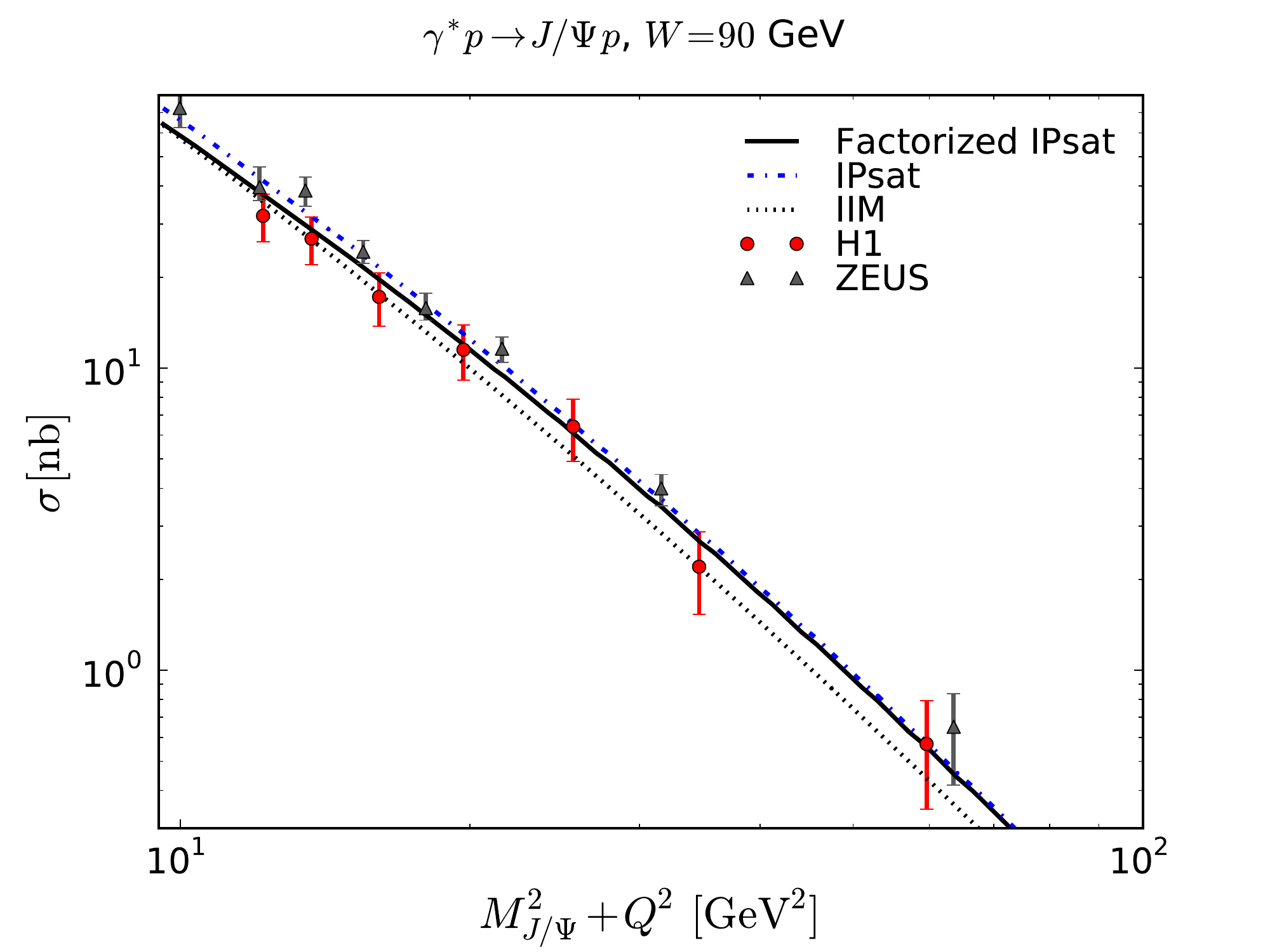}
	\caption{Diffractive J/$\Psi$ production cross section as a function of $Q^2$ compared to different dipole model calculations. Figure from \paper\cite{Lappi:2010dd}.}
	\label{fig:hera-diffraction-comparison}
	\end{minipage}%
	~
	\begin{minipage}[t]{0.5\textwidth}
	\includegraphics[width=1.0\textwidth]{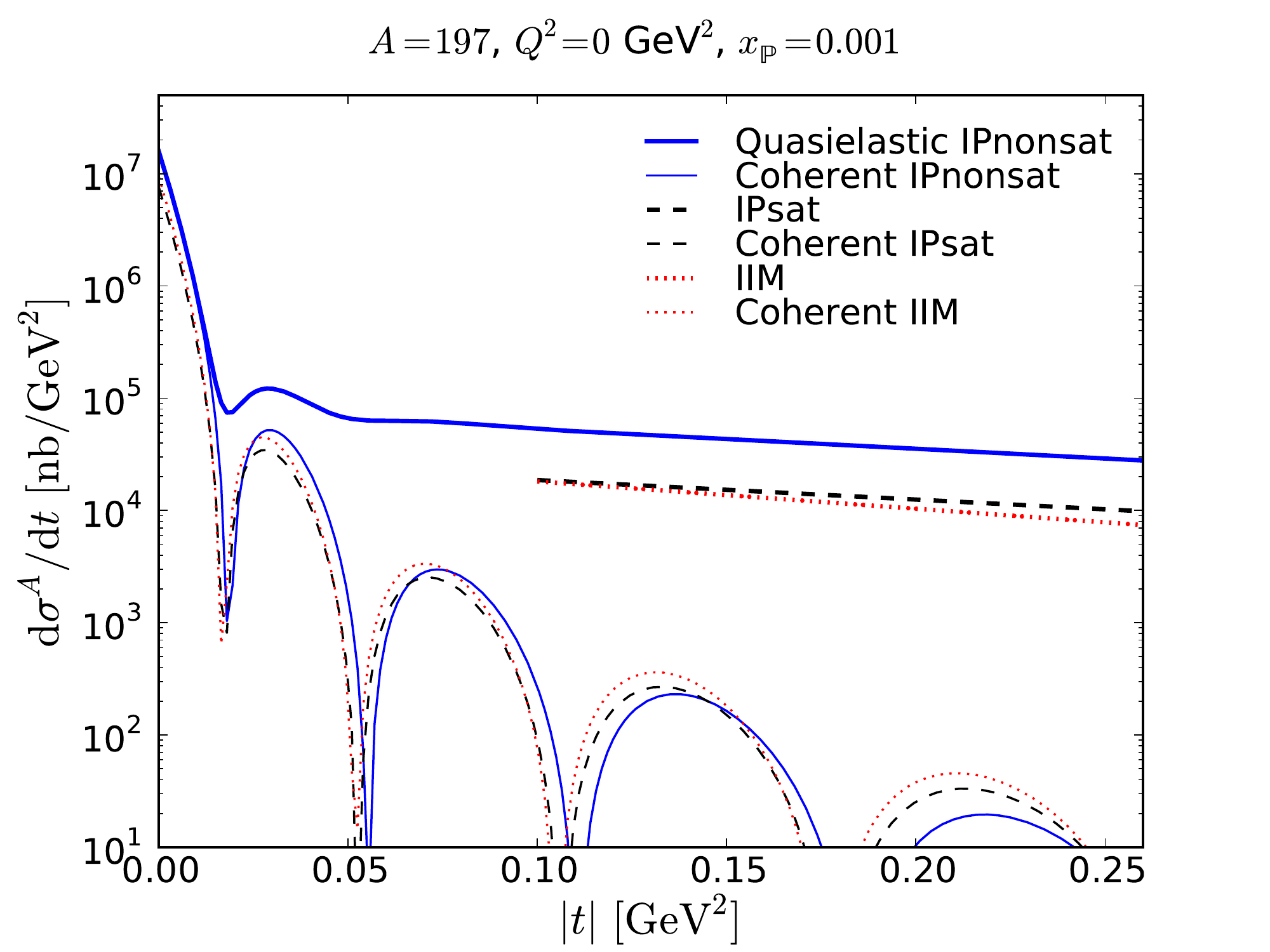}
\caption{The quasielastic and coherent diffractive J/$\Psi$ cross
sections in gold nuclei.
%Shown are the IPsat and IIM parametrizations and ``IPnonsat'' for comparison. 
Figure from \paper\cite{Lappi:2010dd}. %Our approximation \eqref{eq:incohxs} is not valid for small $|t|$; the corresponding part of the distribution has been left out.
}
\label{fig:ddis_jpsi_t}
	\end{minipage}		
\end{figure}

\begin{figure}[tb]
\centering
	\begin{minipage}[t]{0.5\textwidth}
	\includegraphics[width=\textwidth]{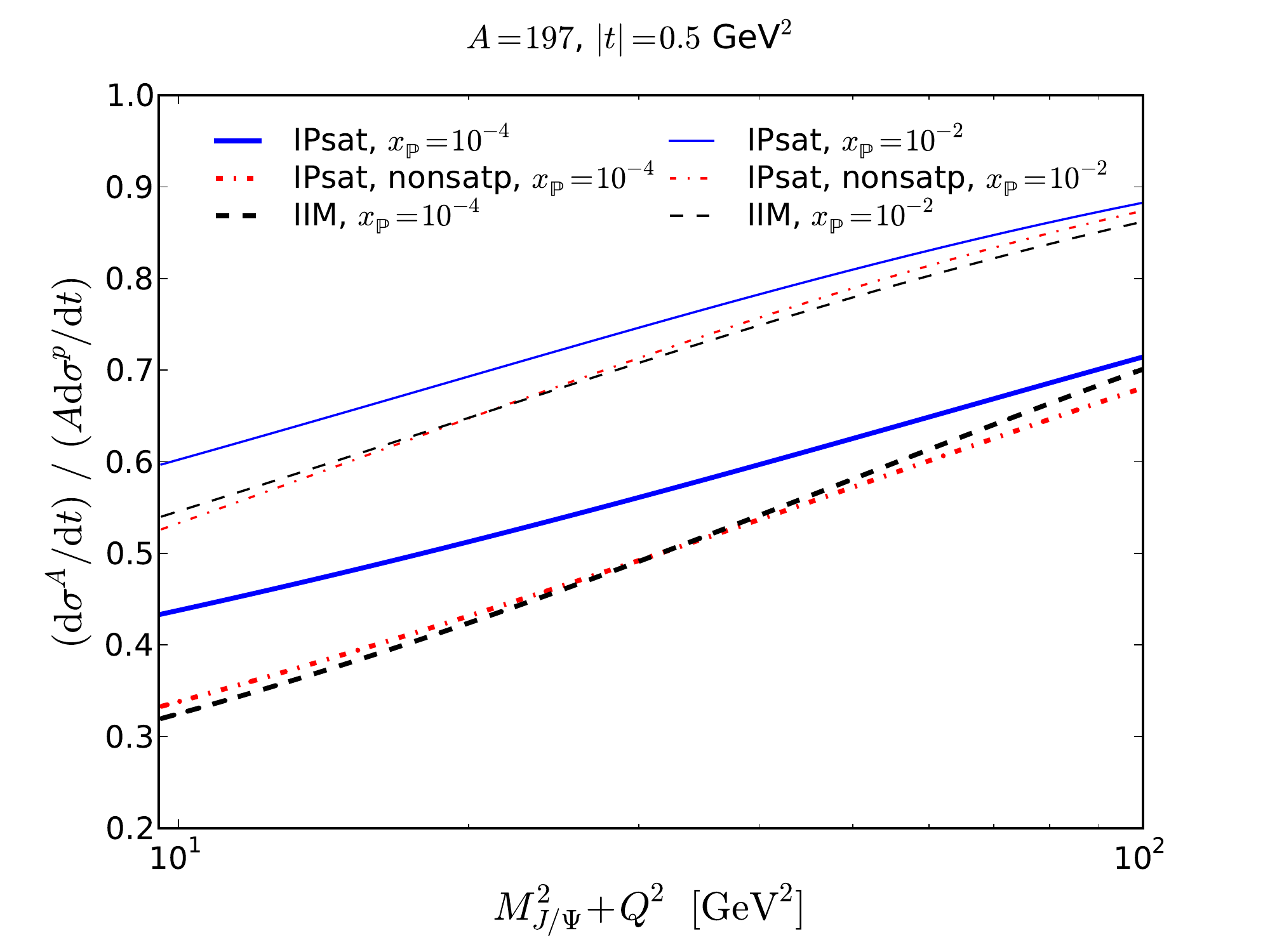}
	\caption{Nuclear suppression factor of the cross sections as a function of $Q^2$ for diffractive J/$\Psi$ production. The upper thin curves correspond to $\xpom=10^{-2}$, and the lower thick curves are calculated at $\xpom=10^{-4}$. Figure from \paper\cite{Lappi:2010dd}.} 
	\label{fig:ea_diff_q}
	\end{minipage}%
	~
	\begin{minipage}[t]{0.5\textwidth}
	\includegraphics[width=1.0\textwidth]{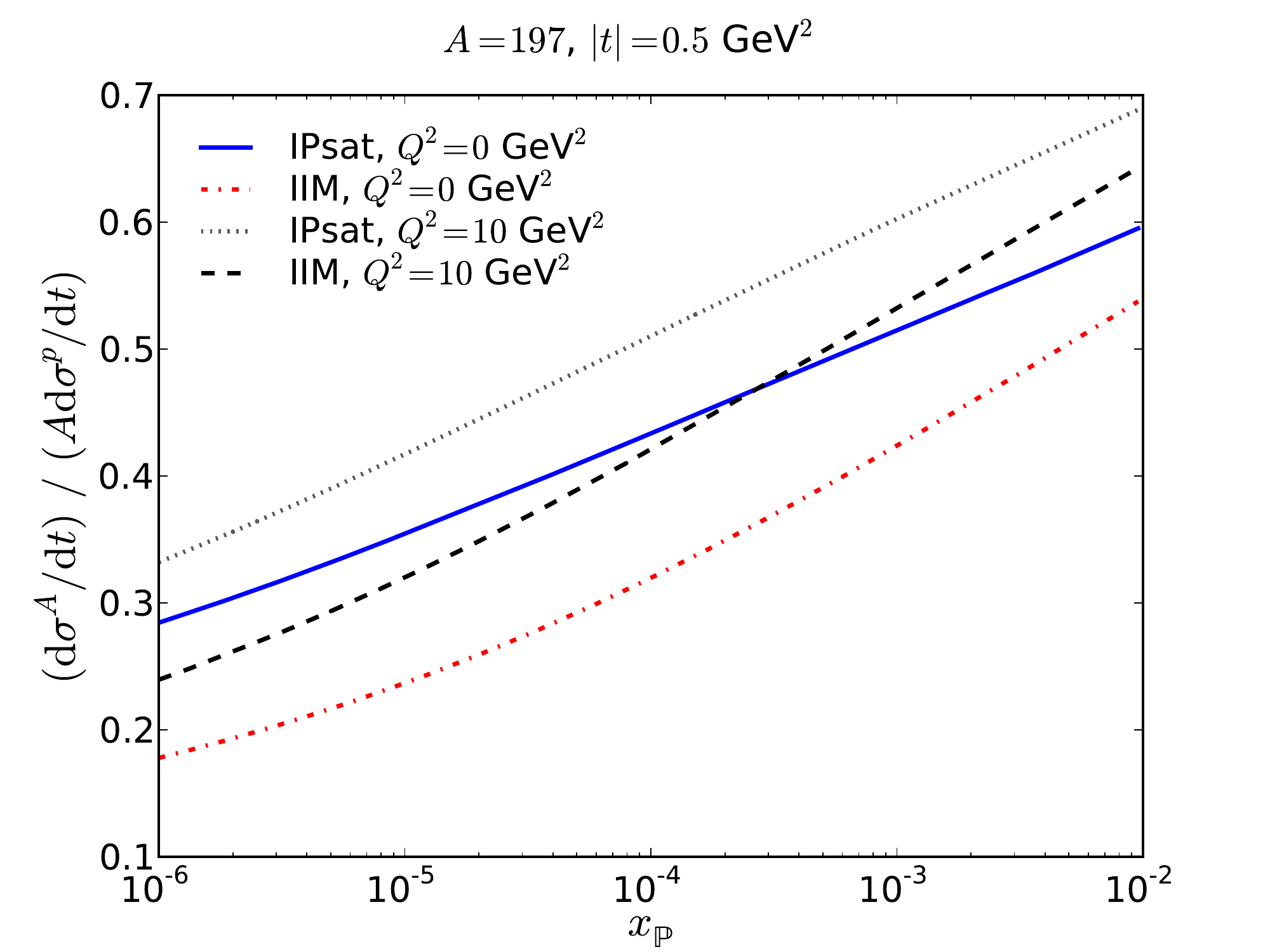}
	\caption{Nuclear suppression factor of the cross sections as a function of $\xpom$ for diffractive J/$\Psi$ production at photoproduction and at $Q^2=10\gev^2$. Figure from \paper\cite{Lappi:2010dd}.}
	\label{fig:ea_diff_x}
	\end{minipage}	
\end{figure}

The calculated incoherent and coherent diffractive $J/\Psi$ photoproduction ($Q^2=0\gev^2$) cross sections in $\gamma^*$-gold scattering are shown in \fig \ref{fig:ddis_jpsi_t}, where the differential cross section is shown as a function of squared transverse momentum transfer $t$.  Note that our approximation for the incoherent cross section is not valid at small $|t|$, and the corresponding part of the distribution has been left out.
 To estimate the saturation effects we also calculate the cross section using the ``IPnonsat'' model, where the IPsat model is linearized such that the dipole amplitude is proportional to $\rt^2$ at all dipole sizes. When this parametrization is used, the incoherent cross section is explicitly $A$ times the dipole-proton cross section (as calculated e.g. in \re \cite{Caldwell:2009ke}).
The two saturation model parametrizations (IPsat and IIM) give comparable results for both coherent and incoherent cross sections, and the incoherent cross section is strongly suppressed by the saturation effects.

The origin of the suppression can be seen from the nuclear attenuation factor \eqref{eq:attenuation}: in the black disk limit $N(\rt, \xpom)=1$ (which is the case when $\rt \gtrsim 1/\qs$), and the nuclear attenuation factor at small impact parameters behaves like $\sim e^{-0.5 A^{1/3}}$, which can be seen by noticing that $T_A(\bt)\sim A^{-2/3}$ when $|\bt|$ is not very large. Thus the contribution from the center of the nucleus to the incoherent cross section is suppressed, and only the scattering from the edges of the nucleus contributes. As the edge of the nucleus has an area $\approx 2 \pi R_A d$ (where $d$ is the thickness of the edge), the incoherent cross section behaves as $\sim A^{1/3}$ in the black disk limit. On the other hand, in the dilute limit (with no saturation effects) the attenuation factor goes to unity as the dipole amplitude is small,  and the cross section is proportional to $A$.

%\begin{figure}[tb]
%\begin{center}
%\includegraphics[width=0.6\textwidth]{figs/coherent_q0.pdf}
%\caption{The quasielastic and coherent diffractive J/$\Psi$ cross
%sections in gold nuclei at $Q^2=0$ and $\xpom = 0.001$. Shown are
%the IPsat and IIM parametrizations and ``IPnonsat'' for compariosn. Our approximation \eqref{eq:incohxs} is not valid for small $|t|$; the corresponding part of the distribution has been left out.}
%\label{fig:ddis_jpsi_t}
%\end{center}
%\end{figure}

To quantify the nuclear effects we calculate the \emph{nuclear suppression factor} defined as the ration of the $\gamma^*$-nucleus and $\gamma^*$-proton cross sections normalized by the number of nucleons $A$ in the gold nucleus. The scale $Q^2$ dependence of the suppression factor is shown in \fig\ref{fig:ea_diff_q}, and the same quantity as a function of $\xpom$ is plotted in \fig\ref{fig:ea_diff_x}. The suppression is larger at small $Q^2$ and at smaller $\xpom$ where the saturation effects should be large. Note that the dominant dipole size is $r\sim 1/Q$, and thus at large $Q^2$ a dilute region of the nucleus is probed where no nuclear effects are expected. Similarly at smaller $\xpom$ the non-linear effects should be larger, which can be clearly seen from \fig\ref{fig:ea_diff_x} where the suppression grows as $\xpom$ decreases. The model uncertainties are quantified by calculating the ratio with both IPsat and IIM parametrizations, the difference being largest at small $Q^2$ and $\xpom$.

In \fig\ref{fig:ea_diff_q} the results are calculated also using a nonsaturated dipole-nucleon cross section (``IPsat, nonsatp''), which corresponds to including unitarity (or equivalently saturation) effects at the level of the nucleus but not for a single nucleon. With this parametrization a much larger suppression is obtained than with the IPsat model. This shows that the nuclear suppression ratio is also sensitive to the saturation effects at the proton level.

The fact that the different parametrizations (IPsat and IIM) differ significantly when calculating spectra or nuclear suppression factor shows that the diffractive electron-nucleus scattering has a discriminating power to separate between different dipole models, and it can be used to constrain the model uncertainties. Note that here both parametrizations are fitted to HERA DIS data\footnote{Although we have made a simplifying assumption for the impact parameter dependence of the IPsat model.}, but the differences in eA scattering can be up to 50\%. The calculations presented here are done in the kinematical region that would be available in a future electron-ion collider.

\section{Centrality in diffractive events at an Electron Ion Collider}

Centrality selection based on event multiplicity has been a powerful tool to learn about the QCD dynamics in heavy ion collisions.
The multiplicity selection has been applied also in proton-nucleus and proton-proton collisions. Recent studies of high multiplicity proton-proton and proton-nucleus events have revealed many interesting phenomena such as long range angular correlations (see e.g. \re\cite{Abelev:2012ola}) that are not always visible in minimum bias events.

In the previous section the diffractive vector meson production cross section was calculated for lepton (or photon)-nucleus scattering. It was also found that coherent diffraction dominates at small $|t|$, whereas  incoherent diffraction is the only component at $|t|\sim 1/R_p$ where the $\pt$ kick is localized on an area comparable to the nucleon area. 
Incoherent diffraction then probes fluctuations at the distance scale of the order of the nucleon size~\cite{Caldwell:2009ke,Miettinen:1978jb}, and the incoherent cross section can be expected to depend on the impact parameter of the photon-nucleus collision\footnote{With $|t|\gtrsim 1/R_p$ the cross section would also be sensitive to sub-nucleon scale fluctuations which are not completely included in this work.}. Triggering on the most central collisions we could probe the densest region of the nucleus where the saturation scale $Q_s$ is expected to be enhanced relative to the minimum bias events. 

In heavy ion collisions the centrality selection can be done relatively easily according to the total multiplicity of the event, such that the events with largest multiplicity are the ones where the two nuclei collided with zero impact parameter. The multiplicity classes in proton-nucleus collisions are discussed later in \ch\ref{ch:sinc}. 
%In proton-nucleus collisions the situation is less clear, as the smaller multiplicities make it difficult to develop centrality selection methods that would not cause biases. In practice, different multiplicities and produced transverse energies in different kinematical regions are used. 
What we propose in \paper\cite{Lappi:2014foa} is to use the proton multiplicities in the ``Roman pot'' detectors as a measure of the impact parameter in incoherent diffraction.

\begin{figure}[tb]
\centering
	\begin{minipage}[t]{0.5\textwidth}
	\includegraphics[width=1.0\textwidth]{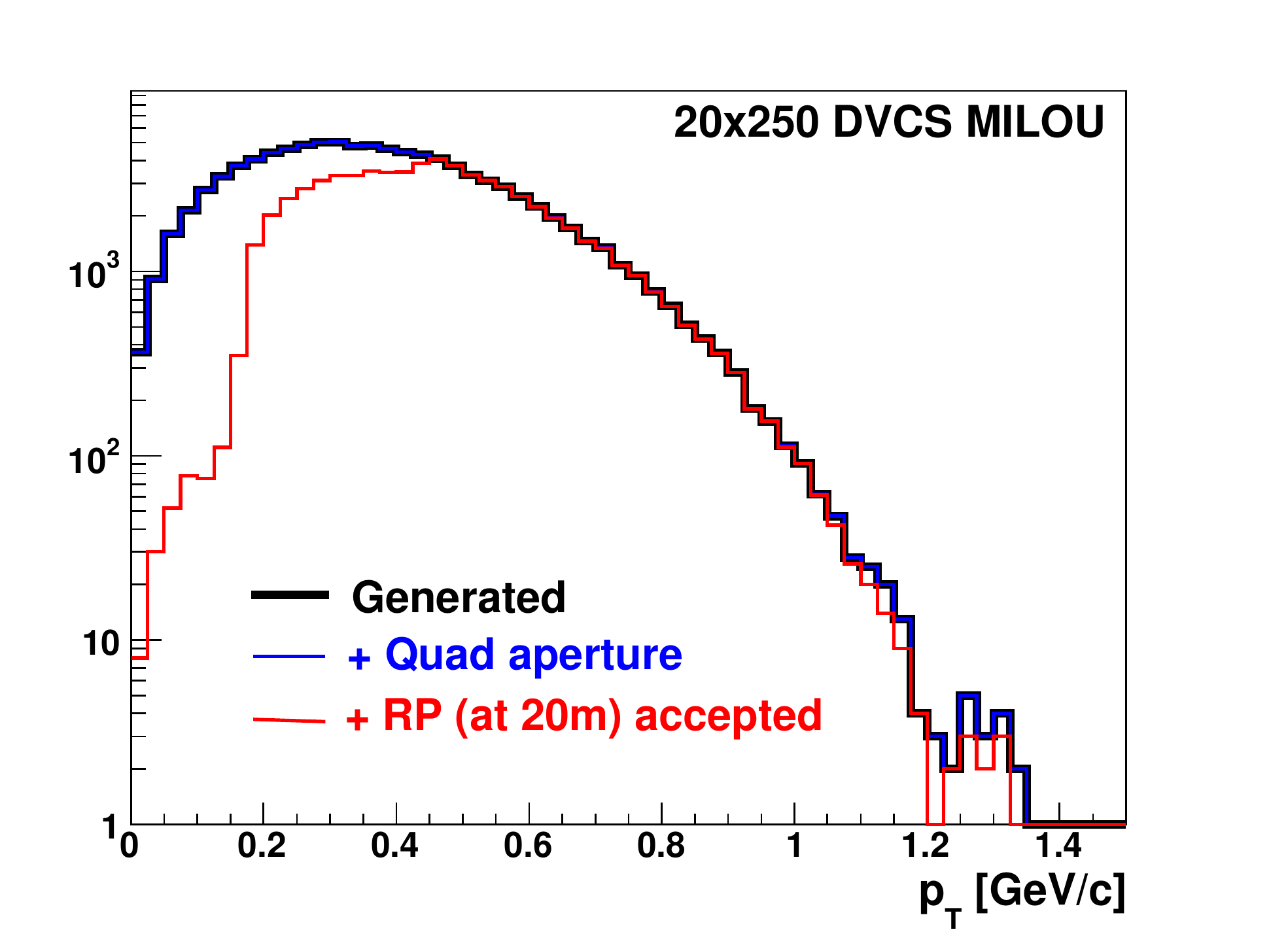} 
		\caption{
EIC acceptance for protons, as a function of $p_T$, in a Roman pot detector. Figure from \re\cite{eicwikidvcs}. 
}
	\label{fig:eic-dvcs}
	\end{minipage}%
	~
	\begin{minipage}[t]{0.5\textwidth}
	\includegraphics[width=1.0\textwidth]{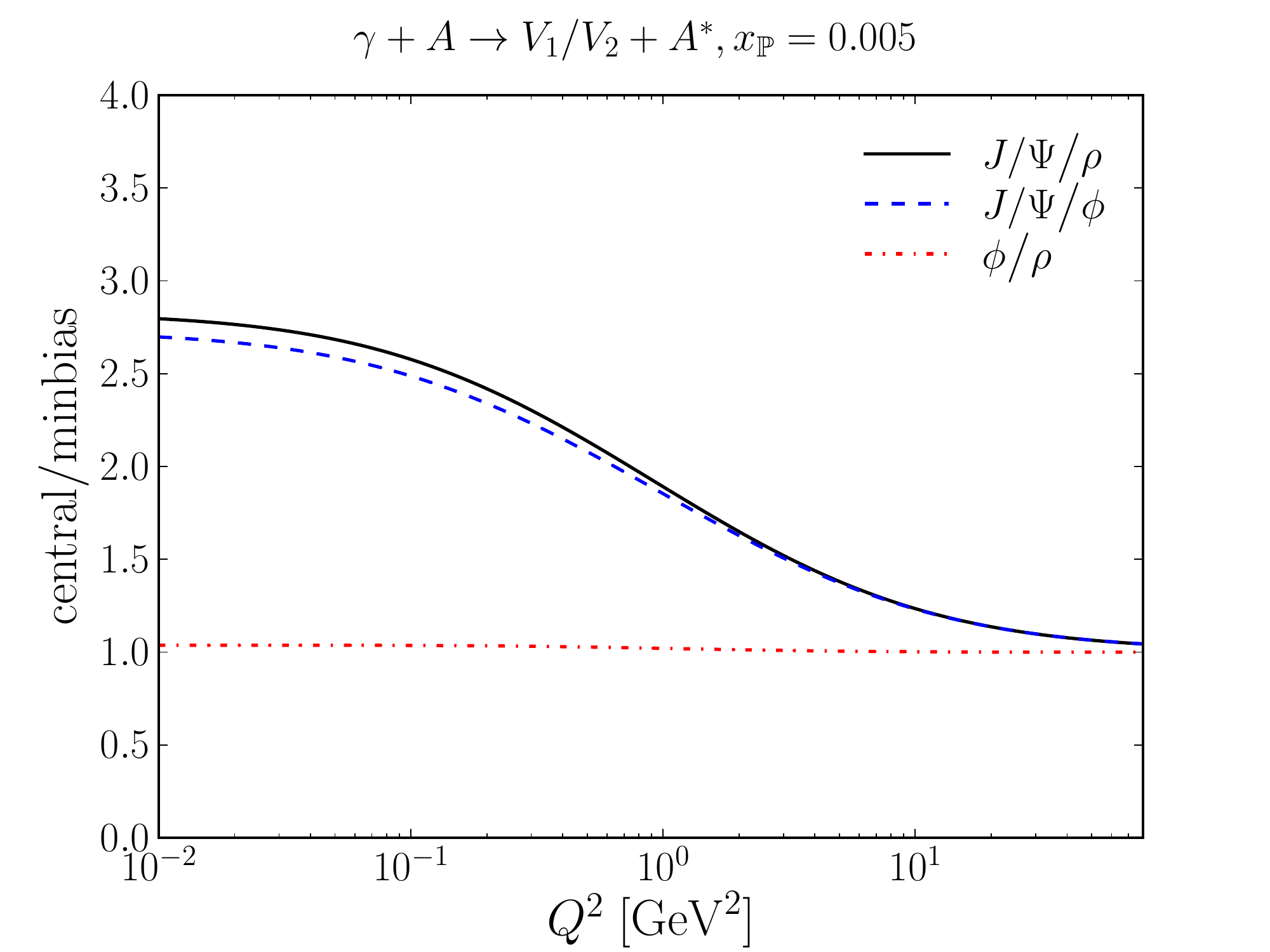}
	\caption{Ratio of two incoherent vector meson production cross sections in central events relative to minimum bias as a function of $Q^2$. Figure from \paper\cite{Lappi:2014foa}.}
	\label{fig:eic-doubleratio}
	\end{minipage}	
\end{figure}

Consider a nucleus participating in a event where a vector meson is produced diffractively. If a nucleon in the nucleus receives a large momentum kick $\sim \sqrt{|t|}$, it escapes from the nucleus and can scatter off other nucleons on its path out. When the scattering takes place close to the center of the nucleus, there are more nucleons off which the participating nucleon can scatter and on average we expect more ``ballistic'' nucleons to be produced than in peripheral events. 
These nucleons travel along the beam pipe but their trajectories differ slightly from that of the original nucleus and can be measured in the Roman pot detectors located in the beam pipe outside the main detectors. 
The simulated acceptance of the EIC~\cite{Accardi:2012qut} Roman pot detector is shown in \fig\ref{fig:eic-dvcs} where it can be seen that the acceptance is good (generated and accepted spectra are the same) in the region of $\pt$ that the ballistic protons would have
\footnote{The simulation is done for Deeply Virtual Compton Scattering in e+p collision, in e+A scattering the acceptance will shift to lower $p_T$ due to the higher magnetic fields required by the different mass-to-charge ratio.}. 
Note that now we can only discuss about ballistic protons, as the neutrons are not bent by the magnetic field and do not reach the Roman pots.

In diffractive events the nucleus is usually left in an excited state. When the nucleus returns to its ground state nucleons can evaporate according to a thermal spectrum. In the laboratory frame these nucleons should have significantly smaller transverse momentum than $400 \mev \dots 1 \gev$ that we expect for many ballistic protons, and are not detected in the Roman pots.

The produced neutrons from both of these channels can be measured in Zero Degree Calorimeters (ZDC), but there the separation between the  ballistic and thermal components is difficult. Thus we do not expect the ZDC measurements to give sufficient information to divide diffractive events into the centrality classes. Note, however, that in other processes such as single inclusive multiplicities and dihadron correlations the ZDC energy has a potential to be a measure of the event centrality~\cite{Zheng:2014cha}.

As a model example we study the centrality dependence of the incoherent diffractive vector meson production cross section using the framework described earlier in this Chapter. ``Central'' events, where the proton multiplicities in the Roman pot should on average be large, are defined by setting the impact parameter to zero when calculating the incoherent cross section from \eq \eqref{eq:incohxs}. The production cross section is compared with the minimum bias cross sections obtained by integrating over all impact parameters. To cancel the theoretical uncertainties associated with e.g. the vector meson wave functions, we consider the $Q^2$ dependence of the double ratio
\begin{equation}
	\frac{ \left. \sigma(\gamma^*A \to V_1 A^*) \Big/  \sigma(\gamma^*A \to V_2 A^*) \right|_{\text{central}} }
	 {\left. \sigma(\gamma^*A \to V_1 A^*) \Big/  \sigma(\gamma^* A \to V_2 A^*) \right|_{\text{minimum bias}} },
\end{equation}
where $V_1$ and $V_2$ refer to different vector meson species. In this analysis, we include J/$\Psi$, $\rho$ and $\phi$. In our approximation the real part and skewedness corrections and $t$ dependence cancel in the double ratio. 

The results are shown in \fig\ref{fig:eic-doubleratio} where the double ratio is calculated at $\xpom=0.005$ which is within the EIC kinematics. A significant enhancement in the $J/\Psi/\rho$ and $J/\Psi/\phi$ ratios is seen at small $Q^2$. This enhancement can be understood as follows. As $J/\Psi$ is more massive, its wave function is peaked at smaller dipoles and the cross section is dominated by configurations with $r^2 \qs^2 \ll 1$ also at low $Q^2$. On the other hand for larger $\rho$ and $\phi$ mesons the typical configurations have $r^2 \qs^2 \ge 1$ at the same $\qs$, and the saturation effects affect the cross section more in central events (where the saturation scale $\qs$ is larger) than in the minimum bias events when $Q^2$ is decreased, which increases the double ratio. 

Another way to think of this is to note that as $r^2 Q_s^2\ge 1$ in the production of larger mesons ($\rho$ and $\phi$), the cross sections in central and minimum bias events are proportional to the geometric area of the interaction and cancel from the double ratio. Moreover, as the $J/\Psi$ production cross section goes like $Q_s^4$ (as the dipole amplitude is proportional to $Q_s^2$ in the dilute region), the double ratio becomes just the ratio of the saturation scales $Q_{s,\text{central}}^4/Q_{s,\text{min.bias}}^4$. At large enough $Q^2$ also $\rho$ and $\phi$ cross sections become dominated by dipole sizes where $r^2\qs^2 \ll 1$ and the numerators and denominators cancel separately from the double ratio, and we obtain unity as can be seen from \fig\ref{fig:eic-doubleratio}.
Since $\rho$ and $\phi$ have roughly the same size, they are simultaneously probing either dilute or dense limits ($r^2\qs^2 \ll 1$ or $r^2\qs^2 \ge 1$) at fixed $Q^2$, and thus almost no $Q^2$ evolution can be seen in the $\phi/\rho$ double ratio.

The calculation shown in \fig\ref{fig:eic-doubleratio} is a direct measure of the nuclear enhancement originating from the non-linear gluon dynamics described in the CGC framework. Even though we can not tell how many protons in the Roman pot corresponds to very central events, experimentally the most central event class can be defined with events that have largest number of ballistic protons in the Roman pot. These events, on average, are more central than the minimum bias results. We expect that the results for the double ratio will be very different when calculating in models where the non-linear QCD dynamics is included differently than in the dipole model. Thus, even though quantitative data comparisons require more work (e.g. combining the presented calculation with a more detailed model for the nuclear breakup), interesting qualitative results can be obtained from the first EIC measurements.

\section{Ultraperipheral heavy ion collisions}
\label{sec:upc}
As discussed in this Chapter, diffraction in lepton-nucleus collision has a potential to be a powerful tool to study the properties of the dense QCD matter, that is, the small-$x$ structure of nuclei and protons. In deep inelastic scattering and lepton-nucleus diffraction the QCD dynamics is encoded in the virtual photon-nucleus scattering and the role of the lepton is to act as a source of virtual photons. 

An electron-nucleus collider would be an ideal tool to study, for example, exclusive vector meson production, but new experimental facilities such as the EIC~\cite{Accardi:2012qut} or the LHeC~\cite{AbelleiraFernandez:2012cc} colliders are needed for these studies. With currently available experiments, one possibility to study photon-nucleus scattering is to consider scattering processes where two hadrons collide with such a large impact parameter that the strong interactions are suppressed and can be neglected. Then, the role of one of the colliding (charged) objects is to act as a moving electromagnetic charge emitting virtual photons that scatter off the second hadron. This is exactly what is done in ultraperipheral heavy ion collisions.

\begin{figure}[tb]
\begin{center}
\includegraphics[width=0.4\textwidth]{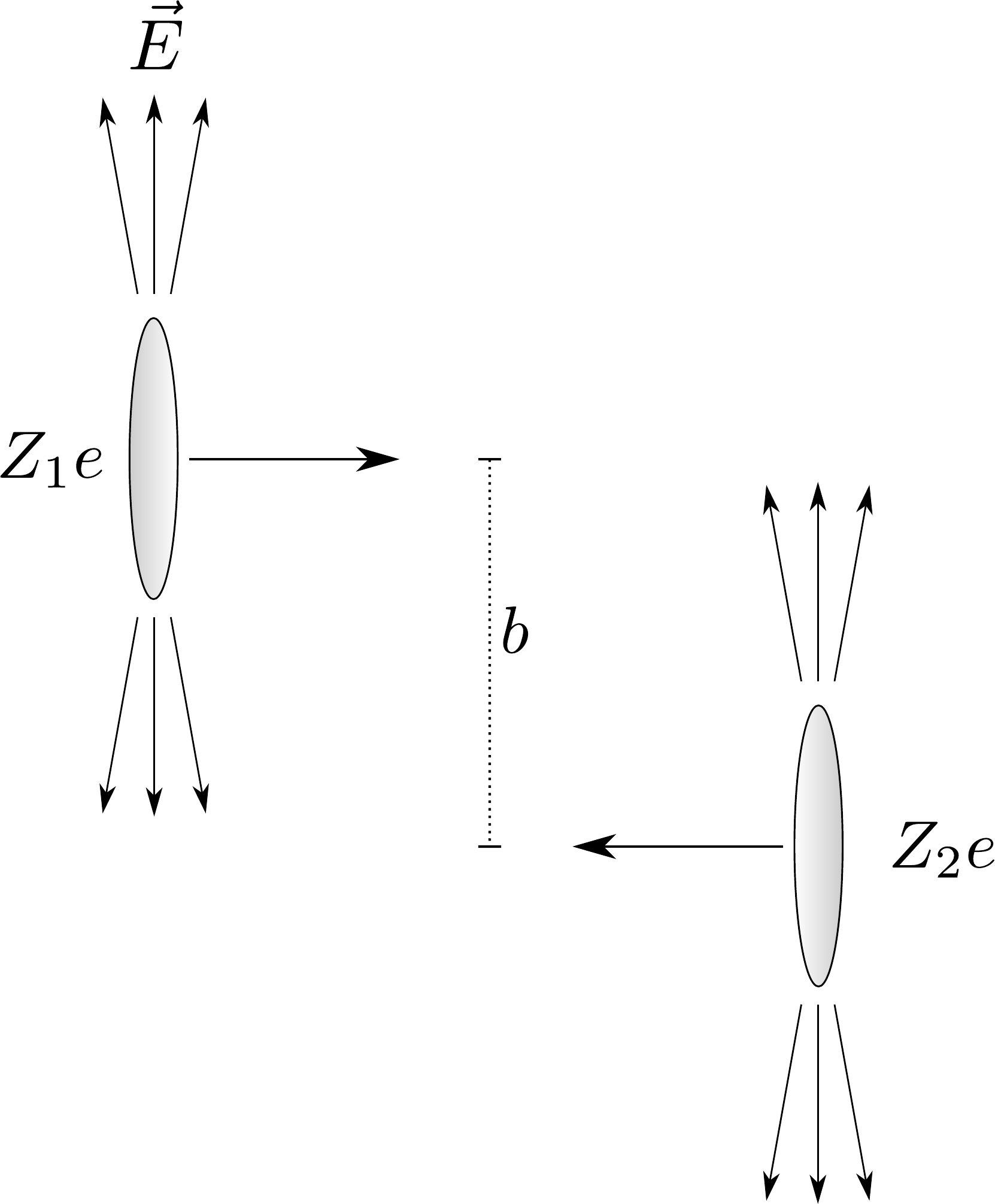}
\caption{Two heavy nuclei colliding with large impact parameter $b$ at high energy.}
\label{fig:ultraperipheral}
\end{center}
\end{figure}

Let us consider a scattering of two heavy nuclei at high energy such that the impact parameter is larger than twice the nuclear radius, $b\gtrsim 2 R_A$, as shown schematically in \fig\ref{fig:ultraperipheral}. In these events it is possible to have photon-nucleus or photon-photon scattering. The later is suppressed by additional powers of electromagnetic coupling $\alphaem$ and is not considered here.

Following the review~\cite{Bertulani:2005ru} the cross section for ultraperipheral hadron-hadron collision $\sigma^{AA}$ can be written as
\begin{equation}
\label{eq:upc-xs}
	\sigma = \int \der \omega \frac{n(\omega)}{\omega} \sigma^{\gamma A},
\end{equation}
where $\sigma^{\gamma A}$ is the photon-hadron cross section. The photon flux integrated over all possible impact parameters $b>b_\text{min}=2R_A$ is
\begin{equation}
	n(\omega)=\frac{2Z^2 \alphaem}{\pi \beta} \left[ \xi K_0(\xi) K_1(\xi) - \frac{\xi^2}{2}(K_1^2(\xi) - K_0^2(\xi) ) \right],
\end{equation}
where $\xi = 2\omega R_A/(\gamma \beta)$, $Z$ is the number of protons in the nucleus and $\gamma$ is the Lorentz boost factor of the beam in the center of mass frame  and $\beta\approx 1$ is the velocity of the incoming hadron.  

Let us consider the production of a vector meson with mass $M_V$, momentum $p_V$ and rapidity $y$ in the laboratory frame in a process where a photon (momentum $q$, energy $\omega$) is emitted from the nucleus that is moving along the $z$ axis in the positive direction, and the nucleon it is scattering off has a momentum $P$. Assuming that the transverse momentum of the diffractively produced meson is small compared to its mass we obtain
\begin{equation}
p_V^+ = \frac{1}{\sqrt{2}} m_T e^y \approx \frac{1}{\sqrt{2}} M_V e^y,
\end{equation}
where $m_T=\sqrt{M_V^2+p_T^2}$ is the transverse mass. On the other hand, momentum conservation gives
\begin{equation}
	p_V^+ = q^+ + x P^+ = \sqrt{2}\omega,
\end{equation}
as $P^+\approx 0$, $q^+=\sqrt{2} \omega$ and $x$ is the longitudinal momentum transfer from the second nucleus. This gives 
\begin{equation}
	\omega = \frac{M_V}{2} e^y.
\end{equation}

Differentiating \eq\eqref{eq:upc-xs} with respect to $\omega$ we find
\begin{equation}
	\frac{\der \sigma^{AA}}{\der \omega} = \frac{n(\omega)}{\omega} \sigma^{\gamma A}
\end{equation}
and noticing that $\der \omega/\omega = \der y$ we can write
\begin{equation}
	\frac{\der \sigma}{\der y} = \int \der t n(\omega) \frac{\der \sigma^{\gamma A}}{\der t}.
\end{equation}
This allows us to calculate the differential cross sections as we know how to compute $\der \sigma^{\gamma A}/{\der t}$, as discussed in previous Sections. Note that experimentally one can not determine which one of the nuclei acted as a source of photons and from which nucleus the photon scattered off. Thus, both processes must be taken into account by calculating
\begin{equation}
	\frac{\der \sigma}{\der y} = n(y) \sigma^{\gamma A_1}(y) + n(-y) \sigma^{\gamma A_2}(-y).
\end{equation}

Let us then limit ourselves into the photoproduction region where $Q^2=0$ and calculate the kinematical invariant $\xpom$. First, we observe that by definition $W^2 = (q+P)^2 \approx 2 q\cdot P \approx 2 q^+P^- = 2\omega \sqrt{s}$, as $q^+=\sqrt{2}\omega$ and $P^- = \frac{1}{\sqrt{2}} 2E$ with $\sqrt{s}=2E$ where $E$ is the energy of the nucleon. Now using the definition of $\xpom$ from \eq\eqref{eq:xpom} gives
\begin{equation}
	\xpom \approx \frac{M^2}{W^2} = \frac{M_V}{\sqrt{s}} e^{-y}.
\end{equation}
Because scattering off both nuclei is possible, in vector meson production at large rapidities both the small and large $x$ structure of the nucleus is probed: a small-$x$ photon can scatter off a large-$x$ gluon or vice versa. The dipole model calculations are only valid at small $\xpom$, which limits the applicability of the model close to central rapidity.

The ALICE collaboration has measured both coherent and incoherent diffractive $J/\Psi$ production~\cite{Abbas:2013oua} differentially in rapidity. Thanks to the large center-of-mass energy, the dipole model results are valid within $|y|\lesssim 2$ for the $J/\Psi$ production, as in this region $\xpom \lesssim 0.02$. In case of RHIC the smaller energy used in the experiment limits the applicability of the framework such that the production cross section at central rapidity can barely be computed, but no rapidity dependence is obtainable.

\begin{figure}[tb]
\centering
	\begin{minipage}[t]{0.5\textwidth}
	\includegraphics[width=\textwidth]{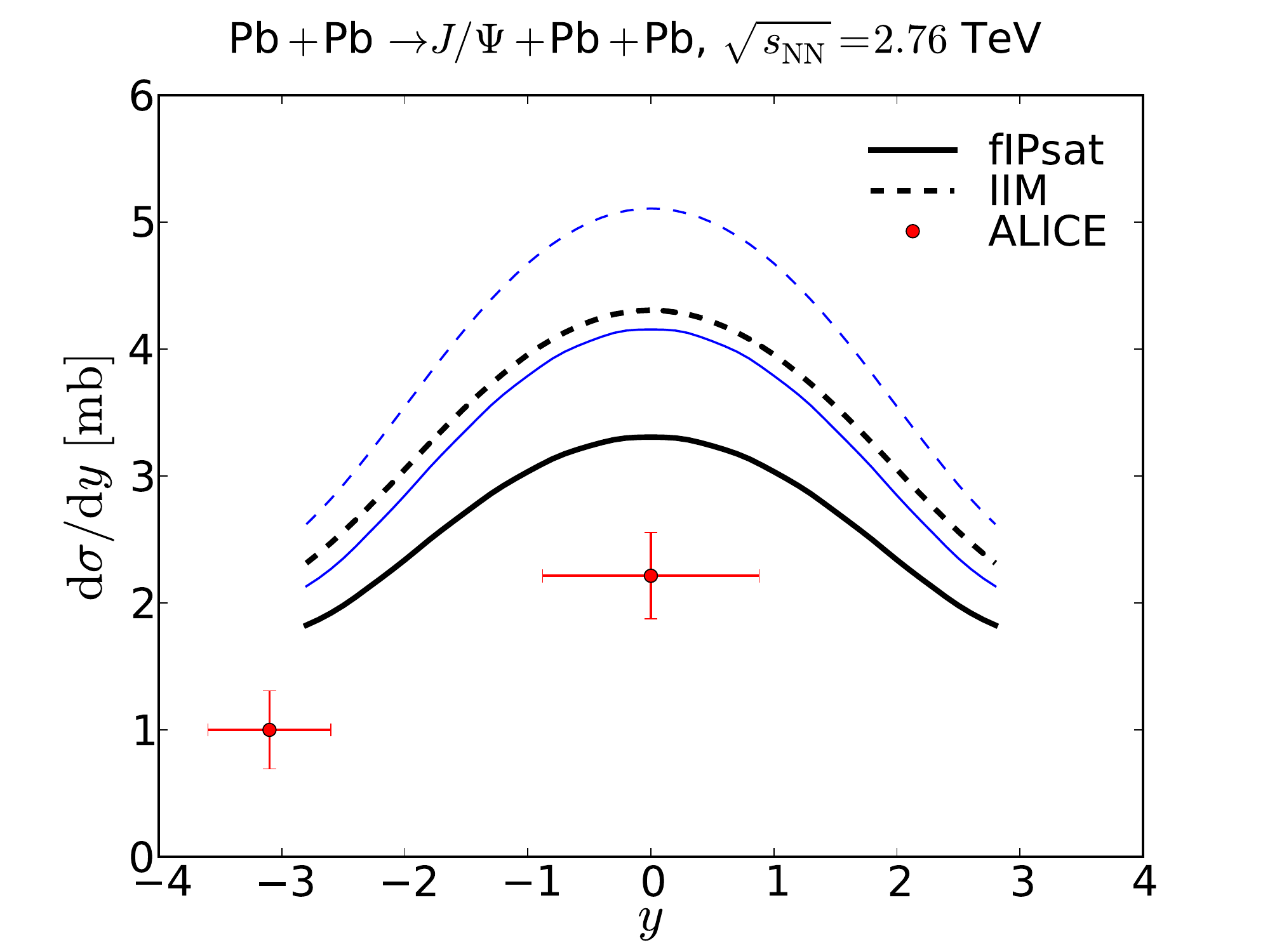}
	\caption{Coherent diffractive J/$\Psi$ production cross section computed using the factorized IPsat (``fIPsat'') and IIM parametrizations compared with the ALICE data~\cite{Abbas:2013oua}. Thin lines are obtained by using the Boosted Gaussian wave function for the J/$\Psi$, and for the thick liens Gaus-LC is used. Figure from \paper\cite{Lappi:2013am}.} 
	\label{fig:coherent-alice}
	\end{minipage}%
	~
	\begin{minipage}[t]{0.5\textwidth}
	\includegraphics[width=1.05\textwidth]{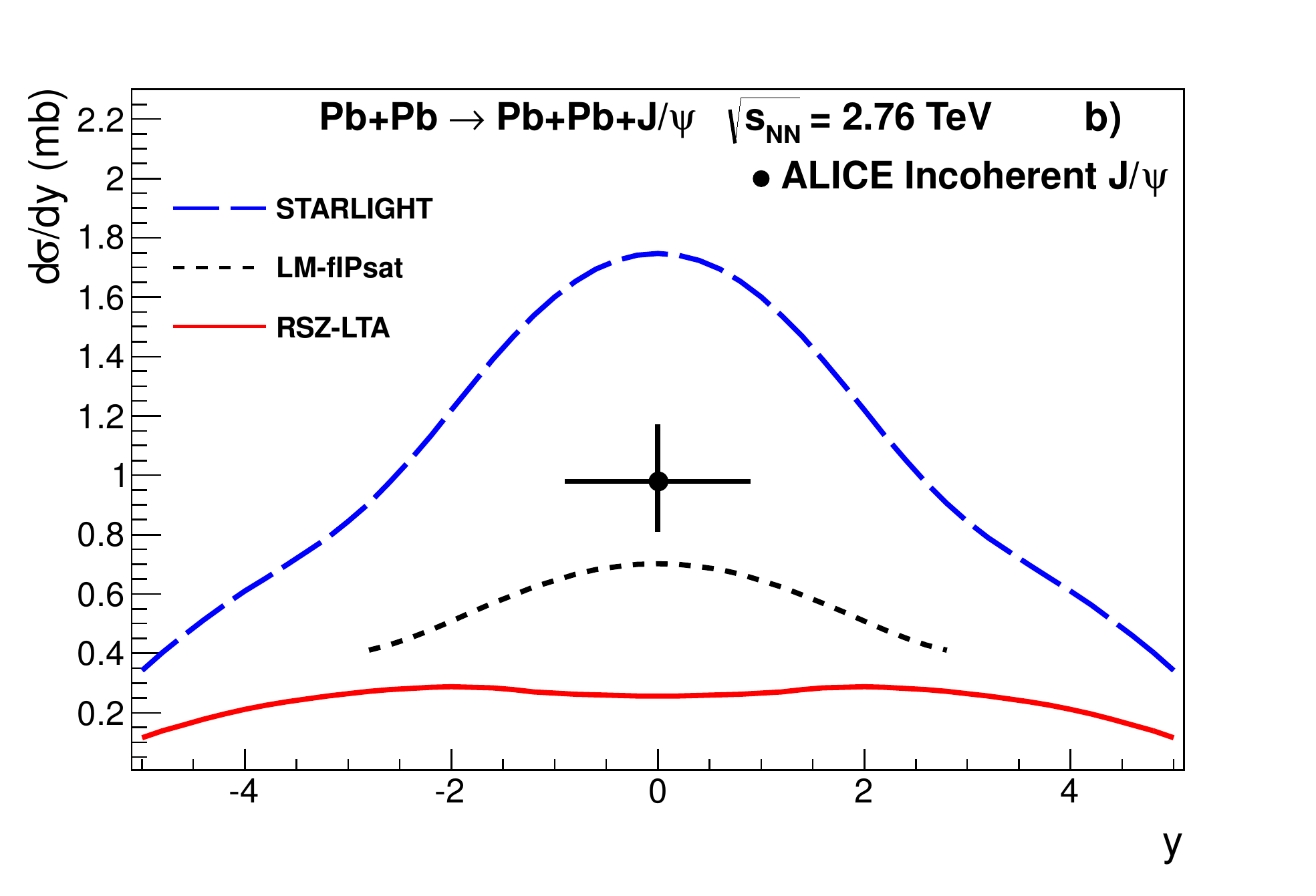}
	\caption{Incoherent diffractive J/$\Psi$ production calculated in dipole model (``LM-fIPsat'') and using a leading twist approximation (``RSC-LTA'') compared to ALICE data~\cite{Abbas:2013oua} and to the STARLIGHT event generator.
	Figure from \re\cite{Abbas:2013oua}.}
	\label{fig:incoherent-alice}
	\end{minipage}	
\end{figure}

In Papers \cite{Lappi:2010dd} and \cite{Lappi:2013am} we compute the diffractive $J/\Psi$ production cross section using the same dipole models (IPsat and IIM parametrizations) as discussed earlier in this Chapter. Recall that the input to these calculations comes from electron-proton DIS measurements combined with standard nuclear geometry. First, we compare our results for coherent diffractive J/$\Psi$ production in ultraperipheral lead-lead collisions at $\sqrt{s_{NN}}=2.76\tev$ to the ALICE data~\cite{Abbas:2013oua}. The results are shown in \fig\ref{fig:coherent-alice}, where the model uncertainty is quantified by calculating the cross section using both IPsat and IIM parametrizations and using two different wave functions for the $\gamma^*$ and J/$\Psi$ overall (Boosted Gaussian and Gaus-LC from \re\cite{Kowalski:2006hc}). 

Our results slightly overshoot the data, but the rapidity dependence comes out correctly, which means that evolution of the dipole cross section as a function of $\xpom$ is roughly correct. All parametrizations give consistently $\der \sigma/\der y|_{y=0} \,/\, \der \sigma/\der y_{y=2} = 1.41\dots 1.46$, thus the prediction for the rapidity dependence is more robust. The absolute normalization is also more strongly model dependent that the shape of the distribution. It is also important to note that the two different wave functions used here have largest difference at $Q^2=0$. 

There are many calculations of the coherent vector meson production cross sections in the market (for references and comparison to the ALICE data see \re\cite{Abbas:2013oua}), but the situation is completely different for incoherent diffraction. In \fig\ref{fig:incoherent-alice} we show the ALICE result for the incoherent diffractive $J/\Psi$ photoproduction cross section in ultraperipheral collisions. The dipole model predictions from \paper\cite{Lappi:2013am} shown in the figure (``LM-fIPsat'') is obtained by using the IPsat model and Gaus-LC wave function. If the Boosted Gaussian wave function is used, the normalization changes and a good agreement with the ALICE data is obtained, but simultaneously the coherent cross section is more overestimated.

PHENIX collaboration at RHIC has also measured coherent diffractive $J/\Psi$ production cross section at midrapidity obtaining $76 \pm 34 \mu$b~\cite{Afanasiev:2009hy}. Using the same parametrizations as when calculating the incoherent cross section in \fig\ref{fig:incoherent-alice} we obtain $109\mu$b, being consistent with the data given the relatively large experimental uncertainty.

As shown above, the absolute normalization depends strongly on the details of the model parametrizations, but the rapidity dependence is a more solid CGC prediction. Simultaneous comparison of the model calculations to experimentally measured coherent and incoherent vector meson production cross sections has a clear potential to constrain the dipole model uncertainties. In addition to $J/\Psi$, it is possible to also consider production of other mesons such as $\Psi(2S)$ that are experimentally measured by ALICE~\cite{Nystrand:2014vra} and calculated using our framework~\cite{Lappi:2014eia}.

\chapter{Single inclusive particle production}
\label{ch:sinc}

\section{Particle spectra}
One of the simplest collider experiments are measurements of the single particle spectra. The total multiplicities in proton-proton and proton-nucleus collisions are dominated by low-$\pt\sim \lqcd$ particles that originate from partons with low transverse momentum $\pt$ whose formation can not be described using perturbative techniques. When the particle production is measured differentially in $\pt$, we can limit ourselves into the kinematical region where perturbative techniques can be applied.

In collider experiments hadrons, not individual quarks or gluons, are collided. Perturbative calculations, on the other hand, can be used to describe only the partonic scattering processes. The partonic density of the proton is given in terms of the parton distribution function (PDF) $xf_i(x,Q^2)$, which gives the probability to pick a parton $i$ from the proton with momentum fraction $x$ at scale $Q^2$. The scale evolution of $f$ is given in terms of the DGLAP equations, and the initial condition is obtained from fits to the HERA deep inelastic scattering data.
In this thesis the CTEQ parton distribution function set~\cite{Pumplin:2002vw} is used.

The partons are not measured in detectors, as the produced parton hadronizes to a cascade of color neutral particles. This transition is described by a fragmentation function $D_{k\to h}(z,Q^2)$, which gives the probability to produce a hadron $h$ which carries the fraction $z$ of the parent parton $k$ momentum, at scale $Q^2$. The scale evolution of $D_{k\to h}$ is again given by the DGLAP equations, but the initial condition must be fit to experimental data (most importantly from electron-positron annihilations) to extract the non-perturbative information. In this work the DSS~\cite{deFlorian:2007aj} fragmentation function set is used\footnote{Recently it has been pointed out in \re\cite{d'Enterria:2013vba} that the gluon-to-hadron fragmentation is too hard in many (NLO) fragmentation functions used in current phenomenological applications.}. The fact that the cross sections in perturbative QCD can be calculated as a convolution of universal parton distribution and fragmentation functions (that include the non-perturbative physics) and the perturbatively calculable parton level cross section is known as the collinear factorization theorem~\cite{Collins:1989gx,Brock:1993sz}.

The kinematics of single inclusive particle production is usually written in terms of the transverse momentum $\pt$ and rapidity $y$ of the produced hadron. Denoting the center-of-mass energy by $\sqrt{s}$, we can define the kinematic variables
\begin{equation}
	x_{h_1} = \frac{\frac{1}{z}p_T}{\sqrt{s}}e^y, \quad x_{h_2} = \frac{\frac{1}{z}p_T}{\sqrt{s}}e^{-y},
\end{equation}
where $x_{h_1}$ is the fraction of the longitudinal momentum of hadron 1 carried by the parton taking part in the scattering process originating from hadron 1, and $x_{h_2}$ has the same interpretation for the second hadron. When the hadron is produced at central rapidity ($y=0$), the situation is symmetric with respect to the collided hadrons. When the hadron is produced at forward or backward rapidities (at large $|y|$), we have a large-$x$ parton from one hadron probing the small-$x$ structure of the second hadron. This discussion is valid in two-to-one kinematics where only one parton (which later fragments into hadrons) is produced. If higher order processes are considered, one has to take into account processes where extra partons are produced in the hard scattering process, and in this case $x_{h_1}$ and $x_{h_2}$ become the lower limits for the momentum fractions.

\section{Particle production cross section from the CGC}

The particle production cross section can be computed in two different kinematical limits in the CGC framework. First, we can study the scattering of a dilute probe off a dense target and derive the so called hybrid formalism. This is applicable in hadron production at forward rapidities, where a large-$x$ parton scatters off a dense target, especially in proton-nucleus collisions where ``forward'' is defined as the proton fragmentation region. The hybrid formalism is discussed in more detail in \se \ref{sec:hybrid}

The second kinematical window where we can study the particle production is the scattering of two dense objects, such that small-$x$ structure of both colliding hadrons is probed. Experimentally this can be realized by measuring particle production at central rapidities where the kinematics is symmetric. In this case the so called $k_T$-factorization can be used. We will return to this in \se \ref{sec:ktfact}.

\subsection{Hybrid formalism}
\label{sec:hybrid}

Let us present a simple derivation for the particle production cross section in the CGC framework first derived in \re\cite{Dumitru:2002qt}.

\begin{figure}[tb]
\begin{center}
\includegraphics[width=0.4\textwidth]{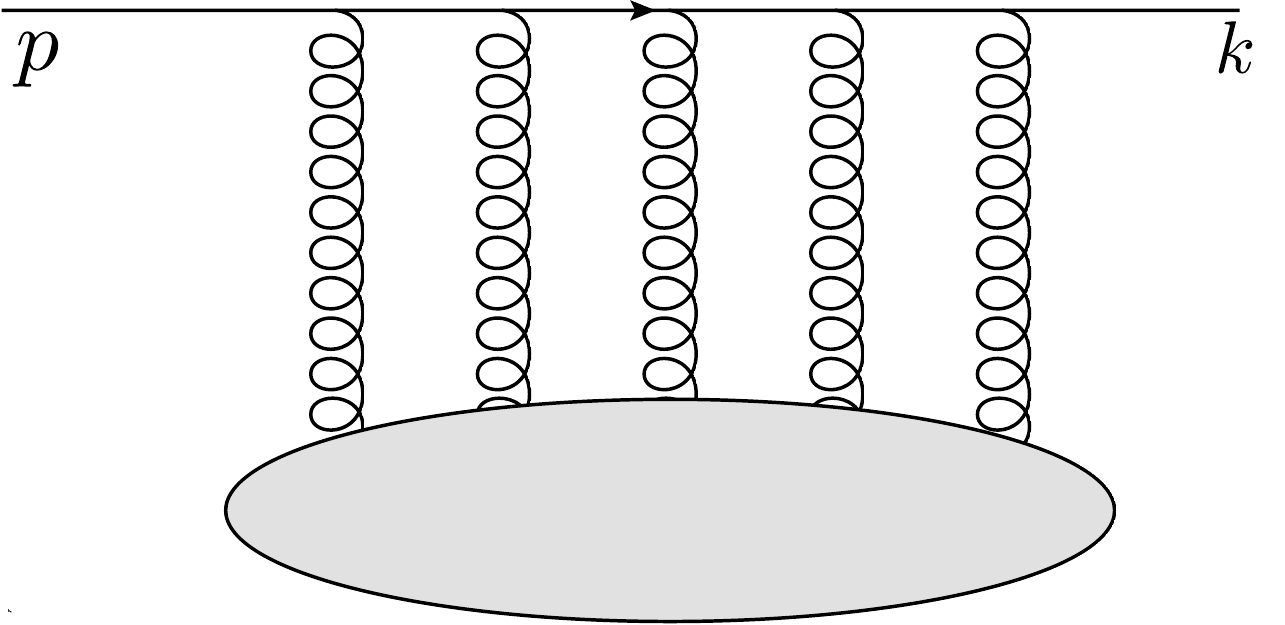}
\caption{High-energy quark scattering off the target color field.}
\label{fig:sinc_scattering}
\end{center}
\end{figure}

We consider a quark\footnote{For simplicity we only consider quark scattering here, the derivation for the incoming gluon is analogous.} travelling along the positive light cone with momentum $p$. The quark scatters off the target color field, and the final state quark momentum is $k$. The situation is shown schematically in \fig\ref{fig:sinc_scattering}. 
As discussed in \se\ref{sec:lightcone} the incoming  quark state can be written as
\begin{equation}
	|\text{in}\rangle = \N |p,i,\alpha\rangle = \N \int \frac{\der^2 \xt}{(2\pi)^2} e^{i \pt \cdot \xt} |p^+,\xt,i,\alpha\rangle . %\otimes |T \rangle,
\end{equation}
where $\N$ is the normalization factor given in \eq\eqref{eq:state-norm}.
%where the target is described by classical color fields $\A$ as
%\begin{equation}
%	|T\rangle = \int D \A \Phi_{x_A}(\A) |\A\rangle.
%\end{equation}
%Here the CGC wave function $\Phi_{x_A}(\A)$ describes the probability distribution of the target color fields at momentum fraction $x_A$. It is normalized such that $\int D\A |\Phi_{x_A}(\A)|^2=1$.

When the quark propagates through the target it acquires a phase given by the Wilson line, see discussion in \se \ref{sec:cgc-intro}. The outgoing state reads
\begin{equation}
	|\text{out}\rangle = \N \int \frac{\der^2 \xt}{(2\pi)^2} e^{i \pt \cdot \xt} V_{ij}(\xt) |p^+, \xt, j, \alpha\rangle  %\otimes |T\rangle.
\end{equation}
The fact that the Wilson line $V$ also depends on the color filed $\A$ of the target is left implicit. 

%To get the cross section we project out from the outgoing state the quark state with momentum $k$, spin $\beta$ and color $k$ and obtain the scattering amplitude
%\begin{multline}
%	M = \N \langle k, j, \beta | \text{out}\rangle \\
%	 =  \N^2 \int \frac{\der^2 \xt}{(2\pi)^2} e^{i \pt\cdot\xt} V_{ij}(\xt) \int \frac{\der^2 \yt}{(2\pi)^2} e^{-i \kt \cdot \yt} \langle 0 | b_{k,\beta}(k^+,\yt) b^\dagger_{i,\alpha}(p^+,\xt) |0\rangle \\
%	 =  \N^2 \delta(k^+-p^+) \int \frac{\der^2 \xt}{(2\pi)^2} e^{i (\pt - \kt)\cdot \xt}  V_{kj} (\xt) \delta_{\alpha\beta}.
%\end{multline}

To obtain the cross section we calculate the number of quarks in the outgoing state by evaluating the expectation value of the quark number operator $N_q(k) = b^\dagger_{j\beta}(k) b_{j\beta}(k)$, which measures the amount of quarks with momenta $k$, spin $\beta$ and color $j$. Summation over repeated indices is left implicit. We get
\begin{multline}
	\der N_\text{out}=\langle \text{out} | N_q(k) | \text{out} \rangle = \N^2 \int \frac{\der^2 \xt}{(2\pi)^2} \frac{\der^2 \xt'}{(2\pi)^2} e^{i \pt \cdot (\xt-\xt')}\\
	\times  \left \langle V_{j'i}^\dagger(\xt') V_{ij}(\xt) \right \rangle  
	 \langle p^+, x',j',\alpha | b_{l,\beta}^\dagger(k) b_{l,\beta}(k) |p^+,\xt,j,\alpha\rangle ,
	\end{multline}
where we have taken the average over the possible target color field configurations, denoted by $\langle \rangle$.
	
The number of quarks can be computed by evaluating
\begin{multline}
\label{eq:sinc-amplitude}
	b_{l,\beta}(k)|p^+,\xt, j, \alpha\rangle = \int \frac{\der^2 \yt}{(2\pi)^2} e^{i \yt \cdot \kt} b_{l,\beta}(k^+,\yt) |p^+,\xt,j,\alpha \rangle  \\
	= \int \frac{\der^2 \yt}{(2\pi)^2} e^{i \yt \cdot \kt} b_{l,\beta}(k^+,\yt)   b^\dagger_{j,\alpha}(p^+,\xt) |0\rangle \\
	= -e^{-i \xt \cdot \kt} \delta_{l,j} \delta_{\beta,\alpha} \delta(p^+-k^+),
\end{multline}
where the last equality was obtained by applying the anticommutator rule \eqref{eq:fermion_anticommutator_coordinate}. This result allows us to write
\begin{multline}
	\der N_\text{out}= \N^2 \int \frac{\der^2 \xt}{(2\pi)^2} \frac{\der^2 \xt'}{(2\pi)^2} e^{i \pt \cdot (\xt-\xt')} \left\langle V_{j'i}^\dagger(\xt') V_{ij}(\xt) \right\rangle \\
	\times e^{-i \kt \cdot(\xt - \xt')} [\delta(p^+-k^+)]^2 \delta_{l,j'} \delta_{\alpha \beta} \delta_{l,j} \delta_{\alpha,\beta}  
\end{multline}
Recall that $\pt$ is the transverse momentum of the quark before scattering, and $\kt$ is the transverse momentum of the scattered quark. The transverse momentum of the incoming quark can be taken to be very small compared to the transverse momentum transfer in the process, and we can set $\pt=0$.

The cross section can be computed from the rate of scattering events (number of quarks in the final state) normalized by the flux of incoming particles ($N_\text{in}=1$ quark in timestep $\der t$, transverse area $S_T$) and by the amount of target particles (here $1$):
\begin{equation}
\label{eq:sinc-xs}
	\der \sigma^{qA \to qA} = \frac{1}{2\nc} \sum_{i \alpha} \frac{\der N_\text{out}}{\der t \frac{N_\text{in}}{S_T \der t} },
\end{equation}
where we have calculated average over the initial state color and spin, and $A$ refers to a proton or a nucleus. The summation over the repeated indices when calculating $\der N_\text{out}$ takes into account the possible spins and colors of the final state quark. Calculating the sums we get
\begin{multline}
	\frac{\der \sigma^{qA \to q+X}}{\der^3 k} =  S_T \N^2 [\delta(p^+-k^+)]^2 \int \frac{\der^2 \xt}{(2\pi)^2} \frac{\der^2 \xt'}{(2\pi)^2} e^{-i \kt \cdot (\xt-\xt')} \\
	\times  \frac{1}{\nc} \left\langle \tr V^\dagger(\xt')V(\xt) \right\rangle .
\end{multline}

To evaluate this  we change integration variables to $\rt=\xt-\xt'$ and $\bt=\frac{1}{2}(\xt+\xt')$. The Wilson line correlator $\langle \tr V^\dagger V\rangle$ now in principle depends on three quantities (as one angle can be integrated over due to rational symmetry): $|\rt|, |\bt|$ and $\rt \cdot \bt$. Next, we neglect the $\rt \cdot \bt$ dependence as it should not contribute after we integrate over the impact parameter $\bt$. 
%Finally, we assume that the $\bt$ dependence can be factorized out from the Wilson line correlator, and we can write $\int \der^2 \bt = S_T$. 
This allows us to write
\begin{equation}
	\frac{\der \sigma^{qA \to q+X}}{\der^3 k} = \N^2 \frac{S_T \delta(p^+-k^+)\delta(p^+=0)}{(2\pi)^4}  \int \der^2 \bt \tilde S(\kt, \bt).
\end{equation}
The Fourier transform of the dipole scattering matrix is defined as
\begin{equation}
	\label{eq:hybrid-ft}
	\tilde S(\kt, \bt) = \int \der^2 \rt e^{-i \kt \cdot \rt} \frac{1}{\nc} \left\langle \tr V(\rt)V^\dagger(0) \right\rangle.
\end{equation}
Here the impact parameter dependence is hidden in the target average, as the color field density (or saturation scale) can be different at different impact parameters.
Finally we note that the second delta function gives the size of the $x^-$ box, as $\delta(p^+=0) = \int \frac{\der x^-}{2\pi} e^{i x^- \cdot 0} = \frac{L^-}{2\pi}$. Substituting the normalization factor from \eq\eqref{eq:state-norm} we finally get
\begin{equation}
\label{eq:hybrid-xs}
 	\frac{\der \sigma^{qA \to q+X}}{\der^3 k} = \frac{1}{(2\pi)^2}  \int \der^2 \bt \tilde S(\kt, \bt).
 \end{equation} 
% The cross section differentially in $\bt$ is the invariant quark production yield
%\begin{equation}
%	\frac{\der N^{qA\to q+X}(\bt)}{\der^3 k} =  \frac{\der \sigma^{qA\to q+X}}{\der^3 k \der^2 \bt} = \frac{1}{(2\pi)^2} \tilde S(\kt, \bt).
%\end{equation}

Let us finally obtain the single particle production cross section for proton-proton collisions. As discussed in \se\ref{sec:fits}, we assume that the impact parameter profile of the scattering matrix $\tilde S$ can be factorized, and replace $\int \der^2 \bt$ by $\sigma_0/2$ which is the proton transverse area measured in deep inelastic scattering. The produced particle is assumed to have a large (positive or negative) rapidity, and  thus
one of the incoming protons can be considered to be dilute such that the collinear factorization is applicable. As shown in  
\re\cite{Dumitru:2005gt}, the DGLAP-evolution of the parton distribution function $f_i$ can be included in the above calculation.  When calculating the cross section we must sum over different parton species. The parton yield is then convoluted with the fragmentation function $D_{i\to h}(z,Q^2)$  which describes the formation of hadron $h$ carrying momentum fraction $z$ of the momentum of the parton $i$. The cross section then becomes
\begin{equation}
	\label{eq:hybrid-formalism}
	\frac{\der \sigma^{pp\to h+X}}{\der^2 \kt \der y} = \sum_i  \frac{\sigma_0/2}{(2\pi)^2} \int \frac{\der z}{z^2} x_1 f_i(x,Q^2) \tilde S_i\left( \frac{\pt}{z}, x_2 \right) D_{i\to h}(z,Q^2).
\end{equation}
Finally the invariant yield is obtained from the cross section as
\begin{equation}
	\frac{\der N^{pp\to h+X}}{\der^2 \kt \der y} = \frac{1}{\sigma_\text{inel}} \frac{\der\sigma^{pp\to h+X}}{\der^2 \kt \der y}
	%  \sum_i  \frac{\sigma_0/2}{(2\pi)^2} \int \frac{\der z}{z^2} x_1 f_i(x,Q^2) S_i\left( \frac{\pt}{z}, x_2 \right) D_{i\to h}(z,Q^2),
\end{equation}
where $\sigma_\text{inel}$ is the total inelastic proton-proton cross section.
Here $x_2$ refers to the Bjorken-$x$ at which the target average is performed in \eq\eqref{eq:hybrid-ft}.
The dipole amplitude is evaluated in adjoint representation when the gluon channel contribution is computed. Note that the factor $\frac{\sigma_0/2}{\sigma_\text{inel}} \sim 0.2\dots 0.3$ arises from the fact that the
proton area measured in DIS, $\sigma_0/2$ is different than the inelastic proton-proton cross section which relates the invariant yield and the cross section to each other.
%dipole-target cross section is proportional to the proton area measured in DIS, $\sigma_0/2$ (see discussion in \se \ref{sec:fits}) and the invariant yield and the cross section are related to each other by the total inelastic cross section $\sigma_\text{inel}$. 

When the different areas are properly included, the $K$ factor that is sometimes used to scale the LO single inclusive spectra to have the same normalization as the data really quantifies how much the leading order calculation is off from the data. 
The yield in proton-nucleus collisions is explicitly calculated at a  fixed impact parameter and no similar area factor appears.

The hybrid formalism has been successfully applied to phenomenology. Combined with the BK evolution, a good description of the RHIC data was obtained in \re \cite{Albacete:2010bs}. The description was improved in \paper \cite{Lappi:2013zma} by including a proper treatment of different geometric areas in the calculation (the proton area measured in DIS and the inelastic proton-proton cross section) to also obtain the LO CGC result for the absolute normalization of the spectra.  

This calculation is done at leading order. The next to leading order cross section in the hybrid formalism is calculated in \res\cite{Chirilli:2012jd,Altinoluk:2014eka}. In principle, combining the NLO cross section with the NLO BK evolved dipole amplitude (see \se\ref{sec:nlobk}) and NLO parton distribution and fragmentation function allows one to calculate the single inclusive cross sections consistently at NLO accuracy. However, in \re\cite{Stasto:2013cha} it was shown (without using the NLO evolution of the dipole) that the NLO cross section becomes negative. Currently it is not known how the NLO calculation should be improved in order to get physically meaningful results.

\subsection{Transverse momentum factorization}
\label{sec:ktfact}

Another approach to the gluon production cross section calculation is to use so called transverse momentum factorization ($\kt$-factorization), which was originally derived in \re\cite{Gribov:1984tu}, and later improved in \re\cite{Kovchegov:2001sc}. It can also be derived from the Classical Yang-Mills theory, as shown in \re\cite{Gyulassy:1997vt}, and in \re\cite{Blaizot:2010kh} it was demonstrated numerically that the $\kt$-factorization formula agrees with the gluon spectrum computed by solving the Classical Yang-Mills equations of motion. For analytical gluon production cross section results in this picture we refer the reader to \re\cite{Dumitru:2001ux}.

The $\kt$-factorization formula for the gluon production cross section is
\begin{equation}
	\frac{\der \sigma^{pp\to g+X}}{\der y \der^2 \kt \der^2 \bt} = \frac{2\as}{\cf\kt^2} \int \der^2 \qt \der^2 \st \frac{\varphi(\qt,\st)}{\qt^2} \frac{\varphi(\kt-\qt, \bt-\st)}{(\kt-\qt)^2},
\end{equation}
where 
\begin{equation}
 \int \der^2 \st \varphi(\kt,\st) = \frac{\cf \sigma_0/2}{8\pi^3 \as} \kt^4 \int \der^2 \rt e^{-i \kt \cdot \rt} S(\rt)
 \end{equation}  
is proportional to the dipole unintegrated gluon distribution: $\varphi(\kt) \sim \kt^2 \varphi^\text{dipole}(\kt)$,
see \eq\eqref{eq:dipole-ugd}. The dipole amplitude is evaluated in the adjoint representation, which is obtained as $N_\text{adjoint}(\rt) = 2N(\rt)-N(\rt)^2$. To get the invariant yield, the cross section is calculated by integrating over $\bt$ and the result is divided by the total inelastic cross section. The result is
\begin{equation}
\label{eq:sinc-yield}
	\frac{\der N^{pp\to g+X}}{\der y \der^2 \kt } = \frac{(\sigma_0/2)^2}{\sigma_\text{inel}} \frac{\cf}{8\pi^4 \kt^2 \as} \int \frac{\der^2 \qt}{(2\pi)^2} \qt^2 \tilde S(\qt) (\kt-\qt)^2 \tilde S(\kt-\qt),
\end{equation}
where $\tilde S$ is the two dimensional Fourier transform of the dipole amplitude $1-N_A$ in the adjoint representation. For the proton DIS area we use the value $\sigma_0/2$ obtained by fitting the HERA DIS data, see \ch\ref{ch:dis}. To get the hadron spectra from the parton level yield \eq\eqref{eq:sinc-yield} must again be convoluted with a fragmentation function.

Assuming then that the transverse momentum of the produced gluon, $|\kt|$, is much larger than the saturation scale the integral in \eq\eqref{eq:sinc-yield} factorizes and the hybrid formalism result \eqref{eq:hybrid-formalism} is obtained:
\begin{equation}
	\frac{\der N^{pp\to g+X}}{\der y \der^2 \kt } = \frac{\sigma_0/2}{\sigma_\text{inel}} \frac{1}{(2\pi)^2} xg(x,\kt^2) \tilde S(\kt).
\end{equation}
Here the gluon distribution function is
\begin{equation}
	xg(x,\kt^2) = \int_0^{\kt^2} \frac{\der \qt^2}{\qt^2} \varphi(\qt).
\end{equation}
%In numerical calculations this can be replaced by the conventional parton distribution function, in this work by the CTEQ leading order PDF~\cite{Pumplin:2002vw}. To get the hadron spectra from the parton level yield is be convoluted with the  DSS leading order fragmentation function~\cite{deFlorian:2007aj} set. 
%This agrees with the hybrid formalism result obtained in \se\ref{sec:hybrid}.
%, with the numerical factor $(\sigma_0/2)/\sigma_\text{inel}\sim 0.2\dots 0.3$ resulting from the fact that the proton area measured in DIS and the total inealstic proton-proton cross section are different. 

\section{Proton-nucleus collisions}
\label{sec:pa}
To describe proton-nucleus collisions the dipole-nucleus scattering amplitude, preferably as a function of the impact parameter, is needed. Here we use the dipole amplitude obtained from the fits to HERA electron-proton data and generalized to nuclei using standard nuclear geometry as discussed in \se\ref{sec:dipole-amplitude-nuke}. For completeness we write again the dipole-nucleus amplitude at $x=x_0$ (which is the initial condition for the BK evolution of the nucleus)
\begin{equation}
	N^A(\rt,\bt) = 1 - \exp \left[ -A T_A(\bt) \frac{\sigma_0}{2} \frac{(\rt^2 \qso^2)^\gamma}{4} \ln \left(\frac{1}{|\rt|\lqcd}+e_c\cdot e\right)\right].
\end{equation}

When the single inclusive yield at fixed impact parameter is calculated using the hybrid formalism, one does not get the different area factors in \eq\eqref{eq:hybrid-formalism}, and the single inclusive yield is
\begin{equation}
	\label{eq:sinc-yield-nuke}
	\frac{\der N(\bt)}{\der y \der^2 \kt} = \frac{1}{(2\pi)^2} xg(x,\kt^2) \tilde S^A(\kt, \bt),
\end{equation}
where $\tilde S^A$ is again the Fourier transform of the dipole-nucleus scattering matrix in adjoint representation. Similarly when using $\kt$-factorization only one proton area is obtained from the unintegrated gluon distribution of the proton, and \eq\eqref{eq:sinc-yield} becomes
\begin{equation}
\label{eq:ktfact-nuke}
	\frac{\der N^{pA\to g+X}}{\der y \der^2 \kt } =  \frac{\cf \sigma_0/2}{8\pi^4 \kt^2 \as} \int \frac{\der^2 \qt}{(2\pi)^2} \qt^2 \tilde S(\qt) (\kt-\qt)^2 \tilde S(\kt-\qt).
\end{equation}
Physically the reason for having different prefactors in proton-proton and proton-nucleus results is due to the consistent usage and separation of the two different proton areas: the small-$x$ gluonic area $\sigma_0/2$ determined in DIS and the ``soft'' area of the proton given by the inelastic proton-proton cross section $\sigma_\text{inel}$, which gets a large contribution from non-perturbative physics.

The nuclear effects in proton-nucleus collisions are quantified in terms of the nuclear suppression factor $R_{pA}$, defined as the ratio of particle production yields in pA and pp collisions normalized by the number of binary nucleon-nucleon collisions $N_\text{bin}$:
\begin{equation}
\label{eq:rpa}
	R_{pA} = \frac{\der N^{pA}}{N_\text{bin} \der N^{pp}} .
\end{equation}
The number of binary collisions can be obtained from the nuclear density distribution $T_A$ using the Optical Glauber model, see \re\cite{Miller:2007ri}.
The nuclear suppression factor can be interpreted such that if the nuclear effects are absent, and the pA collision is the same as a proton colliding with individual nucleons, then $R_{pA}=1$. Having $R_{pA}<1$ can be seen as a signal of saturation effects in the nucleus.

In the high-$\pt$ limit the nucleus is probed at larger $x$ and at smaller scale where the nuclear effects are expected to vanish (except perhaps a small difference originating from the isospin symmetry when computing charged particle spectra). Let us show that in our framework we naturally obtain a value for $R_{pA}$ which goes to unity at high transverse momenta. First, we notice that in this limit the dipole amplitude is evaluated close to the initial condition and the single inclusive yield \eqref{eq:sinc-yield-nuke} becomes
\begin{equation}
\label{eq:sinc-pa-expanded}
	\der N^{pA} \sim xg \int \der^2 \rt e^{i\kt \cdot \rt} \exp\left( -\frac{AT_A}{2} \sigma_0 N^p \right) \sim xg A T_A \frac{\sigma_0}{2} N^p.
\end{equation}
Similarly  the proton-proton yield behaves as
\begin{equation}
	\der N^{pp} \sim \frac{\sigma_0/2}{\sigma_\text{inel}} xg N^p.
\end{equation}
The number of binary collisions is $N_\text{bin}=AT_A \sigma_\text{inel}$, and thus we find that $R_\text{pA}\to 1$ independently of the center of mass energy $\sqrt{s}$. The fact that $R_{pA}\to 1$ thus originates from a proper treatment of the nuclear and nucleon geometry, and it is not a result of the high energy evolution. This can be seen from earlier CGC calculations~\cite{Albacete:2010bs,Albacete:2012xq} where a significantly larger nuclear suppression is predicted for pA collisions, and the main difference between this work and \res\cite{Albacete:2010bs} and \cite{Albacete:2012xq} is the different treatment of the nuclear geometry.

When calculating the minimum bias particle production yields in proton-nucleus collisions we integrate over all impact parameters. At large impact parameters the nucleus becomes very dilute, as shown in \se\ref{sec:dipole-amplitude-nuke}, and the BK evolution which is run separately for all impact parameters reduces to the BFKL equation and causes the nucleus to grow very rapidly in its edges. This unphysical behavior would be cured by an impact-parameter dependent BK equation which is not available, see discussion in \se\ref{sec:bk}.

In practical calculations we use the dipole-nucleus amplitude in the region where the saturation scale in the nucleus is larger than that of the proton. When the saturation scale falls below that at large $|\bt|$, we use the expanded result \eqref{eq:sinc-pa-expanded} and calculate the yield as
\begin{equation}
	\der N^{pA} = N_\text{bin} \der N^{pp},
\end{equation}
which is equivalent to requiring that $R_{pA}=1$ at large impact parameters.

The proton-nucleus collisions can not be divided experimentally into the centrality classes similarly as the heavy ion collisions due to smaller multiplicity and biases introduced by different centrality measures, see e.g. \re\cite{Adam:2014qja}. The optical Glauber model used in these calculations can be used to calculate centrality dependent quantities, but the results can not be compared with experimental measurements divided into ``centrality classes'' based on event multiplicity. However, the minimum bias results obtained by integrating over the impact parameters can be compared with measurements.

%\section{Constraining the initial condition for the BK evolution}
\section{Comparison with experimental data}

\begin{figure}[tb]
\centering
	\begin{minipage}[t]{0.5\textwidth}
	\includegraphics[width=\textwidth]{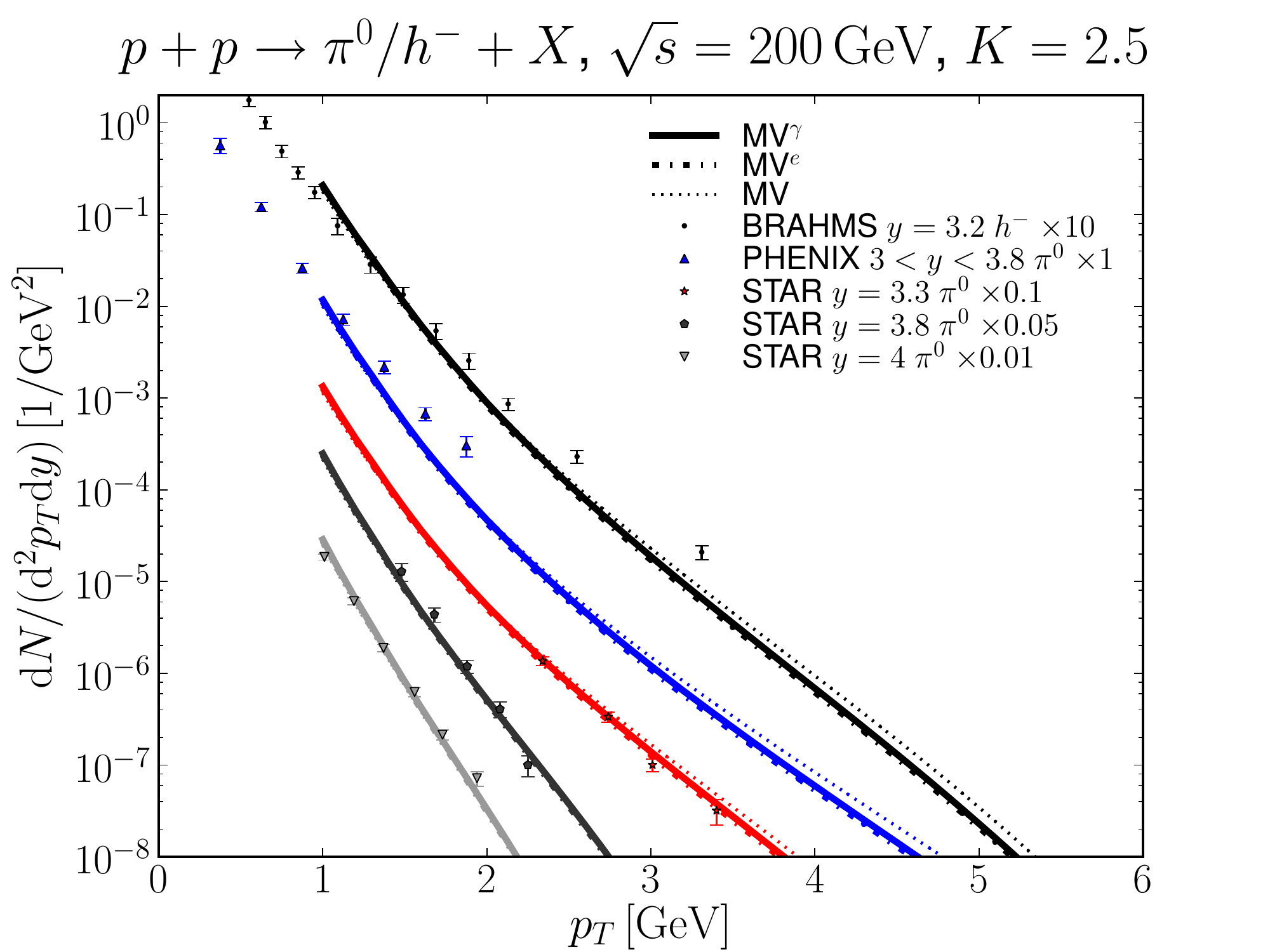}
	\caption{Invariant $\pi^0$ and negative hadron yields at forward rapidity as a function of $\pt$ compared with the RHIC data~\cite{Adams:2006uz,Adare:2011sc,Arsene:2004ux}. Figure from \paper\cite{Lappi:2013zma}.}
	\label{fig:rhic_pp_sinc_modeldep}
	\end{minipage}%
	~
	\begin{minipage}[t]{0.5\textwidth}
	\includegraphics[width=1.0\textwidth]{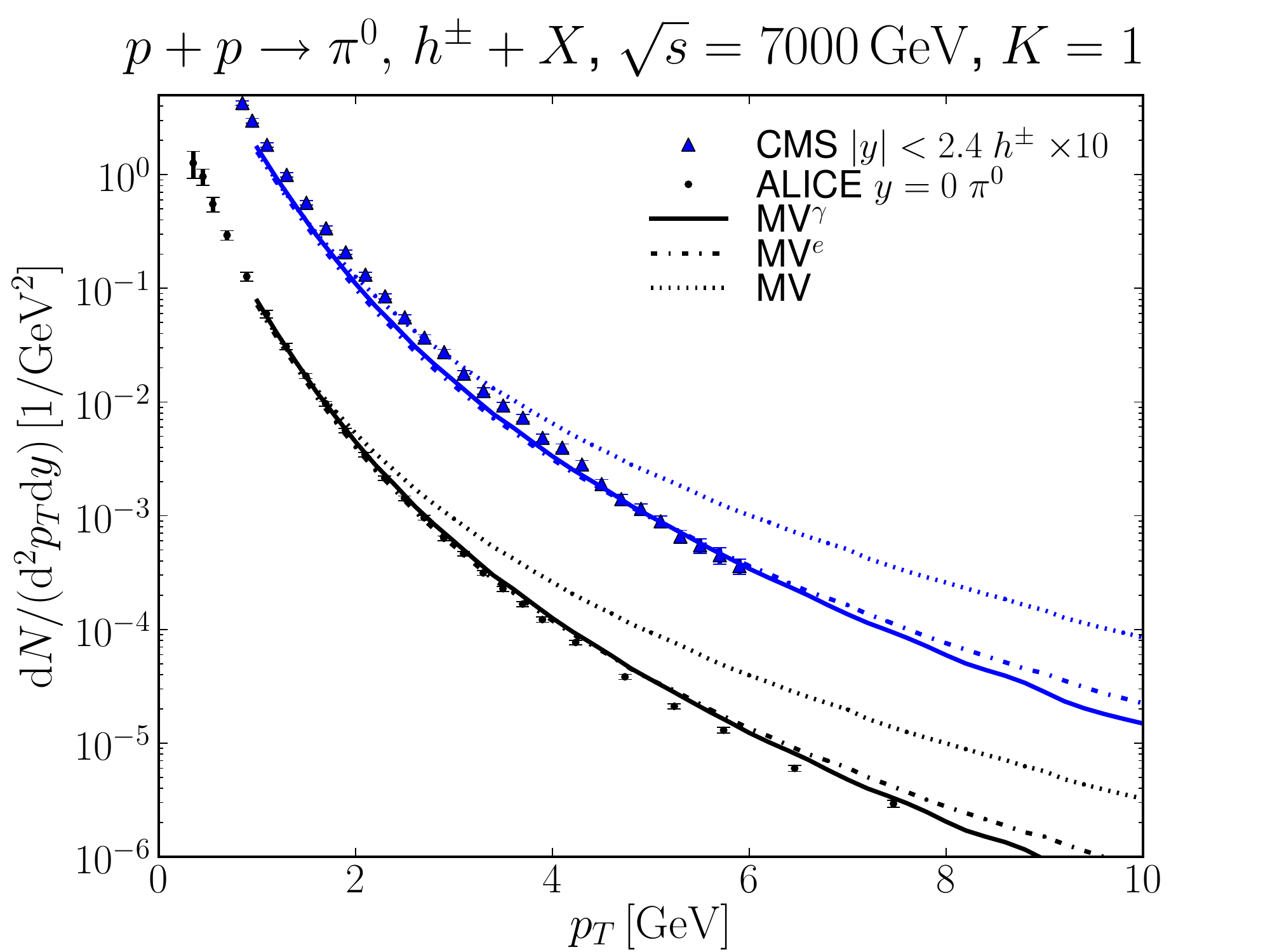}
	\caption{Invariant charged hadron yields at midrapidity as a function of $\pt$ compared with the LHC data~\cite{Abelev:2012cn,Khachatryan:2010us}. Figure from \paper\cite{Lappi:2013zma}.
	}
	\label{fig:lhc_pp_sinc_modeldep}
	\end{minipage}	
\end{figure}

We calculate in \paper\cite{Lappi:2013zma} single inclusive particle spectra in proton-proton collisions using the three different initial conditions for the BK evolution fitted to the HERA data in \ch\ref{ch:dis}. First, we calculate the neutral pion and charged hadron yields  using the hybrid formalism and compare with the RHIC data~\cite{Adams:2006uz,Adare:2011sc,Arsene:2004ux} at $\sqrt{s}=200\gev$. The results are shown in \fig\ref{fig:rhic_pp_sinc_modeldep}. The $\pt$ dependence of the data is quite well described by the calculation, but the absolute normalization must be modified by introducing a factor $K=2.5$ to normalize the calculation to the same level with the data. As the absolute normalization is now consistently calculated from the CGC, the $K$ factor quantifies how much the leading order calculation differs from the data.

A similar comparison with the LHC data~\cite{Abelev:2012cn,Khachatryan:2010us} is shown in \fig\ref{fig:lhc_pp_sinc_modeldep} where the spectra are calculated at midrapidity using $\kt$ factorization. The $\pt$ slope is well reproduced, and no $K$ factor is needed. The unmodified MV model initial condition, which gives a relatively good fit to HERA data and works with the RHIC measurements, is now clearly not favored by the LHC data. The MV$^\gamma$ and MV$^e$ parametrizations give in practice indistinguishable results for the single inclusive yield.

\begin{figure}[tb]
\centering
	\begin{minipage}[t]{0.5\textwidth}
	\includegraphics[width=\textwidth]{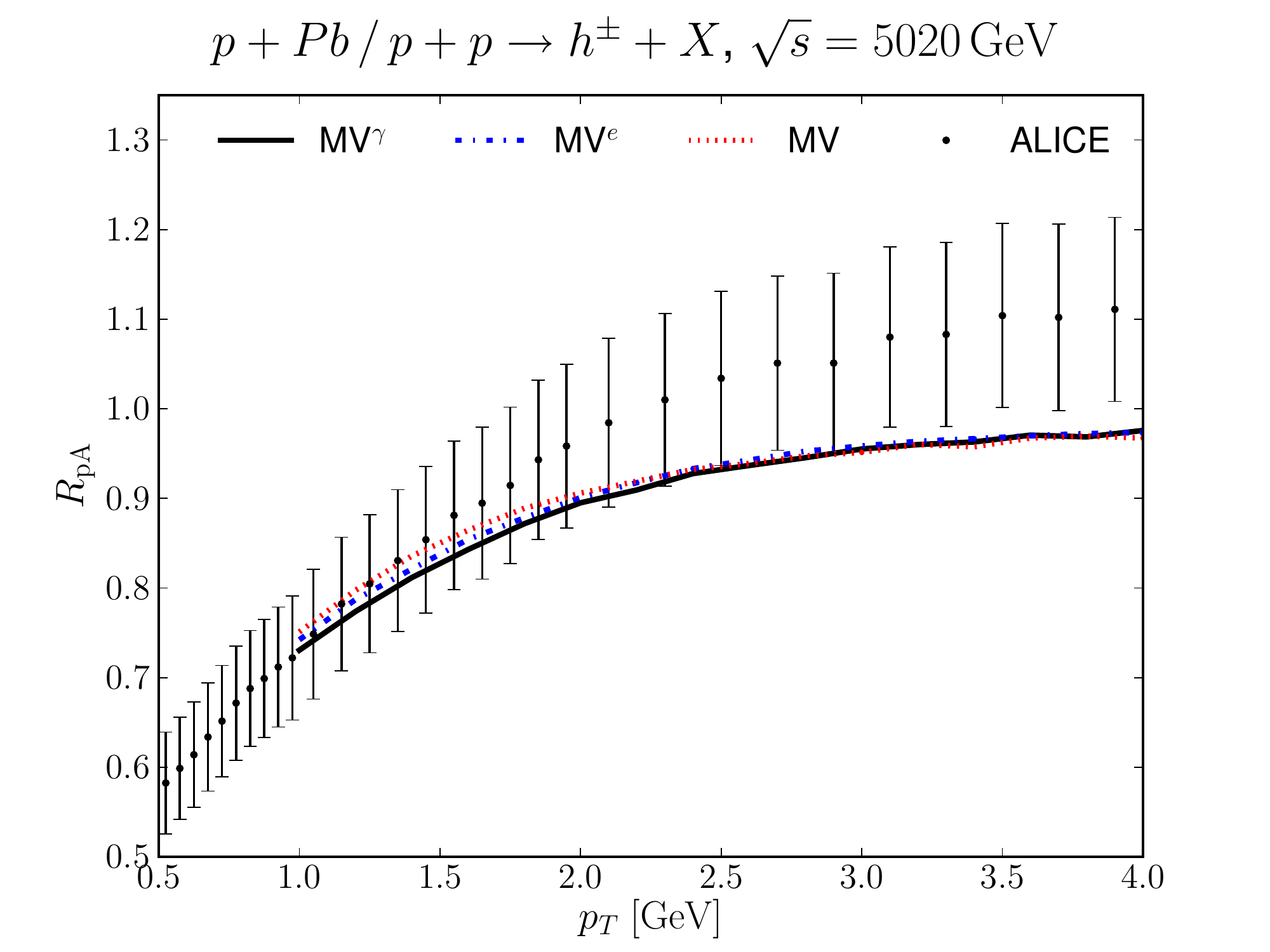}
	\caption{
	Minimum bias nuclear modification factor $R_{pA}(y=0)$ at different centrality classes computed using $k_T$ factorization and MV$^\gamma$, MV$^e$ and MV model initial conditions compared with the minimum bias ALICE data \cite{ALICE:2012mj} at smallest $p_T$ region. Figure from \paper \cite{Lappi:2013zma}.	
	}
	\label{fig:rpa_modeldep}
	\end{minipage}%
	~
	\begin{minipage}[t]{0.5\textwidth}
	\includegraphics[width=1.0\textwidth]{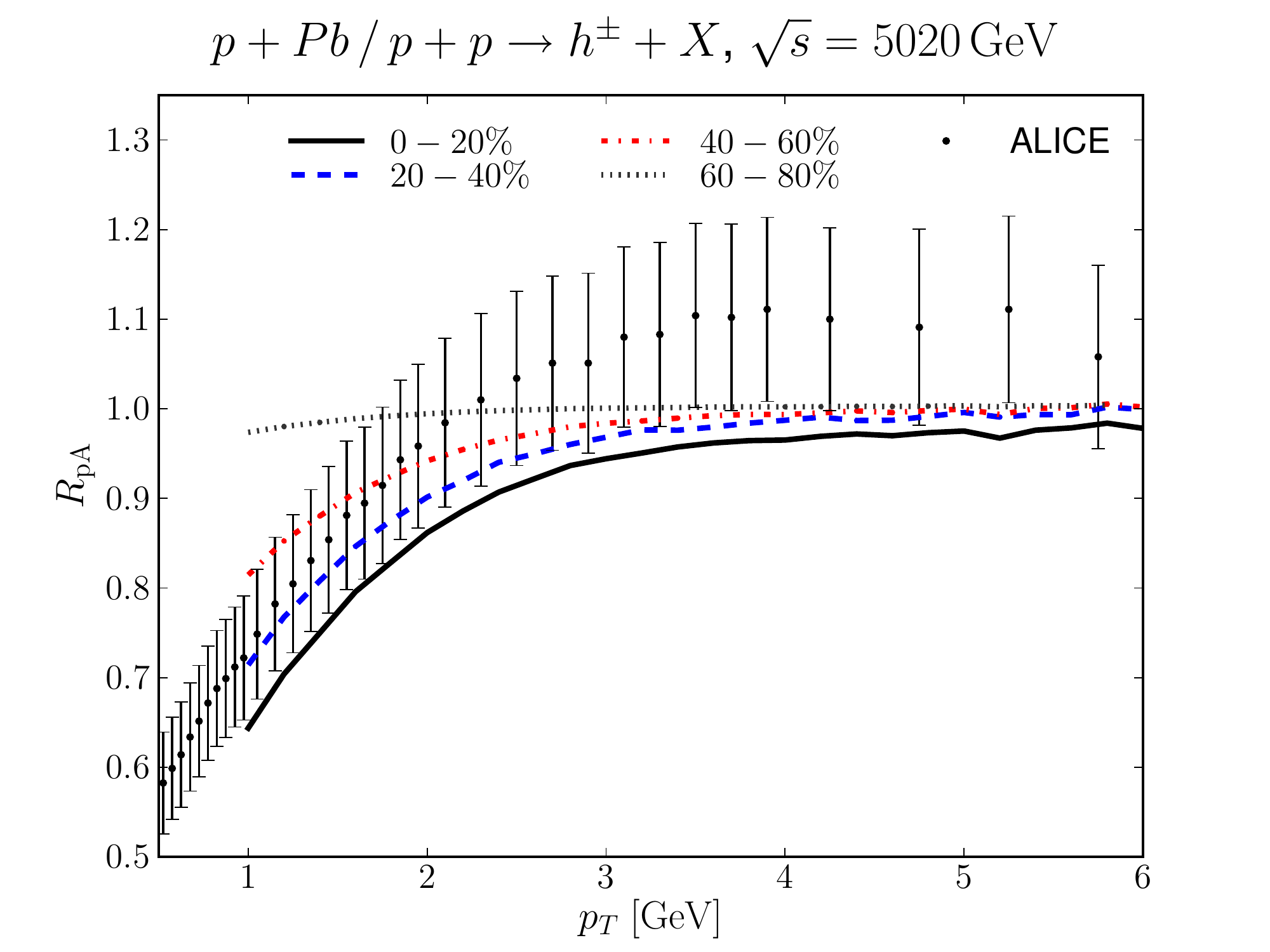}
	\caption{
	Centrality dependence of the midrapidity $R_{pA}$ computed using $k_T$ factorization and MV$^e$ initial condition compared with the ALICE data \cite{ALICE:2012mj}. Figure from \paper \cite{Lappi:2013zma}.	
	}
	\label{fig:rpa_centrality}
	\end{minipage}	
\end{figure}

\begin{figure}[tb]
\centering
	\begin{minipage}[t]{0.5\textwidth}
	\includegraphics[width=\textwidth]{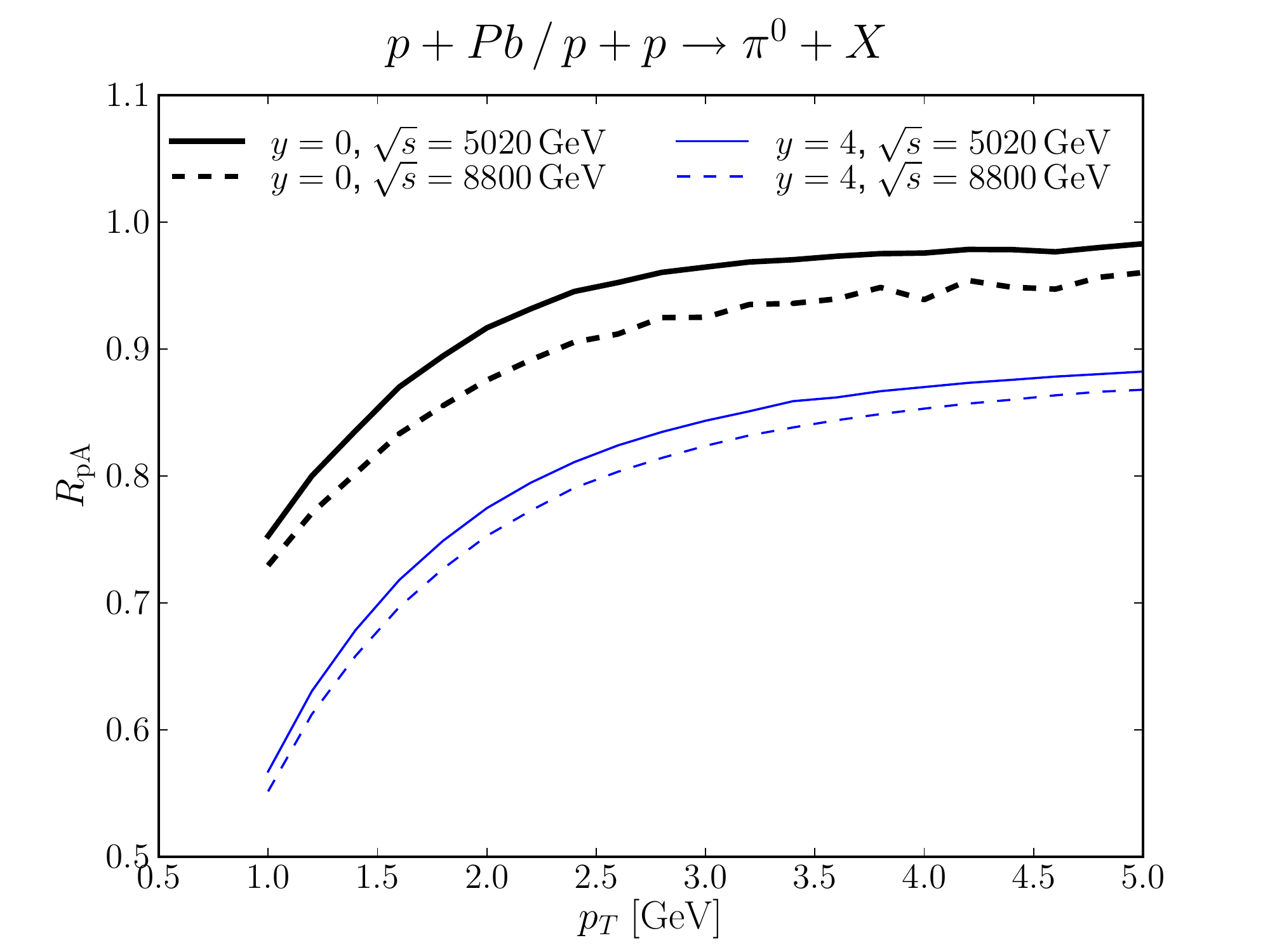}
	\caption{
	Center of mass energy ($\sqrt{s}$) dependence of the nuclear modification factor in neutral pion production in minimum-bias p+Pb collisions computed using the MV$^e$ initial condition. The results at midrapidity $y=0$ are computed using $k_T$-factorisation and at $y=4$ the hybrid formalism is used. Figure from \paper \cite{Lappi:2013zma}.	
	}
	\label{fig:rpa_sqrtsdep}
	\end{minipage}%
	~
	\begin{minipage}[t]{0.5\textwidth}
	\includegraphics[width=1.0\textwidth]{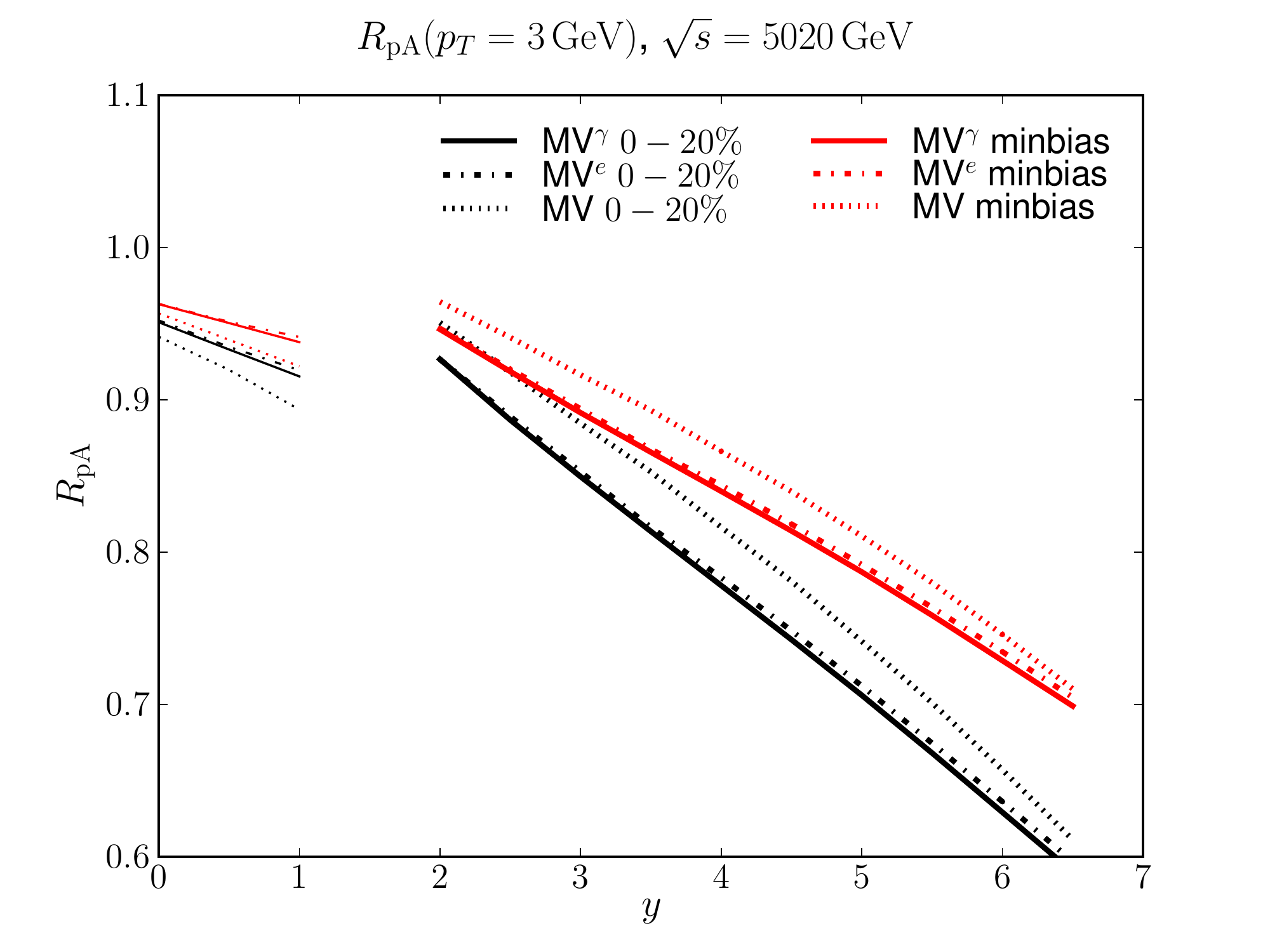}
	\caption{
	Rapidity and centrality dependence of the nuclear modification factor in neutral pion production in $0-20\%$ most central (solid lines) and minimum bias collisions using MV, MV$^\gamma$ and MV$^e$ initial conditions. Thin lines at $y\le1$ are computed using $k_T$-factorization and thick lines at $y\ge 2$ using the hybrid formalism. Figure from \paper \cite{Lappi:2013zma}.
	}
	\label{fig:rpa_ydep}
	\end{minipage}	
\end{figure}

Let us then study proton-nucleus collisions. First we compute, using  $\kt$-factorization, the nuclear suppression factor $R_{pA}$ in central proton-lead collisions at $\sqrt{s}=5020\gev$ and compared with the ALICE minimum bias data~\cite{ALICE:2012mj}. The results shown in \fig\ref{fig:rpa_modeldep} are consistent with the ALICE data especially at low $\pt$, and all the studied initial conditions for the dipole amplitude give the same result for $R_{pA}$ even though the results for the $\pt$ spectra were very different with the MV model initial condition. As the systematical uncertainty in the data is relatively large, it can not be said whether or not $R_{pA}$ approaches unity in the experimental data. 

The question whether $R_{pA}$ goes to unity at large transverse momenta is currently an open question. CMS has observed~\cite{Khachatryan:2015xaa} an enhancement of charged particle $R_{pA}$ at transverse momenta $\gtrsim 50 \gev$, where $R_{pA}\approx 1.4$, and the result is inconsistent with unity given the experimental uncertainties. This behavior is not seen in the ALICE data~\cite{Abelev:2014dsa} which extends up to $\pt\sim 50 \gev$.

Let us then precent our results for the centrality dependence of the nuclear suppression factor. First we emphasize again that the centrality classes defined using the optical Glauber model do not correspond to the experimental centrality classes in pA collisions, and the comparison with the data should be done with extreme care. The midrapidity $R_{pA}$ calculated using the MV$^e$ initial condition is shown in \fig\ref{fig:rpa_centrality}. We find a weak centrality dependence, and the results start to deviate only at the most peripheral centrality classes. Note that as discussed in \se\ref{sec:pa}, we in practice set $R_{pA}=1$ in the peripheral collisions when the nuclear saturation scale becomes smaller than that of the proton, which happens at centralities $\gtrsim 70\%$.

To study the center-of-mass energy $\sqrt{s}$ dependence of the nuclear modification factor we show in \fig\ref{fig:rpa_sqrtsdep} the midrapidity $R_{pA}$ calculated using $\kt$-factorization and $R_{pA}(y=4)$ calculated using the hybrid formalism at two different center of mass energies, corresponding the energies available at the LHC before and after the first long shutdown. We find a weak $\sqrt{s}$ dependence. Note that we get explicitly $R_{pA}\to 1$ at high $\pt$ at all center of mass energies. 

Finally we plot in \fig\ref{fig:rpa_ydep} the rapidity dependence of the nuclear suppression factor for neutral pions, namely $R_{pA}(\pt=3\gev)$. 
As $\pt$ is fixed and the Bjorken-$x$ probed in the scattering goes like $e^{-y}$, this quantity measures the $x$ evolution of the gluon density in the nucleus (where the saturation effects are expected to be larger) compared with the proton.
 The nuclear suppression factor close to midrapidity is computed using $\kt$-factorization, and at forward rapidities the hybrid formalism is used. The hybrid formalism calculation is done using the CTEQ parton distribution function instead of an unintegrated gluon distribution function (see discussion in \se\ref{sec:cgc-intro}), and the quark channel is included in the calculation, thus the curve is not exactly continuous. The evolution speed of $R_{pA}$ close to midrapidity is slower than at more forward rapidities. We also find that all dipole model parametrizations for the initial condition give roughly the same evolution speed, and it can be concluded that $R_{pA}$ is not sensitive to the details of the initial dipole amplitude.

 %What the evolution of $R_{pA}$ probes is the high-energy evolution of the gluon density. The evolution speed is faster than what is obtained from NLO pQCD calculations using the EPS09s nuclear parton distribution functions~\cite{Helenius:2012wd,helenius2013rapidity}.

We can conclude that $R_{pA}$ does not distinguish between the dipole model initial condition parametrizations, as it is mostly sensitive to the BK evolution of the dipole amplitude.
It would be interesting to compare the CGC calculations for $R_{pA}$ at different rapidities to experimental data and calculations done using different approaches, for example perturbative QCD with nuclear parton distribution functions~\cite{Eskola:2009uj,Helenius:2012wd}. Unfortunately, currently the only forward rapidity $R_{pA}$ data comes from RHIC where the lower center-of-mass energy limits the utility of the data for this purpose.

The $\pt$ spectra seem to be sensitive to the details of the initial condition by clearly disfavoring the MV model. Even though the MV model did not give equally good fit to the HERA data as the modified MV model parametrizations, comparison with the LHC spectra in proton-proton collisions gives an additional constraint which calls for a modification to the MV model initial condition. 

In addition to charged particles and pions studied here, it is also possible to study for example inclusive $J/\Psi$ production cross section. This is done by calculating the cross section for $c\bar c$ quark pair production~\cite{Blaizot:2004wv,Fujii:2013gxa}, and then modelling the $c\bar c \to J/\Psi$ transition using for example nonrelativistic QCD (as done e.g. in \re\cite{Ma:2014mri}) or the so called color evaporation model, where it is assumed that fixed fraction of the $c\bar c$ pairs produces the vector meson, and the additional color is ``evaporated'' away as a soft gluon. We have shown in \re\cite{Ducloue:2015gfa} that if the nuclear geometry is properly taken into accoung similarly as described in \se\ref{sec:pa}, the CGC calculation is in agreement with the LHC data in pp and pA collisions.

\chapter{Double inclusive scattering and correlations}
\label{ch:dihad}
\section{Azimuthal correlations as a signature of saturation}

So far in this thesis we have discussed how it is possible to describe single particle production in inclusive and diffractive deep inelastic scattering, ultraperipheral heavy ion collisions and hadronic collisions within the CGC framework. Going beyond the single inclusive spectrum can provide additional tests for the theory, as the multiparton correlations are expected to be more sensitive to the detailed dynamics of the colliding particles.

\begin{figure}[tb]
\begin{center}
\includegraphics[width=1.0\textwidth]{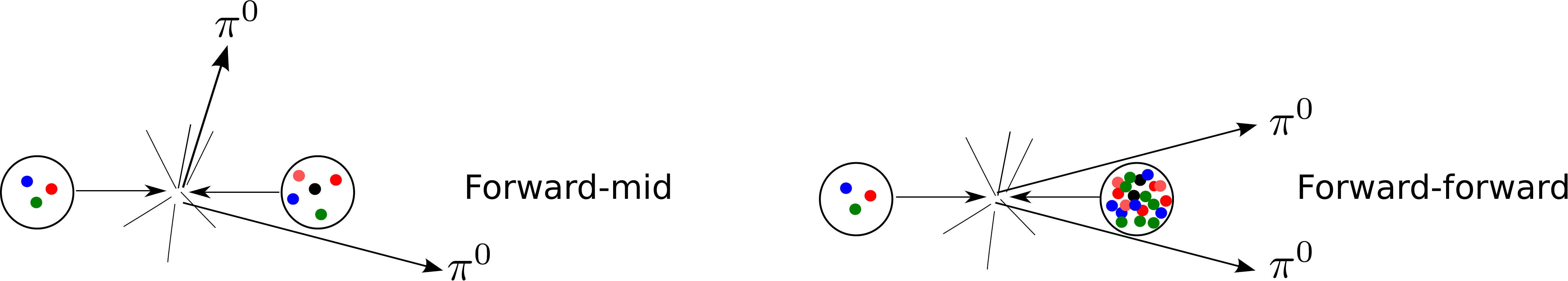}
\caption{Evolution of the parton densities from forward-midrapidity dihadron production to the case where both hadrons are produced at forward rapidity.}
\label{fig:dihad-kin}
\end{center}
\end{figure}

The azimuthal angle correlations between two pions in the forward rapidity have been measured by STAR~\cite{Braidot:2010ig,Braidot:2011zj} and PHENIX~\cite{Adare:2011sc} collaborations at RHIC. The striking result from these measurements is that in proton-proton collisions the produced hadrons are clearly back-to-back on the transverse plane, as expected from  naive momentum conservation, but this correlation vanishes when the same measurement is done for central deuteron-gold collisions.

In the kinematical region where one of the pions is produced at midrapidity and the second one at forward rapidity (``forward'' defined as the direction of the projectile), the back-to-back correlation does not disappear. To understand this difference, let us consider the two different kinematical windows shown schematically in \fig\ref{fig:dihad-kin}. If one of the produced pions is produced at midrapidity, a large amount of longitudinal momentum is needed from the target which means that the large-$x$ structure of the target is probed. On the other hand, if both pions are produced at forward rapidity only a tiny amount of longitudinal momentum can be taken from the target, and thus the significantly more dense small-$x$ part of the target wave function is probed.
More precisely, if two partons with transverse momenta $p_{T,i}$ are produced at rapidities $y_1$ and $y_2$, then the Bjorken-$x$ of the targed probed in the collision is
\begin{equation}
	x_A  = \frac{p_{T,1}e^{-y_1}}{\sqrt{s}} + \frac{p_{T,2}e^{-y_2}}{\sqrt{s}}.
\end{equation}

\begin{figure}[tb]
\begin{center}
\includegraphics[width=0.8\textwidth]{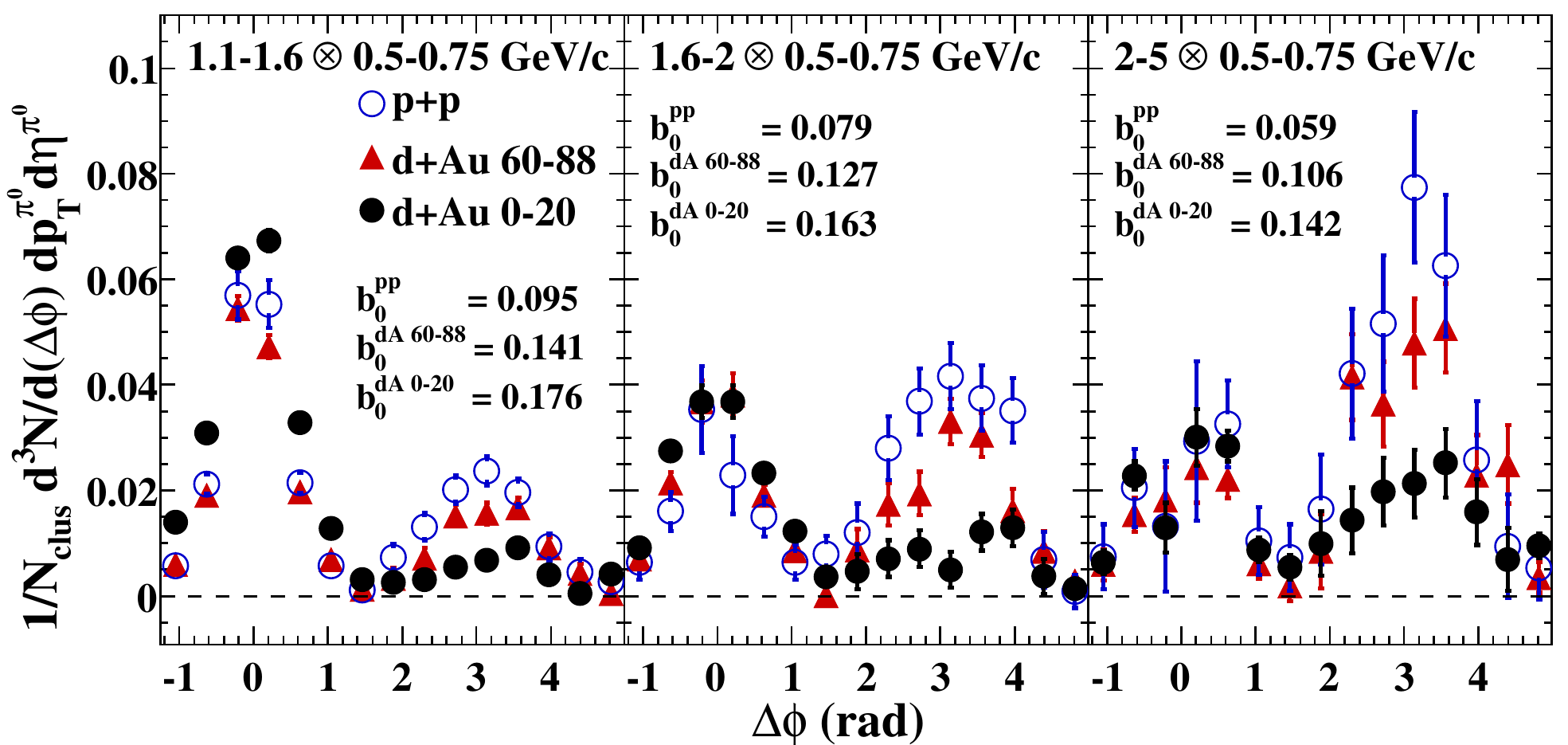}
\caption{Dihadron correlation as a function of azimuthal angle difference. The trigger and associate particle transverse momenta are shown at the top of the panels. Deuteron-gold results are shown for most central (0-20\%) and peripheral (60-80\%) collisions based on charge deposited in the gold-going direction. Figure reprinted with permission from \re\cite{Adare:2011sc}. Copyright (2011) by the American Physical Society.}
\label{fig:phenix-dihad}
\end{center}
\end{figure}

In the Color Glass Condensate the disappearance of the back-to-back correlation peak can be naturally explained. The incoming (valence) quark from the projectile emits a gluon either before or after it interacts with the target. After the emission the two partons would be back-to-back in the transverse plane, but when interacting with the target the partons obtain a transverse momentum kick of the order of the saturation scale $Q_s$. When the saturation scale becomes of the same order as the transverse momentum scale of the produced partons, the back-to-back correlation is washed out. As the saturation scale at the center of the nucleus is much larger than that of the proton (or in minimum bias collisions of the nucleus), it is natural to expect the correlation to vanish only in central deuteron-gold collisions. Similarly, in this picture we would expect the back-to-back peak to return when the transverse momentum of the hadrons is required to be large enough, which is exactly what can be seen from the RHIC data in \fig\ref{fig:phenix-dihad}.

\section{Two-particle production cross section}

\begin{figure}[tb]
\begin{center}
\includegraphics[width=0.8\textwidth]{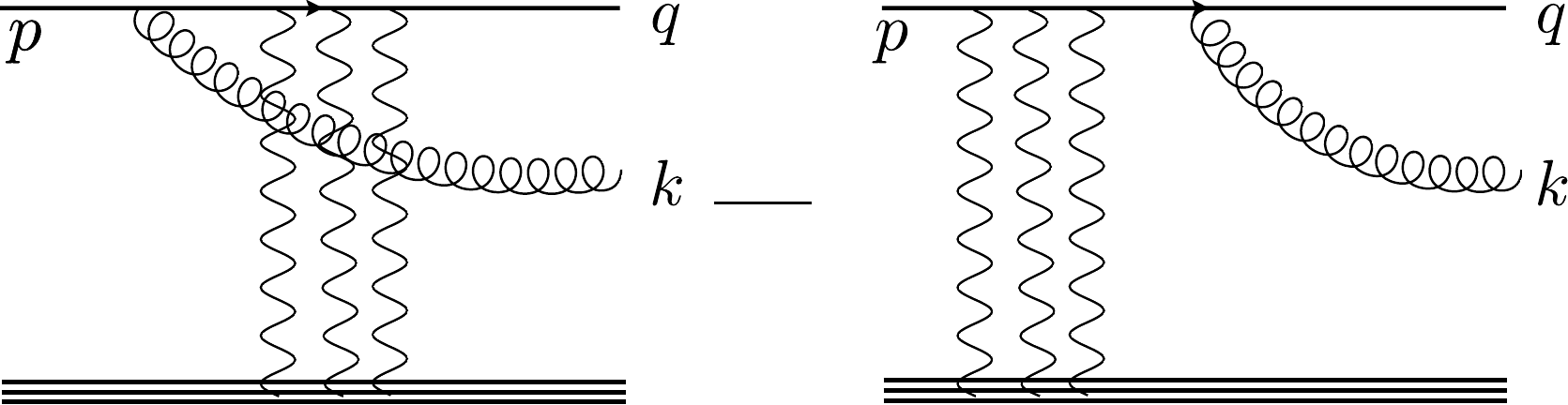}
\caption{Quark-gluon production cross section in the scattering of a quark off the target. The wavy lines represent multiple interactions with the target. }
\label{fig:dihad-vuorovaikutus}
\end{center}
\end{figure}

The cross section for the production of a quark with momentum $q$ and a gluon with momentum $k$, can be computed by calculating the gluon emission from a quark as discussed in \se\ref{sec:lightcone}, and calculating the quark-target and gluon-target scattering as in \se\ref{sec:cgc-intro}. The situation is shown schematically in \fig\ref{fig:dihad-vuorovaikutus}, where we note that the gluon can be emitted either before or after the interaction with the target. The cross section in coordinate space is derived in the CGC framework in \re\cite{Marquet:2007vb} (see also \re\cite{JalilianMarian:2004da} for an original CGC derivation in the momentum space): 
\begin{multline}
\label{eq:dihadron-xs}
\frac{\der \sigma^{qA\to qgX}}{\der k^+ \der^2 \kt \der q^+ \der^2 \qt } 
= \as \cf \delta(p^+-k^+-q^+) 
\int \frac{\der^2 \xt}{(2\pi)^2} \frac{\der^2 \xt'}{(2\pi)^2} 
\frac{\der^2 \bt}{(2\pi)^2} \frac{\der^2 \bt'}{(2\pi)^2} 
\\ \times  e^{i \kt \cdot(\xt' - \xt)} e^{i\qt \cdot(\bt'-\bt)} 
 \sum_{\alpha\beta \lambda} 
\phi_{\alpha\beta}^{\lambda*}(\xt'-\bt') \phi_{\alpha\beta}^\lambda(\xt- \bt) 
\{ S^{(4)}(\bt,\xt,\bt',\xt') \\
- S^{(3)}(\bt,\xt,\zt') 
- S^{(3)}(\zt,\xt',\bt') 
+ S(\zt,\zt') \}.
\end{multline}
Here $\zt = z\xt + (1-z)\bt$ and $\zt'=z\xt' + (1-z)\bt'$ and $z$ is the longitudinal momentum fraction of the original quark carried by the gluon, $z=k^+/p^+$. When calculating the hadron level cross section, the result must be convoluted with a parton distribution function and a fragmentation function. In this work we use the leading order CTEQ parton distribution function~\cite{Pumplin:2002vw} and DSS fragmentation function~\cite{deFlorian:2007aj} sets.

The Wilson line correlators in \eq\eqref{eq:dihadron-xs} are
\begin{align}
\label{eq:s4}
S^{(4)}(\bt,\bt',\xt,\xt') &= \frac{2}{\nc^2-1}\left \langle 
\tr \left(V(\bt) V^\dagger(\bt') t^d t^c \right) [U(\xt) U^\dagger(\xt')]^{cd}\right \rangle 
\\
\label{eq:s3}
S^{(3)}(\bt,\xt,\zt') &= \frac{2}{\nc^2 -1 } \left \langle
 \tr \left( V^\dagger(\zt') t^c V(\bt) t^d \right) U^{cd}(\xt) \right \rangle 
\\
\label{eq:s2}
	S(\zt,\zt') &= \frac{1}{\nc} 
\left \langle \tr \left( V(\zt)V^\dagger(\zt')\right)\right \rangle.
\end{align}
Here $U$ again stands for the Wilson line in the adjoint representation and $\phi_{\alpha\beta}^\lambda(\xt- \bt)$ is the wave function for the emission of a gluon with polarization $\lambda$  from a quark whose spin changes from $\alpha$ to $\beta$. Note that, as opposite to single inclusive cross section and DIS, the cross section can not be expressed in terms of the dipole operators (correlator of two Wilson lines) only. The different terms in equation \eqref{eq:dihadron-xs} have a clear physical interpretation: if the gluon is emitted before the interaction with the target, both the quark and the gluon interact with the target at coordinates $\xt$ and $\bt$ (and in the complex conjugate amplitude at $\xt'$ and $\bt'$), and this scattering is described by function $S^{(4)}$. If the gluon is emitted after the interaction, only the quark is scattering at transverse coordinate $\zt$ ($\zt'$ in the conjugate). The operators $S^{(3)}$ correspond to interference between the two channels.

In addition to the quark-initiated dihadron production channel, there is also a possibility that the incoming parton from the probe is a gluon which splits to two gluons or quark-antiquark pair. The cross sections for these channels can be calculated similarly and are available in the literature~\cite{Dominguez:2011wm,Iancu:2013dta}. However, as we are here considering only production of two forward rapidity hadrons at RHIC, the Bjorken-$x$ for the probe is very large $\sim 0.5$ and the valence quark distribution dominates. If the same analysis is done in case of the LHC, the gluon channel should be taken into account. Currently, no data on dihadron correlation at forward rapidity at the LHC exists. The correlations are measured at midrapidity, but even though the Bjorken-$x$ of the target is there similar as in the forward RHIC kinematics, there are other diagrams in addition to the ones discussed above that are important in the LHC kinematics, see e.g. \re\cite{Dusling:2013oia}.

\section{Double parton scattering}

The expression for the dihadron production cross section, \eq\eqref{eq:dihadron-xs}, is infrared divergent due to a double parton scattering contribution (DPS) included in the calculation. In this picture, the DPS contribution corresponds to the case where the incoming quark emits a gluon with small transverse momentum long before the interaction takes place, and the two parton subsequently scatter independently off the target. As the initial transverse momenta of the partons are small, the DPS scattering cross section does not depend on the azimuthal angle difference $\Delta \varphi$ between the particles and contributes to the two-particle correlation measurement a $\Delta \varphi$ independent background.

To see that this contribution is indeed included in the calculation, let us consider the case where the quark and the gluon are far away from each other: $|\ut|=|\bt-\xt|\gg 1/\qs$, $|\ut'|=|\bt'-\xt'|\gg 1/\qs$. Note that the quark coordinate in the amplitude and in the complex conjugate can not be far away in order to have a non-zero cross section (and similarly for the gluon), thus $|\xt-\xt'|\sim |\bt-\bt'|\sim 1/\qs$. We call this case the DPS limit. 
Now the quark and the gluon are not correlated, and the expectation value in \eq\eqref{eq:s4} factorizes. Using the fact that the expectation values must be color singlets we obtain
\begin{multline}
\label{eq:s4dps}
S^{(4)}(\bt,\xt,\bt',\xt') 
\underset{\textrm{DPS}}{\approx}
S^{(4)}_\textrm{DPS}(\bt,\xt,\bt',\xt') 
\\
\equiv \frac{2}{\nc^2-1} 
\left \langle  \tr \left( V(\bt) V^\dagger(\bt') t^d t^c \right) \right \rangle
\left \langle[U(\xt) U^\dagger(\xt')]^{cd} \right \rangle
\\
=
\frac{\nc^2}{\nc^2-1}  S(\bt,\bt')
\left(  S(\xt,\xt')^2 - \frac{1}{\nc^2}\right).
\end{multline}
Here we used the Fierz identities to express the adjoint representation Wilson lines in terms of the fundamental representation ones, see discussion in \se\ref{sec:multipoint}. This can be interpreted such that the quark scattering is given by the correlators $S(\bt,\bt')$, and
\begin{equation}
 \frac{\nc^2}{\nc^2-1}\left( S(\xt,\xt') - \frac{1}{\nc^2}\right) 
 \end{equation} 
 is the two point function in the adjoint representation normalized to unity at $\xt=\xt'$. Note that $S^{(4)}$ is finite in the DPS limit, and the other dipole operators $S^{(3)}$ and $S$ in \eq\eqref{eq:dihadron-xs} vanish.
 
In the massless limit the wave function product, that describes the gluon emission is
\begin{equation}
\label{eq:dihad-wavef-prod}
\sum_{\alpha\beta\lambda} \phi_{\alpha\beta}^{\lambda*}(\ut')
			  \phi_{\alpha\beta}^\lambda(\ut) = 
\frac{8\pi^2}{k^+} \frac{\ut \cdot \ut'}{|\ut|^2|\ut'|^2}(1+(1-z)^2).
\end{equation}
Changing the integration variables to $\ut$ and $\ut'$ in \eq \eqref{eq:dihadron-xs} it becomes clear that the integral over these transverse coordinates is infrared divergent, as the $S^{(4)}$ correlator is finite in the DPS limit, and a logarithmic divergence is obtained from the wave function product.

Physically, this divergence must be regulated by confinement scale physics in the projectile wave function. However, it is not calculable in our perturbative framework. Thus we will subtract the DPS contribution from the calculated dihadron production cross section \eq\eqref{eq:dihadron-xs} and calculate it separately by introducing a simple model for the double parton distribution function. In practice, we replace the correlator $S^{(4)}$ in \eq\eqref{eq:dihadron-xs} by
\begin{multline}
\label{eq:s4sub}
	S^{(4)}_\text{sub}(\bt,\xt,\bt',\xt') = S^{(4)}(\bt,\xt,\bt',\xt') \\
	- \theta\left( |\xt-\bt|-\frac{1}{\lqcd} \right)\theta\left( |\xt'-\bt'|-\frac{1}{\lqcd} \right) S^{(4)}_\text{DPS}(\bt,\xt,\bt',\xt'),
\end{multline}
whith $S^{(4)}_\text{DPS}$ defined in \eq\eqref{eq:s4dps}. Note that with this subtraction the integral in \eq\eqref{eq:dihadron-xs} vanishes in the DPS limit. It was shown explicitly in \paper\cite{Lappi:2012nh} that without this subtraction the cross section \eqref{eq:dihadron-xs} explicitly reduces to a product of quark and gluon scattering matrices and a piece that can be identified as the double parton distribution function in the DPS limit.

To calculate the subtracted double parton scattering contribution we first notice that in deuteron-gold collisions there are different processes contributing to the $\Delta \phi$ independent background.
First, we can have a contribution where the two partons are taken from the same proton (or neutron), described by the double parton distribution function DPDF. In this case, a kinematical constraint $x_1+x_2<1$, where $x_1$ and $x_2$ are the momentum fractions of the partons, must be satisfied. This is implemented by modelling  the double parton distribution function (DPDF) following \re\cite{Strikman:2010bg}:
\begin{equation}
 	D_{ij}^{(1)}(x_i,x_j,Q^2) = \frac{1}{2}\left[ f_i(x_i) f_j\left(\frac{x_j}{1-x_i}\right) + f_i\left(\frac{x_i}{1-x_j}\right)f_j(x_j)\right].
 \end{equation} 
 Here $i$ and $j$ denote parton species and the single particle PDFs are evaluated at scale $Q^2$. 
 
The second DPS contribution involves taking one parton from the neutron and the other one from the proton. In this case the same kinematical constraint does not apply, and for the deuteron we can write the DPDF as
\begin{equation}
	D_{ij}^{(2)}(x_i,x_j,Q^2) = \left[ f_i^p(x_i)f_j^n(x_j) + f_i^n(x_i) f_j^p(x_j)\right].
\end{equation}
Here $f^p$ is the proton PDF and $f^n$ is the same for the neutron. When proton-nucleus collisions are considered, this contribution is not included. 

\section{Evaluating the multi-point correlators}
\label{sec:multipoint}
In order to calculate the cross section \eqref{eq:dihadron-xs} (with the subtracted DPS contribution from \eq\eqref{eq:s4sub}) one has to be able to evaluate the higher point functions $S^{(3)}$ and $S^{(4)}$ written in \eqs\eqref{eq:s3} and \eqref{eq:s4}.

Using the Fierz identities 
\begin{align}
	[V(\xt)]_{ij}][V^\dagger(\xt)]_{kl} &= \frac{1}{\nc} \delta_{il}\delta_{jk} + 2U^{cd}(\xt) t^c_{il}t^d_{kj} \\
	t_{ij}^ct_{kl}^c = \frac{1}{2} \delta_{il}\delta_{jk} - \frac{1}{2\nc} \delta_{ij}\delta_{kl}
\end{align}
one can show that $U^{cd}(\xt) = 2\tr [V^\dagger(\xt)t^cV(\xt)t^d]$. Using this result, the adjoint representation dipoles can be expressed in terms of the fundamental dipoles and the higher point functions can be written as
\begin{multline}
\label{eq:s4-fierz}
	S^{(4)}(\bt,\xt,\bt',\xt') = \frac{\nc}{2\cf} \Bigg\{ -\frac{1}{\nc^2} S(\bt,\bt') 
	  \\
	+  S(\xt,\xt') 	
	 \frac{1}{\nc} \left\langle  \tr \left( V(\bt)V^\dagger(\bt') V(\xt')V^\dagger(\xt)\right)
	  \right\rangle \Bigg\}
\end{multline}
and
%\begin{multline}
%\label{eq:s3-fierz}
%	S^{(3)}(\bt,\xt,\bt') = \frac{\nc}{2\cf} \Bigg\{ \left \langle \frac{1}{\nc} \tr \left( V^\dagger(\bt')V(\xt)\right) \frac{1}{\nc} \tr \left( V(\bt)V^\dagger(\xt) \right) \right. \\
%	\quad - \left. \frac{1}{\nc^3} \tr \left( V^\dagger(\bt')V(\bt) \right) \right\rangle \Bigg\}.
%\end{multline}
\begin{equation}
\label{eq:s3-fierz}
	S^{(3)}(\bt,\xt,\bt') = \frac{\nc}{2\cf} \Bigg\{ S(\bt',\xt) S(\bt,\xt) - \frac{1}{\nc^2} S(\bt',\bt) \Bigg\}
\end{equation}
Note that here we have assumed that the expectation value of a product of traces factorizes into a product of expectation values. The three-point function $S^{(3)}$ can be calculated using only the dipole operator \eqref{eq:n-def} that can be obtained by solving the BK equation. On the other hand, for the four-point function an additional correlator of four Wilson lines, known as the quadrupole 
\begin{equation}
	Q(\xt,\yt,\ut,\vt) = \frac{1}{\nc} \left \langle \tr V(\xt)V^\dagger(\yt)V(\ut)V^\dagger(\vt) \right \rangle
\end{equation}
is needed. 

To calculate the quadrupole operator (and its energy evolution) one would in principle have to solve the JIMWLK evolution equation. As this would be numerically a formidable task, we apply the so called Gaussian approximation as presented in \re\cite{Dominguez:2011wm}, which allows one to write any higher point function in terms of the dipole operators alone. It was shown in \re\cite{Dumitru:2011vk} that the Gaussian approximation for the quadrupole is close to the solution obtained by solving the full JIMWLK equation. Using the Gaussian approximation for the quadrupole and taking the large-$\nc$ limit we obtain
\begin{multline}
\label{eq:s4-gaussian}
S^{(4)}(\bt,\xt,\bt',\xt') = S(\xt,\xt') \Big[ S(\bt,\xt)S(\xt',\bt') \\
-\frac{F(\bt,\xt,\xt',\bt')}{F(\bt,\xt',\xt,\bt')} \left[ S(\bt,\xt)S(\xt',\bt') - S(\bt,\bt')S(\xt',\xt)\right] \Big],
\end{multline}
where the auxiliary function is
\begin{equation}
	\frac{F(\bt,\xt,\xt',\bt')}{F(\bt,\xt',\xt,\bt')} = \frac{\ln S(\bt,\xt') - \ln S(\bt,\bt') + \ln S(\xt,\bt') - \ln S(\xt,\xt')
	}{\ln S(\bt,\xt) - \ln S(\bt,\bt') + \ln S(\xt',\bt') - \ln S(\xt',\xt)}.
\end{equation}

In previous phenomenological calculations~\cite{Marquet:2007vb,Albacete:2010pg} before \paper\cite{Lappi:2012nh} only a so called ``naive large-$\nc$ limit'' for the four point function was used, which would give
\begin{equation}
	S^{(4)}(\bt,\xt,\bt',\xt') = S(\xt,\bt)S(\xt',\bt') S(\xt,\xt'),
\end{equation}
which is the same as the first term in the Gaussian approximation \eqref{eq:s4-gaussian}. Following \re\cite{Dominguez:2011wm} we refer to this term as ``elastic''. The second term in \eqref{eq:s4-gaussian} is called ``inelastic'', and it is exactly this term that remains finite in the DPS limit making the integral in \eq\eqref{eq:dihadron-xs} infrared divergent. 
Because the inelastic part is not included in previous works~\cite{Marquet:2007vb,Albacete:2010pg}, the infrared-divergent DPS contribution was not found previously.
The two-particle correlations were also calculated in \re\cite{Stasto:2011ru} by taking the so called back-to-back correlation limit where the requirement of having large transverse momenta suppresses the DPS contribution.

%This allows us to write the cross section as {\bf tämän voisi kirjoittaa muuttujanvaihdon jälkeen}
%\begin{multline}
%\frac{\der \sigma}{\der^3k\der^3 q} = \as \cf \delta(p^+-k^+-q^+) \int \frac{\der^2 \xt}{(2\pi)^2} \frac{\der^2 \xt'}{(2\pi)^2} \frac{\der^2 \bt}{(2\pi)^2} \frac{\der^2 \bt'}{(2\pi)^2} e^{ik_T\cdot(\xt'-\xt)} e^{i(\qt-\pt)\cdot(\bt'-\bt)} \\
%\quad\times \sum_{\alpha\beta \lambda} \phi_{\alpha\beta}^{\lambda*}(\xt'-\bt')\phi_{\alpha\beta}^\lambda(\xt-\bt) \frac{\nc}{2\cf} \Big\{ S(\xt-\xt')Q(\bt,\bt',\xt',\xt) \\
% - S(\xt-\bt)S(\xt - \zt') - S(\xt' -\zt)S(\xt'-\bt') 
% + \frac{2\cf}{\nc} S(\zt-\zt') \\
%  + \frac{1}{\nc^2} \left[ S(\zt-\bt') + S(\zt'-\bt) - S(\bt-\bt') \right] \Big\} .
%\end{multline}

\section{Numerical results}

\begin{figure}[tb]
\centering
	\begin{minipage}[t]{0.5\textwidth}
	\includegraphics[width=1.0\textwidth]{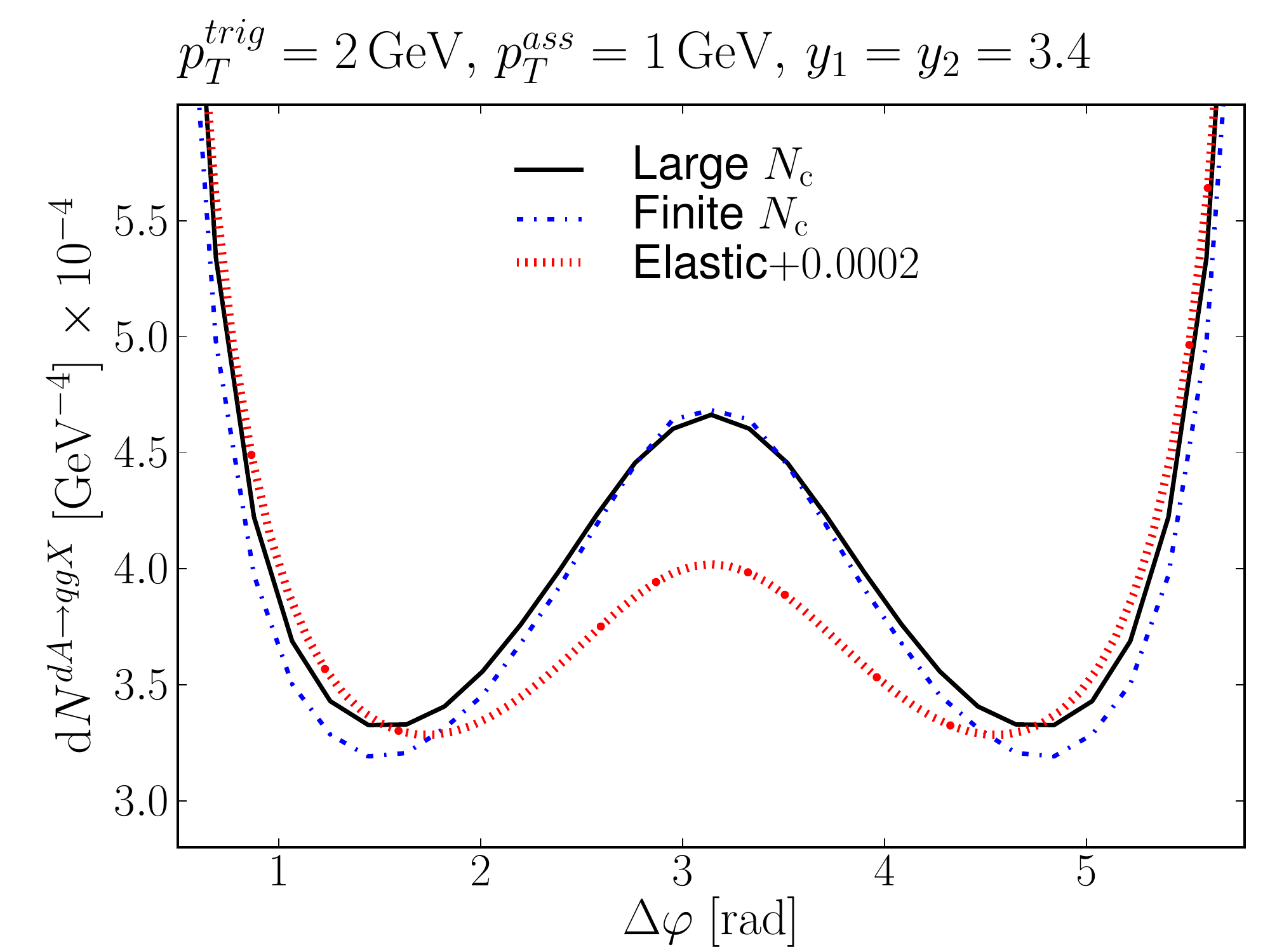}
	\caption{
	The quark-gluon  parton level azimuthal correlation in dAu collisions
at forward RHIC kinematics. A $\Delta \varphi$-independent pedestal has been added to
the ``elastic'' approximation for purposes of visualization. 
	Figure from \paper \cite{Lappi:2012nh}.	
	}
	\label{fig:parton_finite_large_nc}
	\end{minipage}%
	~
	\begin{minipage}[t]{0.5\textwidth}
	\includegraphics[width=\textwidth]{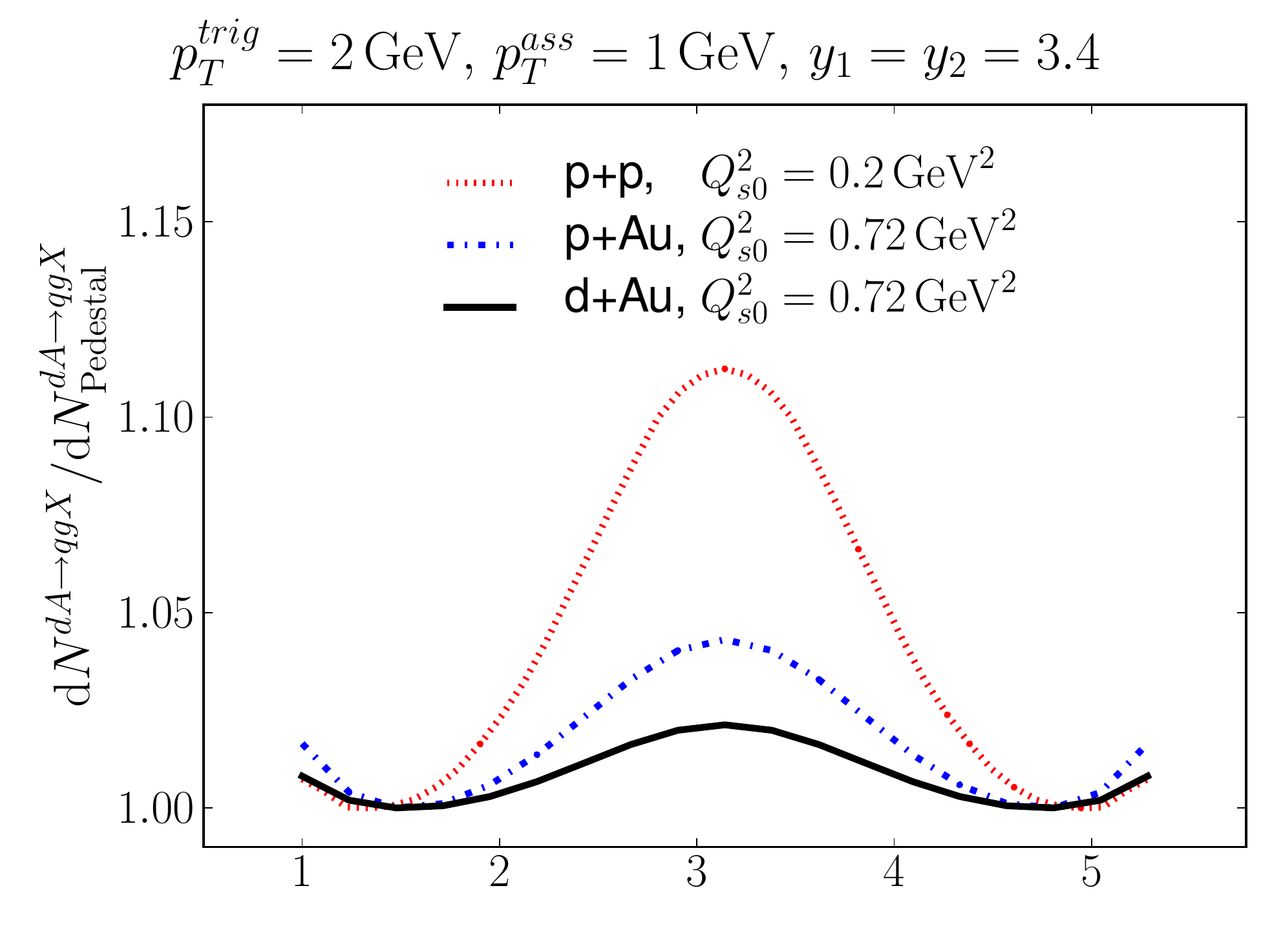}
	\caption{
	The quark-gluon  parton level azimuthal correlation at forward RHIC kinematics
in proton-proton, proton-nucleus and deuteron-nucleus
collisions normalized by the pedestal contribution.
Shown is only the large $\nc$ result.	
	Figure from \paper \cite{Lappi:2012nh}.	
	}
	\label{fig:parton_qs}
	\end{minipage}
\end{figure}

\begin{figure}[tb]
\centering
	\begin{minipage}[t]{0.5\textwidth}
	\includegraphics[width=\textwidth]{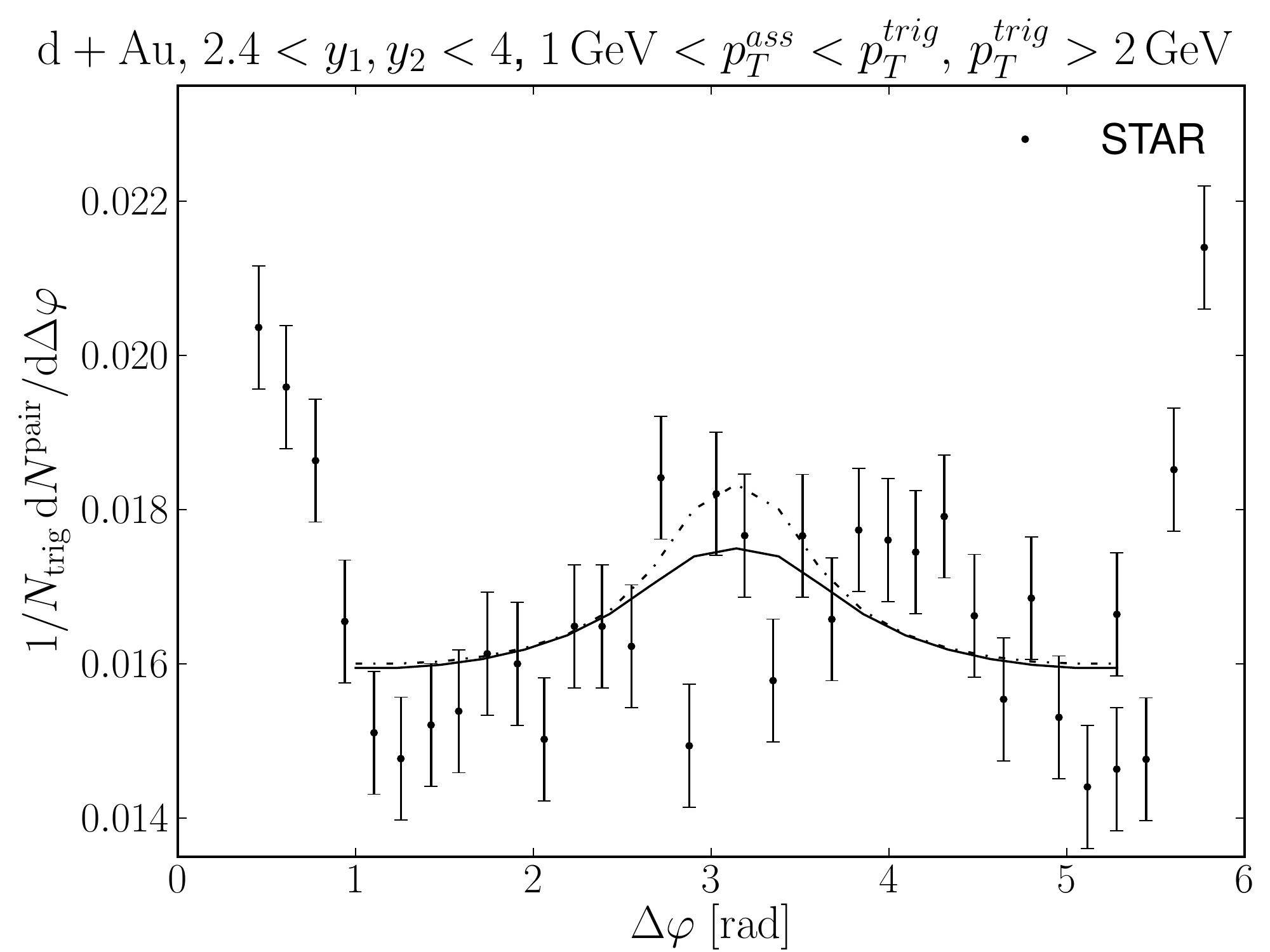}
	\caption{
	The $\pi^0$ azimuthal correlation compared to the 
preliminary STAR~\cite{Braidot:2011zj} result. 
The initial saturation scales are $\qso^2=1.51 \gev^2$ (solid line) and
$\qso^2=0.72 \gev^2$ (dashed line). Figure from \paper \cite{Lappi:2012nh}.	
	}
	\label{fig:rpa_cp_star}
	\end{minipage}%
	~
	\begin{minipage}[t]{0.5\textwidth}
	\includegraphics[width=1.0\textwidth]{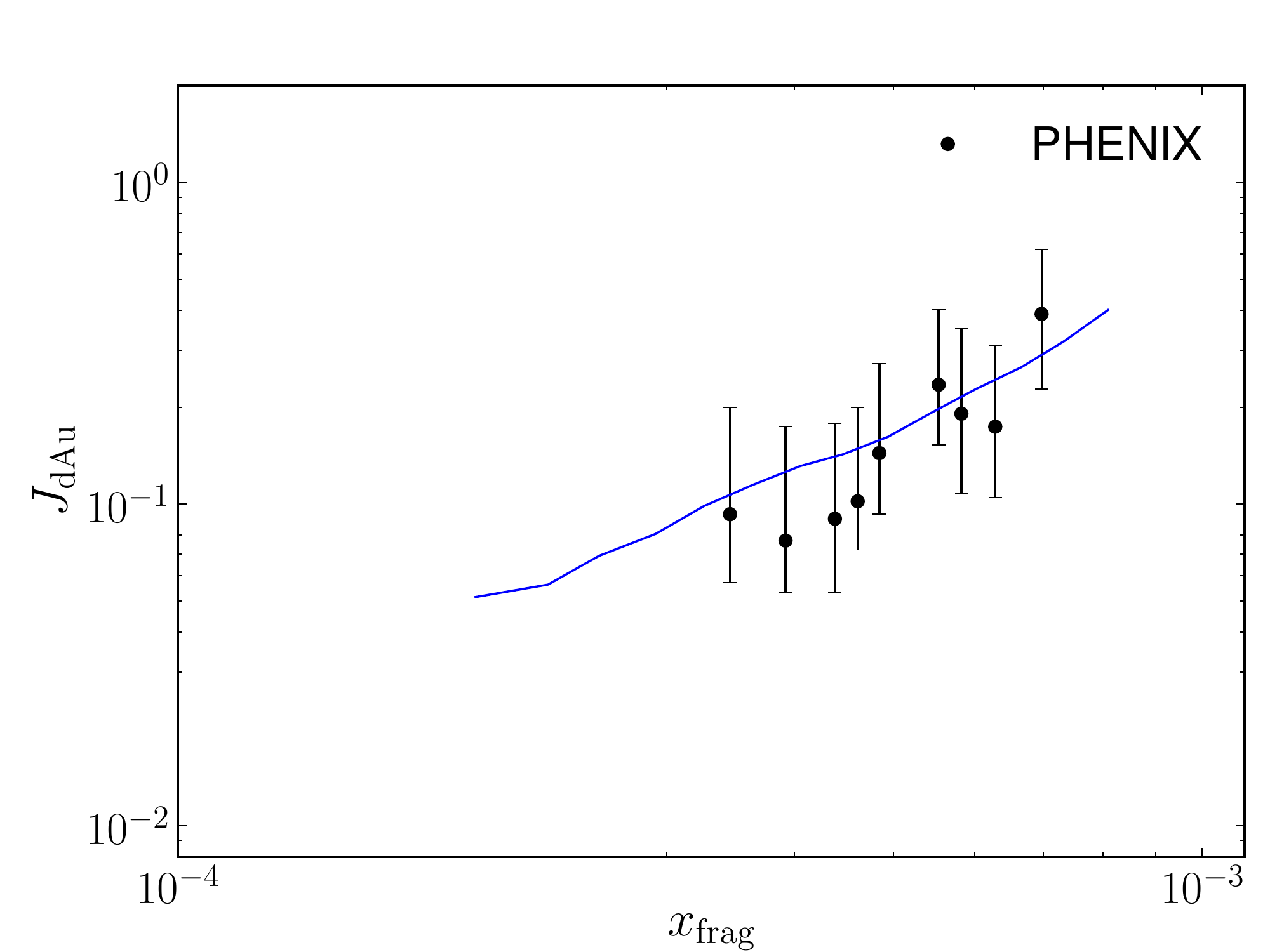}
	\caption{
	Integrated yield under the away side peak in central dAu collision
divided by the corresponding yield in pp compared
to the forward rapidity part of the PHENIX data~\cite{Adare:2011sc}. Figure from \paper \cite{Lappi:2012nh}.	
	}
	\label{fig:jdau}
	\end{minipage}	
\end{figure}

Let us then present our numerical results for the dihadron correlations at RHIC from \paper\cite{Lappi:2012nh}.
The DPS-subtracted cross section \eqref{eq:dihadron-xs}, where the quadrupole operator is evaluated using the Gaussian approximation discussed in the previous Section, can be evaluated once the dipole scattering amplitude $N(\rt,x)$ is known. In the calculation we use a BK evolved dipole amplitude. For simplicity, and because this work was done before \paper\cite{Lappi:2013zma} where we fit the dipole amplitude to the HERA data, the calculation does not use the best fit results for the initial condition of the BK evolution. Instead, we just use the MV model \eqref{eq:mv} with the initial saturation scale $\qso=0.2\gev^2$ for the proton, and scale the saturation scale when changing the proton to a nucleus.

First we show in \fig\ref{fig:parton_finite_large_nc} the effect of the ``inelastic'' contribution to the dihadron production cross section (without the DPS contribution). In order to get an estimate for the $\nc$ suppressed corrections we show both large-$\nc$ and finite-$\nc$ results for the cross section. For comparison, the result obtained by using only the elastic part of the $S^{(4)}$ correlator (see \eq\eqref{eq:s4-gaussian}) is shown. Including the inelastic contribution increases the away side peak by a factor $\sim 2$ (if the $\Delta \varphi$ independent pedestal is subtracted). Thus, inclusion of both elastic and inelastic contributions is essential for a meaningful comparison with the experimental data.

To study the difference between proton-nucleus, deuteron-nucleus and proton-proton collisions we show in \fig\ref{fig:parton_qs} the two-parton production cross section with DPS contribution included divided by the minimum of the $\Delta \varphi$ distribution. Now the larger DPS contribution makes the away side peak smaller in deuteron-gold collisions compared to the case where a proton is the probe. The disappearance of the away side peak when $\qs$ is increased (target is changed from a proton to a heavy nucleus) can be clearly seen from the figure. This is a solid prediction from the CGC calculation, and the observation of the depletion of the peak in experimental data supports the concept of having gluons with larger transverse momenta (of the order of $\qs$) in the heavy nucleus than in the proton.

To compare with the experimental data we show in \fig\ref{fig:rpa_cp_star} the two-particle production yield normalized by the number of trigger particles, known as the conditional yield. The calculation is done with two different values for the initial saturation scale $\qso^2$ of the nucleus calculated by scaling the proton saturation scale $0.2\gev^2$ by the number of binary collisions $N_\text{bin}^{pA}$ in proton-nucleus collisions. For the minimum bias collisions we obtain $\qso^2=0.72\gev^2$ and for $0-20\%$ centrality class the saturation scale is $\qso^2=1.51\gev^2$. The agreement with the STAR data~\cite{Braidot:2011zj} is reasonably good taking into account the large experimental uncertainties. Note that the pedestal contribution is fixed by hand in order to compare the $\Delta \varphi$ dependence. As the calculation seems to slightly underestimate the peak height, it is clear that using only the elastic contribution when evaluating $S^{(4)}$ would make the agreement with the data worse\footnote{A good description with the STAR data by using only the elastic contribution was obtained in \re\cite{Albacete:2010pg}. This agreement was due to a numerical error in the calculation.}.

The PHENIX collaboration has measured the nuclear modification factor to the integrated conditional yield under the away side peak. We calculate this by integrating the area under the peak in dAu collisions and divide the result by the same area in proton-proton scattering.
The ratio of areas under the away side peak, $J_{dAu}$, is compared with the PHENIX data~\cite{Adare:2011sc} in \fig\ref{fig:jdau}. The experimental data is measured as a function of
\begin{equation}
	x_\text{frag} = \frac{\langle \ptrig \rangle e^{-\langle \eta_1 \rangle} + \langle \pass \rangle e^{-\langle \eta_2 \rangle}}{\sqrt{s}},
\end{equation}
which, at parton level, has an interpretation as the Bjorken-$x$ of the target in the event.  We calculate this quantity numerically by calculating the ratio of the conditional yields in proton-proton and deuteron-gold collisions with different hadron momenta and rapidity, and then average the obtained $J_{dAu}$ values in each $x_\text{frag}$ bin. The suppression increases as $x_\text{frag}$ decreases, as expected from the saturation picture, and the theory calculation is in good agreement with the PHENIX data. On the other hand, as shown in \paper \cite{Lappi:2012nh}, the numerical results slightly underestimate the PHENIX data for the conditional yield, especially in case of the proton-proton collisions. When calculating the conditional yield ratios uncertainties in e.g. single inclusive baseline calculations partially cancel, which makes the $J_{dAu}$ calculation perhaps more solid than the results for the conditional yields.

\begin{table}
\begin{tabular}{|l|l||r|r|}
\hline
Data & $\pt$ range & pedestal & exp  \\
\hline\hline
PHENIX pp & $1.1\gev  < \ptrig < 1.6\gev$ &  0.04 & 0.095\\ 
\hline
PHENIX pp & $1.6\gev  < \ptrig < 2.0\gev$ &  0.02 & 0.079\\ 
\hline
PHENIX dAu & $1.1\gev  < \ptrig < 1.6\gev$ &  0.10 & 0.176\\ 
\hline
PHENIX dAu & $1.6\gev  < \ptrig < 2.0\gev$ &  0.08 & 0.163\\ 
\hline
STAR dAu & $2\gev  < \ptrig,\ 1\gev < \pass < \ptrig$ &  0.02 & 0.0145\\ 
\hline
\end{tabular}
\caption{Calculated estimates for the pedestal height compared to
the experimental values. In PHENIX results the units are $\gev^{-1}$.
The dAu values are for central
collisions where $\qso^2=1.51\gev^2$. Table from \paper \cite{Lappi:2012nh}.
}
\label{tab:ped}
\end{table}

Finally we calculate an estimate for the $\Delta \varphi$ independent pedestal contribution by summing the DPS contribution and the $\Delta \varphi$ independent part of the dihadron production cross section \eqref{eq:dihadron-xs}. The results are shown in \tab\ref{tab:ped} and are compared with the experimental data. As our model to calculate the DPS contribution is quite rough and there are theoretical uncertainties also related to the calculation of the single inclusive spectra, we do not expect to get a perfect match with the data, the results being off by roughly a factor of $2$.

To conclude, we emphasize that the disappearance of the away side peak in central deuteron-gold collisions compared with the proton-proton ones is naturally described within the CGC framework both qualitatively and (at least semi) quantitatively. In the CGC calculations the inclusion of the ``inelastic contribution'' is essential, and it is also necessary to correctly separate the double parton scattering contribution.

A measurement of the correlation of two forward rapidity particles at the LHC would provide interesting additional set of data. The advantage of the LHC is that with significantly higher center-of-mass energy the two-particle production takes place fare away from the kinematical boundary, in contrast to the RHIC kinematics. This would require an inclusion of the gluon channel in the calculation, as the projectile proton would also be probed at relatively small $x$. This channel is calculated in \res\cite{Dominguez:2011wm,Iancu:2013dta}, but no numerical predictions exist so far. 
%Note that even tough the kinematics is similar at forward rapidity RHIC and central rapidity LHC dihadron production, the calculation presented here can not be applied to describe central rapidity LHC data as there are other diagrams that are important in this kinematics but neglected in the calculation, see \re\cite{Dusling:2013oia}.

In addition to correlations of two hadrons, a powerful probe of QCD dynamics could be photon-hadron correlations at forward rapidities. The photon could be used to fix the parton level kinematics, in contrast to dihadron production where the hadronization procedure smears the process. The second advantage would be that no higher-point functions are needed to evaluate the cross section, which makes the theory calculation more solid as one does not have to use e.g. Gaussian approximation to approximate the quadrupole operator.
For CGC predictions of photon-hadron correlations, the reader is referred to \res\cite{JalilianMarian:2012bd,Rezaeian:2012wa}.

\chapter{Conclusions and outlook}
\label{ch:outlook}
The work presented this thesis shows, when combined with the other works cited in the text, that the Color Glass Condensate is in agreement with the currently available experimental data. The picture can also be systematically improved in the future.

Particle production cross sections calculated for inclusive and exclusive processes in \papers\cite{Lappi:2012nh,Lappi:2013am,Lappi:2013zma} are compatible with the available experimental data. The precise combined HERA deep inelastic scattering measurements are in agreement with the CGC calculations; taking the non-perturbative input from the DIS measurements, it is possible to obtain a good description of single particle production in proton-proton and proton-nucleus collisions (\paper\cite{Lappi:2013zma} and \ch\ref{ch:sinc}). However, in perturbative QCD calculations the next to leading order corrections to the cross section are known to be large and so far we have only done calculations at leading order. It is therefore an important goal for CGC phenomenology to calculate the single particle spectra and nuclear suppression factors at next to leading order accuracy. First steps in this direction have been taken, as discussed in \ch\ref{ch:sinc}.

In addition to single particle spectra, two-particle correlations can be used to obtain more detailed information of the strong color fields of the nucleus at small-$x$. The fact that the RHIC forward dihadron correlation measurements are naturally explained qualitatively, and even semi-quantitatively, in the CGC framework (as discussed in \paper\cite{Lappi:2012nh} and in \ch\ref{ch:dihad}) strongly suggest that we are able to see the gluon saturation already at RHIC energies. Measuring the correlation at the LHC as well would be extremely interesting, since at the LHC we would be far away from the kinematical boundary, in contrast to RHIC. A much cleaner process would be photon-hadron correlations, and if an electron-ion collider will be built in the future, it will also open many new possibilities for correlation measurements. 

Exclusive vector meson production, as discussed in \ch\ref{ch:ddis}, can be used as a direct probe of the gluon distribution. If an electron-ion collider will be realized, diffractive production of vector mesons can be used to probe gluon densities and density fluctuations in the nucleus, giving access to spatial distribution of gluons as well. Towards this goal, we have calculated cross sections for incoherent diffractive vector meson production in \paper\cite{Lappi:2010dd} and studied the centrality selection in \paper\cite{Lappi:2014foa}. Our proposal for the centrality estimator is the multiplicity of the ``ballistic'' protons produced in the scattering process. Inclusion of a more detailed description for the nuclear breakup has a potential to make this estimator more accurate. Incoherent and coherent diffraction can also be studied with the current experimental facilities in ultraperipheral heavy ion collisions, as discussed in \se\ref{sec:upc}. In \paper\cite{Lappi:2013am} we emphasized the importance of the simultaneous description of both coherent and incoherent vector meson production when comparing the theory calculations with the data. An important next step for the CGC calculations in this field is to figure out how to do these calculations properly with a BK evolved dipole amplitude and simultaneously describe production of different mesons. 

An important ingredient in the next to leading order CGC phenomenology will be a small-$x$ evolution equation at next to leading order accuracy. Towards this goal, we have made a detailed analysis of the NLO  BK equation (see \se\ref{sec:nlobk} and \paper\cite{Lappi:2015fma}) and proposed a method to include running coupling corrections in the JIMWLK equation (\se\ref{sec:jimwlk}, \paper\cite{Lappi:2012vw}). The NLO BK equation is shown to be unstable when solved with phenomenologically relevant initial conditions, which suggests that the equation would require a proper resummation of higher order contributions before it is ready for applications.

Signatures of the saturation phenomena described by the Color Glass Condensate are seen in the experimental data and the CGC calculations are in qualitative (and in many cases quantitative) agreement with the measurements. Taking the theory to next to leading order accuracy, while simultaneously comparing the saturation physics calculations with new interesting measurements from current and especially future accelerators, is an important and compelling task for the community.

\bibliography{../refs}

\end{document}